\documentclass[11pt,a4paper,british]{article}

\newcommand{\blind}{0}

\newcommand{\BW}{0}

\usepackage{scalerel,tikz,xcolor,hyperref} 

\definecolor{lime}{HTML}{A6CE39}
\DeclareRobustCommand{\orcidicon}{%
    \thinspace\scalerel*{\includegraphics{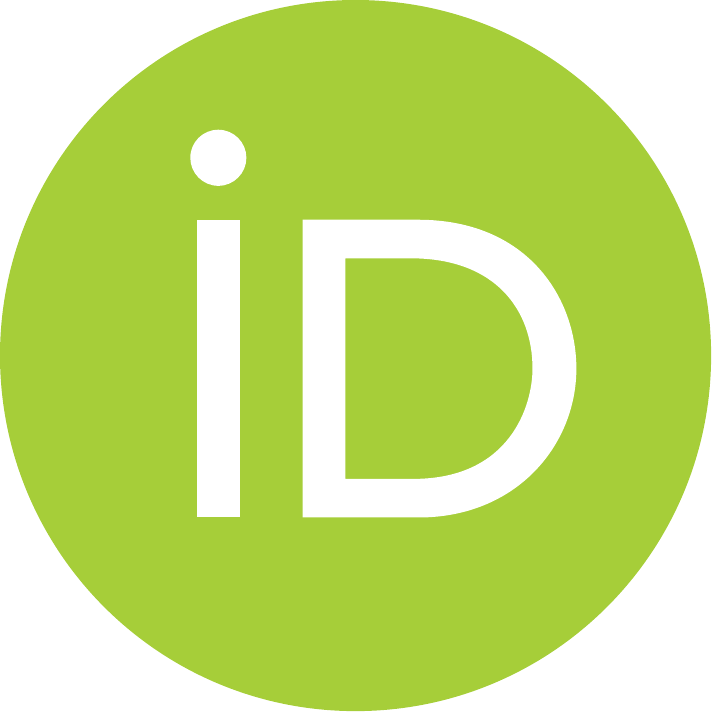}}{A}\!\! 
}

\foreach \x in {A, ..., Z}{%
	\expandafter\xdef\csname orcid\x\endcsname{\noexpand\href{https://orcid.org/\csname orcidauthor\x\endcsname}{\noexpand\orcidicon}}
}

\usepackage[T1]{fontenc}
\usepackage[latin9]{inputenc}
\usepackage{lmodern}
\usepackage{graphicx}
\usepackage{amsthm}
\usepackage{babel}
\usepackage{bm}

\usepackage{epstopdf}
\DeclareGraphicsExtensions{.eps,.pdf}
\if1\BW
	\graphicspath{{./Figures/Black_White/BW_}}
\else
	\graphicspath{{./Figures/Color/}}
\fi

\usepackage{authblk,bibentry,natbib}
\usepackage{mathrsfs}
\usepackage[titletoc,title]{appendix}
\usepackage{amsmath}
\usepackage{amssymb}
\usepackage{setspace}
\usepackage{float}
\usepackage[labelformat=simple]{subcaption}
\usepackage{hhline}
\usepackage{booktabs}
\usepackage{mathtools}
\usepackage{hyperref}
\usepackage{cleveref}
\usepackage{nameref}
\usepackage{thmtools}
\usepackage{pdflscape}
\usepackage[stable]{footmisc}
\usepackage{relsize}
\usepackage{enumitem}
\usepackage{tabularx}
\usepackage{bbm}
\usepackage{datetime}
\usepackage[ruled,vlined,linesnumbered]{algorithm2e}

\usepackage{tensor}

\usepackage{colonequals}

\usepackage{textcomp} 
\usepackage{chngcntr} 

\usetikzlibrary{matrix}
\usetikzlibrary{calc, arrows, positioning, arrows.meta, matrix}
\usetikzlibrary{shapes,decorations,arrows,calc,arrows.meta,fit,positioning}
\tikzset{
    auto,node distance =1 cm and 1 cm,semithick,
    state/.style ={circle, draw, minimum width = 1.2 cm},
    el/.style = {inner sep=2pt, align=left, sloped}
}
\usetikzlibrary{bayesnet}
\usetikzlibrary{arrows,decorations.pathreplacing}

\usepackage{geometry}
\PassOptionsToPackage{normalem}{ulem}
\usepackage{ulem}

\usepackage{color}
\definecolor{webbrown}{rgb}{0.65, 0.16, 0.16}
\definecolor{RoyalBlue}{rgb}{0.0, 0.14, 0.4}
\definecolor{webgreen}{rgb}{0.0, 0.5, 0.0}
\hypersetup{
colorlinks=true, linktocpage=true, pdfstartpage=1, pdfstartview=FitV,
breaklinks=true, pdfpagemode=UseNone, pageanchor=true, pdfpagemode=UseOutlines,
plainpages=false, bookmarksnumbered, bookmarksopen=true, bookmarksopenlevel=1,
hypertexnames=true, pdfhighlight=/O, urlcolor=webbrown, linkcolor=RoyalBlue, citecolor=webgreen}

\renewcommand{\theequation}{\thesection.\arabic{equation}}
\numberwithin{equation}{section}

\newtheorem*{assumption*}{Assumption}
\crefname{equation}{equation}{equations}
\Crefname{algocf}{Algorithm}{Algorithms}

\newcommand{\nonl}{\renewcommand{\nl}{\let\nl\oldnl}}

\newcommand*{\doi}[1]{\url{https://doi.org/#1}}

\newcommand{\double}{0}

\newcommand{\finalv}{0}

\newcommand{\tikzpdf}{0}

\if1\double
    
\fi

\begin{document}

\title{Frequency-Severity Experience Rating based on Latent Markovian Risk Profiles}
\if0\blind
    \author{\large Robert Matthijs Verschuren\orcidA{}\thanks{
    \if1\double
        \protect\linespread{1.5}\protect\selectfont 
    \fi
    Corresponding author: Amsterdam School of Economics, University of Amsterdam, Roetersstraat 11, 1018 WB, Amsterdam, The Netherlands. E-mail: \href{mailto:r.m.verschuren@uva.nl}
    {\tt r.m.verschuren@uva.nl}}}
    \affil{\it Amsterdam School of Economics, University of Amsterdam}
\fi
\if1\blind
    \author{}
\fi
\if0\finalv
    \date{This version is released on \usdate\today.}
\fi
\if1\finalv
    \date{}
\fi
\maketitle

\begin{abstract}

\noindent Bonus-Malus Systems traditionally consider a customer's number of claims irrespective of their sizes, even though these components are dependent in practice. We propose a novel joint experience rating approach based on latent Markovian risk profiles to allow for a positive or negative individual frequency-severity dependence. The latent profiles evolve over time in a Hidden Markov Model to capture updates in a customer's claims experience, making claim counts and sizes conditionally independent. We show that the resulting risk premia lead to a dynamic, claims experience-weighted mixture of standard credibility premia. The proposed approach is applied to a Dutch automobile insurance portfolio and identifies customer risk profiles with distinctive claiming behavior. These profiles, in turn, enable us to better distinguish between customer risks.

\end{abstract}


\section{Introduction} \label{Section1}

Bonus-Malus Systems (BMSs) are nowadays widely employed in automobile insurance to dynamically adjust a premium based on a customer's claims experience. The intuition behind these posterior ratemaking systems is that as we observe more claiming behavior, we learn more about the underlying risk profile. These systems are therefore a commercially attractive form of experience rating, in which we correct the prior premium for past claims to reflect our updated beliefs about a customer's risk profile. Moreover, they traditionally consider a customer's number of claims irrespective of their sizes and thus implicitly assume independence between the claim counts and sizes \citep{hey1970,denuit2007,boucher2014,verschuren2021}. Alternative Bayesian forms of experience rating typically depend only on the frequency component as well or consider the two components separately (see, e.g., \citet{denuit2004,buhlmann2005,mahmoudvand2009,bermudez2011,bermudez2017}).

While it makes commercial sense to only include a customer's number of claims in these experience rating systems, recent studies show that the independence assumption is actually violated in practice \citep{czado2012,kramer2013,frees2014,baumgartner2015,shi2015,garrido2016,park2018,lee2019kor,valdez2021}. Customers who rarely claim may, for instance, have much smaller claims than those who claim relatively often. They may even avoid reporting small claims when their size does not matter, leading to the phenomenon of bonus hunger. Several methods have therefore been proposed in the literature to capture this dependence. Conditional two-part models, for instance, allow the number of claims to enter the (average) claim severity regression as an additional covariate (see, e.g., \citet{gschlosl2007,frees2008,shi2015yang,shi2015,park2018,yang2019,lee2019}). Copula models, on the other hand, construct a joint distribution for the dependent claim counts and sizes by applying a copula function to their marginal distributions (see, e.g., \citet{czado2012,kramer2013,frees2016,shi2020,oh2021scand,ahn2021}). Alternatively, the two components may incorporate shared or bivariate random effects to account for the frequency-severity dependence (see, e.g., \citet{pinquet1997,dimakos2002,hernandez2009,baumgartner2015,lu2019,oh2020,oh2021ime}).

Despite the growing literature on frequency-severity models, the frequency-severity dependence has barely been considered in the context of experience rating \citep{park2018,oh2020}. One of the few examples is given by \citet{gomezdeniz2016} and \citet{gomezdeniz2018}, who distinguish between different types of claims and adjust the prior premium based on the observed number of claims of each type. However, they specify these claim types beforehand and do not infer their (relative) importance or severity from the data. From a modeling perspective, we would prefer to set these choices optimally based directly on our data through, for instance, a joint, frequency-severity mixture model. Mixture models have been considered previously in the literature to describe the total claim amount (see, e.g., \citet{miljkovic2016,hong2017,miljkovic2018,blostein2019,hu2019}). Numerous other insurance studies have also adopted mixture models to represent a separate frequency and/or severity component (see, e.g., \citet{brown2015,tzougas2014,tzougas2018,fung2019,pocuca2020}). Nonetheless, no study to date appears to have considered a joint, frequency-severity mixture model to capture the frequency-severity dependence observed in practice.

To bridge these gaps in the literature, this paper proposes a novel joint experience rating approach based on dynamic mixture models to allow for an individual frequency-severity dependence. More specifically, we let the mixture weights evolve over time to account for any updates in a customer's claims experience, as these may influence the frequency-severity dependence \citep{park2018}. These weights can therefore be interpreted as latent risk profiles and are described by a Hidden Markov Model (HMM). Conditional on these Markovian risk profiles, the claim frequencies and severities are modeled separately, in order to closely follow standard practices in the insurance industry, while unconditionally they can be either positively or negatively correlated. We show that the resulting risk premia lead to a dynamic, claims experience-weighted mixture of standard credibility premia, where the profile assignment probabilities account for all potential evolutions of a profile and include a customer's observed claims experience \textit{a posteriori}. We additionally allow a customer's risk characteristics to affect these credibility premia, and consequently their posterior weights, and efficiently estimate all (prior) parameters through an empirical Bayes version of the Expectation-Maximization (EM) algorithm. This approach allows us to identify data-driven customer risk profiles with distinctive claiming behavior and the flexibility to incorporate any observed frequency-severity dependence.

The remainder of this paper is organized as follows. In \Cref{Section2}, we derive the novel frequency-severity experience rating approach and show how to optimize it using EM. \Cref{Section3} describes the Dutch automobile insurance portfolio and the exact optimization procedure, whereas we elaborate on the results of applying this methodology in \Cref{Section4}. The final section of this paper concludes with a discussion of the most important findings and implications. 
\section{Theoretical framework} \label{Section2}

\subsection{Latent Markovian risk profiles} \label{Section2.1}

Within non-life insurance, risk premia are traditionally based on a prediction of the total claim amount relative to a policy's exposure to risk. Decompose the total claim amount $L_{i, t} \in \mathbb{R}_{+}$ for policyholder $i$ in period $t$, for instance, as the sum of $N_{i, t} \in \mathbb{N}$ claims, with $X_{i, t, n} \in \mathbb{R}_{+}$ the size of the $n$-th claim, $X_{i, t, 0} = 0$, and $e_{i, t} \in (0, 1]$ the deterministic exposure to risk of the policy. The risk premia $\pi_{i, t}$ are in this case given by
\begin{equation*}
    \pi_{i, t} = \mathbb{E}\left[\frac{L_{i, t}}{e_{i, t}}\right] = \mathbb{E}\left[\frac{1}{e_{i, t}} \mathbb{E}\left(\sum_{n = 0}^{N_{i, t}} X_{i, t, n} \Bigg| N_{i, t}\right)\right] = \mathbb{E}\left[\frac{N_{i, t}}{e_{i, t}} \mathbb{E}\left(X_{i, t, 1} \big| N_{i, t}\right)\right] = \mathbb{E}\left[\frac{N_{i, t}}{e_{i, t}}\right] \mathbb{E}\left[X_{i, t, 1}\right]
\end{equation*}
for $i = 1, \dots, M$ and $t = 1, \dots, T_{i}$, where the third and fourth equality follow from the classical assumption of identically distributed claim sizes $X_{i, t, n} \sim X_{i, t, 1}$ and independent claim counts $N_{i, t}$ and sizes $X_{i, t, 1}$, respectively \citep{scheel2013,garrido2016,verbelen2018,verschuren2021}.

While it is standard practice among non-life insurers to assume this independence, it is usually violated in practice. A more logical approach would be to relax this assumption of independence, but it turns out to be difficult to construct an explicit joint distribution for a discrete frequency component on the one hand and a continuous severity component on the other hand. As a compromise, we therefore assume that these two claim components remain independent, but only conditional on some latent process $Z_{i, t} \in \{1, \dots, K\}$, with $K \in \mathbb{N}_{+}$. By assuming independence conditional on this latent process, the theoretical framework closely follows the classical approach and allows us to intuitively account for dependencies between the number and size of claims. More specifically, under this conditional independence assumption, the risk premium is defined as a $K$-components mixture of risk premia, namely as
\begin{equation*}
    \pi_{i, t} = \sum_{j = 1}^{K} \mathbb{P}\left[Z_{i, t} = j\right] \mathbb{E}^{(j)}\!\left[\frac{N_{i, t}}{e_{i, t}}\right] \mathbb{E}^{(j)}\!\left[X_{i, t, 1}\right] \equalscolon \sum_{j = 1}^{K} \mathbb{P}\left[Z_{i, t} = j\right] \pi_{i, t}^{(j)}
\end{equation*}
with $\mathbb{E}^{(j)}\!\left[\cdot\right]$ the expectation conditional on $Z_{i, t} = j$ and $\pi_{i, t}^{(j)}$ the risk premium for type $j$. The $K$ conditional frequency and severity components can be modeled separately using standard actuarial techniques. However, unconditionally the two components can be either positively or negatively correlated since the covariance between the claim frequency and the claim severity equals
\begin{align*}
    \mathrm{Cov}&\!\left[\frac{N_{i, t}}{e_{i, t}},  X_{i, t, 1}\right] = \mathbb{E}\left[\frac{N_{i, t}}{e_{i, t}} X_{i, t, 1}\right] - \mathbb{E}\left[\frac{N_{i, t}}{e_{i, t}}\right] \mathbb{E}\left[X_{i, t, 1}\right] \\
    &\hspace{5pt}= \sum_{j = 1}^{K} \mathbb{P}\left[Z_{i, t} = j\right] \mathbb{E}^{(j)}\!\left[\frac{N_{i, t}}{e_{i, t}}\right] \mathbb{E}^{(j)}\!\left[X_{i, t, 1}\right] - \sum_{j = 1}^{K} \mathbb{P}\left[Z_{i, t} = j\right] \mathbb{E}^{(j)}\!\left[\frac{N_{i, t}}{e_{i, t}}\right] \sum_{h = 1}^{K} \mathbb{P}\left[Z_{i, t} = h\right] \mathbb{E}^{(h)}\!\left[X_{i, t, 1}\right] \\
    &\hspace{5pt}= \sum_{j = 1}^{K} \mathbb{P}\left[Z_{i, t} = j\right] \mathbb{E}^{(j)}\!\left[\frac{N_{i, t}}{e_{i, t}}\right] \left[\mathbb{E}^{(j)}\!\left(X_{i, t, 1}\right) - \sum_{h = 1}^{K} \mathbb{P}\left(Z_{i, t} = h\right) \mathbb{E}^{(h)}\!\left(X_{i, t, 1}\right) \right]
\end{align*}
To demonstrate this, suppose that we have only $K = 2$ types, with $\mathbb{P}\left[Z_{i, t} = 1\right] = w_{i, t}$ and $\mathbb{P}\left[Z_{i, t} = 2\right] = 1 - w_{i, t}$, and that $\lambda_{i, t} = \mathbb{E}^{(1)}\!\left[N_{i, t} / e_{i, t}\right]$ and $\mu_{i, t} = \mathbb{E}^{(1)}\!\left[X_{i, t, 1}\right]$ denote the expected claim frequency and severity for type $1$, respectively. Without loss of generality, we can then express the expectations for type $2$ as a scaled version of the expectations for type $1$, as $\mathbb{E}^{(2)}\!\left[N_{i, t} / e_{i, t}\right] = r_{i, t} \lambda_{i, t}$ and $\mathbb{E}^{(2)}\!\left[X_{i, t, 1}\right] = s_{i, t} \mu_{i, t}$ with $r_{i, t} > 0$ and $s_{i, t} > 0$. The resulting covariance is in this case given by
\begin{align}\label{eq2.1}
    \mathrm{Cov}\!\left[\frac{N_{i, t}}{e_{i, t}},  X_{i, t, 1}\right] &= w_{i, t} \lambda_{i, t} \left\{ \mu_{i, t} - \left[w_{i, t} \mu_{i, t} + \left(1 - w_{i, t}\right) s_{i, t} \mu_{i, t}\right]\right\} \nonumber\\
    &\quad+ \left(1 - w_{i, t}\right) r_{i, t} \lambda_{i, t} \left\{ s_{i, t} \mu_{i, t} - \left[w_{i, t} \mu_{i, t} + \left(1 - w_{i, t}\right) s_{i, t} \mu_{i, t}\right]\right\} \nonumber\\
    &= w_{i, t} \left(1 - w_{i, t}\right) \lambda_{i, t} \mu_{i, t} \left[s_{i, t} \left(r_{i, t} - 2\right) + 1\right]
\end{align}
and will be larger than zero when (i) $r_{i, t} > 2$ or (ii) $r_{i, t} < 2$ and $s_{i, t} < 1/(2 - r_{i, t})$ and smaller than zero when (iii) $r_{i, t} < 2$ and $s_{i, t} > 1/(2 - r_{i, t})$. 
Note that the covariance is always larger than zero when $r_{i, t} > 2$ since claims sizes cannot be negative and so $s_{i, t} > 0 > 1/(2 - r_{i, t})$ holds by definition. The claim counts and sizes in this example will thus be positively correlated when we expect at least twice as many claims (i) or at most twice as many and sufficiently large claims (ii) and negatively correlated in all other cases (iii). Moreover, this example shows that the type of dependence, namely positive or negative correlation, is independent of the assumption on the assignment probabilities $w_{i, t}$ but that these probabilities do of course have an effect on the degree of correlation.

While the above covariance captures the linear dependence between the claim counts and sizes, we can consider other, non-linear forms of dependence as well. There may, for instance, be a roughly monotone, rather than linear, relationship between the number and size of claims such that when the number of claims increases the claim size does not increase linearly but does still tend to increase. Two measures of monotone association, or non-linear dependence, in particular are commonly used in the insurance literature on dependence modeling, namely Spearman's rho and Kendall's tau (see, e.g., \citet{diers2012,kramer2013,joe2014,frees2016,yang2020,oh2021scand}). Spearman's rho \citep{spearman1904} and Kendall's tau \citep{kendall1938} are defined as
\begin{align*}
    \rho\!\left[\frac{n_{i, t}}{e_{i, t}},  x_{i, t, 1}\right] &= 12 \int_{0}^{1} \int_{0}^{1} C\!\left(u_{i, t}, v_{i, t}\right) \mathrm{d}u_{i, t} \, \mathrm{d}v_{i, t} - 3, \\
    \tau\!\left[\frac{n_{i, t}}{e_{i, t}},  x_{i, t, 1}\right] &= 4 \int_{0}^{1} \int_{0}^{1} C\!\left(u_{i, t}, v_{i, t}\right) \mathrm{d}C\!\left(u_{i, t}, v_{i, t}\right) - 1,
\end{align*}
respectively, with $u_{i, t} = \mathbb{P}\left[N_{i, t} \leq n_{i, t}/e_{i, t}\right]$ and $v_{i, t} = \mathbb{P}\left[X_{i, t, 1} \leq x_{i, t, 1}\right]$ and where both measures depend on the joint distribution of the claim counts and sizes through the copula $C(\cdot)$. In general, this copula function specifies how to find the joint cumulative distribution function from the (two) marginal distributions \citep{nelsen2006,shi2014,joe2014}. Although a copula has not been specified for the claim counts and sizes, we know from Sklar's theorem \citep{sklar1959} that for every joint distribution with known marginal distributions a certain copula exists \citep{nelsen2006,kramer2013,frees2016,yang2022}. The frequency-severity framework introduced in this paper, for instance, therefore implies the copula and copula density function given by
\begin{align*}
    C\!\left(u_{i, t}, v_{i, t}\right) & = \mathbb{P}\left[N_{i, t} \leq \frac{n_{i, t}}{e_{i, t}}, X_{i, t, 1} \leq x_{i, t, 1}\right] = \sum_{j = 1}^{K} \mathbb{P}\left[Z_{i, t} = j\right] \mathbb{P}^{(j)}\!\left[N_{i, t} \leq \frac{n_{i, t}}{e_{i, t}}\right] \mathbb{P}^{(j)}\!\left[X_{i, t, 1} \leq x_{i, t, 1}\right], \\
    c\!\left(u_{i, t}, v_{i, t}\right) &= \frac{\mathbb{P}\left[N_{i, t} = \frac{n_{i, t}}{e_{i, t}}, X_{i, t, 1} = x_{i, t, 1}\right]}{\mathbb{P}\left[N_{i, t} = \frac{n_{i, t}}{e_{i, t}}\right] \mathbb{P}\left[N_{i, t} = \frac{n_{i, t}}{e_{i, t}}\right]} \\
    &= \frac{\sum_{j = 1}^{K} \mathbb{P}\left[Z_{i, t} = j\right] \mathbb{P}^{(j)}\!\left[N_{i, t} = \frac{n_{i, t}}{e_{i, t}}\right] \mathbb{P}^{(j)}\!\left[X_{i, t, 1} = x_{i, t, 1}\right]}{\sum_{j = 1}^{K} \mathbb{P}\left[Z_{i, t} = j\right] \mathbb{P}^{(j)}\!\left[N_{i, t} = \frac{n_{i, t}}{e_{i, t}}\right] \sum_{h = 1}^{K} \mathbb{P}\left[Z_{i, t} = h\right] \mathbb{P}^{(h)}\!\left[X_{i, t, 1} = x_{i, t, 1}\right]},
\end{align*}
respectively. Even though the focus of this paper is not on copula models, we can use the copula and copula density function implied by our frequency-severity model to investigate whether any non-linear dependence exists between the claim counts and sizes. The insurance application in this paper will additionally show that the proposed model can lead to either a positive or negative Spearman's rho and Kendall's tau.

The latent processes thus provide a very flexible and data-driven way of capturing the frequency-severity dependence through a joint mixture model. As such, these processes can be interpreted as a customer's underlying risk profile, since they indicate which frequency-severity distribution best reflects the customer's claiming behavior. We additionally allow these latent risk profiles to evolve over time in a time-homogeneous HMM to account for updates in a customer's claims experience, as these updates may lead to changes in the frequency-severity dependence \citep{park2018}. The latent profiles $Z_{i, t}$ are categorically distributed in this HMM with probabilities $\bm{w}_{0}$ in the initial period and $\bm{w}_{Z_{i, t - 1}}$ in every successive period given the previous profile $Z_{i, t - 1}$, where the rows of the transition matrix $\bm{W} = \left(\bm{w}_{1}, \dots, \bm{w}_{K}\right)^{\prime}$ sum to one. The Markov property of these profiles enables us to learn more over time about future possible profile assignments from observations from the past. The marginal assignment probabilities in period $t > 0$ are characterized recursively by
\begin{equation*}
    \mathbb{P}\left[Z_{i, t} = j\right] = \sum_{h = 1}^{K} \mathbb{P}\left[Z_{i, t - 1} = h\right] \mathbb{P}\left[Z_{i, t} = j \big| Z_{i, t - 1} = h\right] = \left( \bm{w}_{0} \bm{W}^{t - 1}\right)_{j}
\end{equation*}
with $\left( \bm{w}_{0} \bm{W}^{t - 1}\right)_{j}$ the $j$-th element of the row vector $\bm{w}_{0}$ times the $(t - 1)$-th power of $\bm{W}$. We will show in the next section that this Markov property allows us to account for a customer's claims experience \textit{a posteriori}. The joint Markovian mixture approach proposed here thus strongly resembles the classical approach under independence, but enables a dynamic, posterior adjustment of the positive or negative individual frequency-severity dependence.

\subsection{Bayesian frequency-severity experience rating} \label{Section2.2}

While we require a HMM for the latent Markovian risk profiles, we can adopt classical pricing models for the frequency and severity components. Traditionally, a Generalized Linear Model (GLM) is employed to describe both the frequency and severity component \citep{nelder1972,lee1996}. The claim counts are often assumed to be Poisson distributed, whereas a Gamma distribution is usually assumed for the claim sizes \citep{ohlsson2010,tzougas2014,frees2016}. These distributions are typically considered as a baseline approach in insurance pricing. Within Bayesian experience rating it is customary to add further heterogeneity to the two components by including a stochastic heterogeneity factor in each distribution (see, e.g., \citet{klugman1992,buhlmann2005,ohlsson2010,antonio2012}). The intuition behind this factor is that as we observe more of a customer's claims history over time, we learn more about his/her claiming behavior and incorporate this information into our beliefs about a customer's riskiness. Hence, this approach allows us to improve our predictions based on information from the past while leading to a form of experience rating.

In this paper, the frequency and severity components will largely follow this hierarchical structure but now conditional on a HMM. We will show, and use in this paper, that by including the stochastic heterogeneity factors in a Poisson and Gamma distribution we effectively obtain a Negative Binomial (NB) and Generalized Beta of the second kind (GB2) distribution, which are more appropriate for the, in general, overdispersed claim counts and heavy tailed claim sizes, respectively. Suppose therefore that the claim counts are Poisson distributed with rate parameter $e_{i, t} \lambda_{i, t}^{(j)} U^{(j)}$ and that the claim sizes obey a Gamma distribution with shape parameter $\mu_{i, t}^{(j)}$ and rate parameter $\varphi^{(j)} / V^{(j)}$ for each risk profile $j$, conditional on $Z_{i, t} = j$ and heterogeneity factor $U^{(j)}$ or $V^{(j)}$, respectively. The individual and serial dependencies are now captured by a GLM with conditional mean predictors
\begin{align*}
    \mathbb{E}^{(j)}\!\left[N_{i, t} \big| U^{(j)}\right] &= e_{i, t} \lambda_{i, t}^{(j)} U^{(j)} = e_{i, t} \exp\!\left( \bm{A}_{i, t}^{\prime} \bm{\delta}^{(j)}_{A} \right) U^{(j)}, \\
    \mathbb{E}^{(j)}\!\left[X_{i, t, 1} \big| V^{(j)}\right] &= \frac{\mu_{i, t}^{(j)}}{\varphi^{(j)}} V^{(j)} = \exp\!\left( \bm{B}_{i, t}^{\prime} \bm{\delta}^{(j)}_{B} \right) V^{(j)},
\end{align*}
where $\bm{A}_{i, t}$ and $\bm{B}_{i, t}$ denote customer $i$'s risk characteristics in period $t$ with unknown deterministic effects $\bm{\delta}^{(j)}_{A}$ and $\bm{\delta}^{(j)}_{B}$ and where we assume a logarithmic link function for both GLMs to facilitate a common, multiplicative rate structure.

\begin{figure}[t!]
\centering
\if1\tikzpdf
    \vspace{-3pt}
\fi
\if0\tikzpdf
    \vspace{-9pt}
\fi
\scalebox{0.952}{
\if1\tikzpdf
    \includegraphics[width=1.044\textwidth]{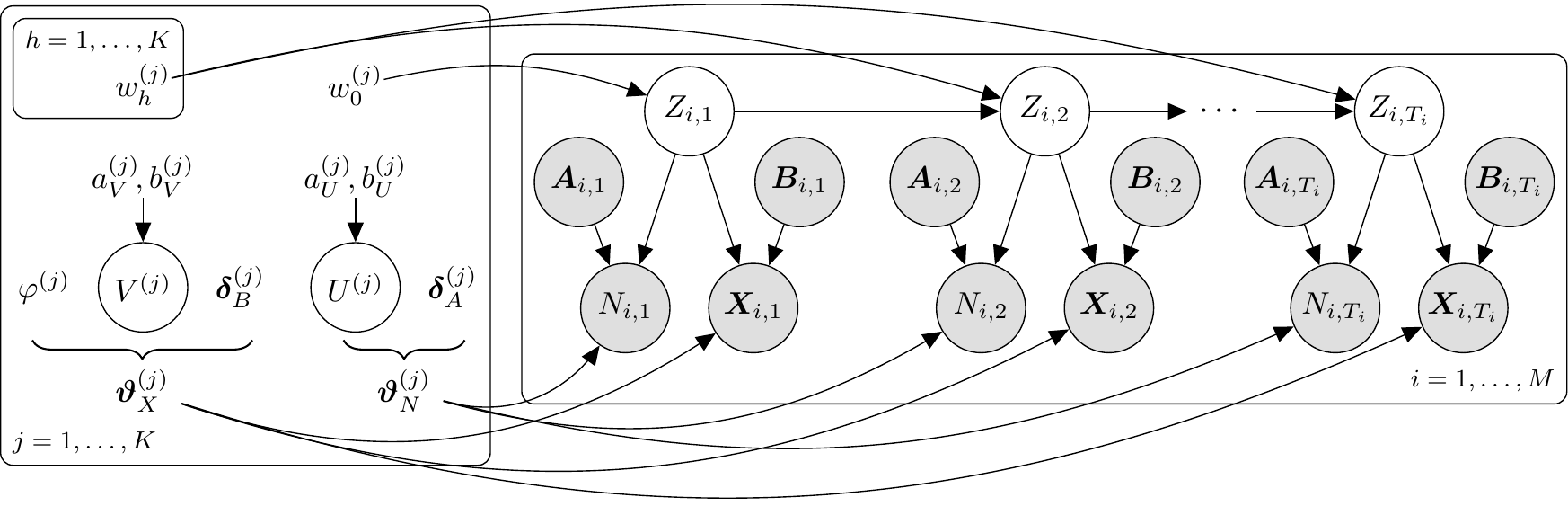}
\fi
\if0\tikzpdf
    \begin{tikzpicture}[decoration={brace, amplitude=6pt},latent/.append style={minimum size=10.1mm},obs/.append style={minimum size=10.1mm}]
    \node[latent] (Z1) {$Z_{i, 1}$};
    \node[latent] (Z2) [right =3.0cm of Z1] {$Z_{i, 2}$};
    \node (Z3) [right =1.1cm of Z2] {$\cdots$};
    \node[latent] (ZT) [right =1.1cm of Z3] {$Z_{i, T_{i}}$};
    \node[obs] (N1) [below left =1.5cm and 0cm of Z1] {$N_{i, 1}$};
    \node[obs] (A1) [above left =0.7cm and -0.2cm of N1] {$\bm{A}_{i, 1}$};
    \node[obs] (X1) [below right =1.5cm and 0cm of Z1] {$\bm{X}_{i, 1}$};
    \node[obs] (B1) [above right =0.7cm and -0.2cm of X1] {$\bm{B}_{i, 1}$};
    \node[obs] (N2) [below left =1.5cm and 0cm of Z2] {$N_{i, 2}$};
    \node[obs] (A2) [above left =0.7cm and -0.2cm of N2] {$\bm{A}_{i, 2}$};
    \node[obs] (X2) [below right =1.5cm and 0cm of Z2] {$\bm{X}_{i, 2}$};
    \node[obs] (B2) [above right =0.7cm and -0.2cm of X2] {$\bm{B}_{i, 2}$};
    \node[obs] (NT) [below left =1.5cm and 0cm of ZT] {$N_{i, T_{i}}$};
    \node[obs] (AT) [above left =0.7cm and -0.2cm of NT] {$\bm{A}_{i, T_{i}}$};
    \node[obs] (XT) [below right =1.5cm and 0cm of ZT] {$\bm{X}_{i, T_{i}}$};
    \node[obs] (BT) [above right =0.7cm and -0.2cm of XT] {$\bm{B}_{i, T_{i}}$};
    \edge {Z1} {N1, X1, Z2};
    \edge {A1} {N1};
    \edge {B1} {X1};
    \edge {Z2} {N2, X2, Z3};
    \edge {A2} {N2};
    \edge {B2} {X2};
    \edge {Z3} {ZT};
    \edge {ZT} {NT, XT};
    \edge {AT} {NT};
    \edge {BT} {XT};
    {\tikzset{plate caption/.append style={below left=5pt and 0pt of #1.south east}}
    \plate {i} {(Z1)(A1)(N1)(ZT)(BT)(XT)} {\scriptsize$i = 1, \dots, M$};}
    
    \node[const] (dA) [above left =-0.37cm and 1.3cm of N1] {\small$\bm{\delta}_{A}^{(j)}$} coordinate[below=0.3cm of dA.south] (theta_N);
    \node[latent] (U) [left =0.3cm of dA] {\small$U^{(j)}$};
    \node[const] (abU) [above =0.5cm of U] {\small$a_{U}^{(j)}\!, b_{U}^{(j)}$};
    \node [fit=(dA.south) (U.south),yshift = 0.05cm] (theta_N_j) {};              
    \draw [decorate,line width=0.6pt] (theta_N_j.south east) -- (theta_N_j.south west) node [midway, below=0.2cm] (theta_N) {\small$\bm{\vartheta}^{(j)}_{N}$};
    \node[const] (dB) [left =0.5cm of U] {\small$\bm{\delta}_{B}^{(j)}$} coordinate[below=0.3cm of dB.south] (theta_X);
    \node[latent] (V) [left =0.3cm of dB] {\small$V^{(j)}$};
    \node[const] (abV) [above =0.5cm of V] {\small$a_{V}^{(j)}\!, b_{V}^{(j)}$};
    \node[const] (varphi) [left =0.3cm of V] {\small$\varphi^{(j)}$};
    \node [fit=(dB.south) (V.south) (varphi.south),yshift = 0.05cm] (theta_X_j) {};
    \draw [decorate,line width=0.6pt] (theta_X_j.south east) -- (theta_X_j.south west) node [midway, below=0.2cm] (theta_X) {\small$\bm{\vartheta}^{(j)}_{X}$};
    \node[const] (w0) [above =0.55cm of abU] {\small$w_{0}^{(j)}$};
    \node[const] (W) [above =0.55cm of abV] {\small$w_{h}^{(j)}$};
    \path (w0) edge [bend left=16,->,>={triangle 45}] (Z1);
    \path (W) edge [bend left=15,->,>={triangle 45}] (Z2);
    \path (W) edge [bend left=14,->,>={triangle 45}] (ZT);
    \edge {abU} {U};
    \path (theta_N) edge [bend right=35,->,>={triangle 45}] (N1);
    \path (theta_N) edge [bend right=23,->,>={triangle 45}] (N2);
    \path (theta_N) edge [bend right=19,->,>={triangle 45}] (NT);
    \edge {abV} {V};
    \path (theta_X) edge [bend right=26,->,>={triangle 45}] (X1);
    \path (theta_X) edge [bend right=23,->,>={triangle 45}] (X2);
    \path (theta_X) edge [bend right=21,->,>={triangle 45}] (XT);
    {\tikzset{plate caption/.append style={above left=3pt and 0pt of #1.north east}} \plate {h} {(W)} {\scriptsize$h = 1, \dots, K$}}
    {\tikzset{plate caption/.append style={below right=2pt and 0pt of #1.south west}} \plate {theta_left} {(w0)(W)(U)(abU)(dA)(V)(abV)(dB)(varphi)(h)(theta_N)(theta_X)} {\scriptsize$j = 1, \dots, K$}}
    \end{tikzpicture}
\fi
}
\if1\tikzpdf
    \vspace{-22pt}
\fi
\if0\tikzpdf
    \vspace{-36pt}
\fi
\caption{A graphical model of the proposed Bayesian HMM for capturing the frequency-severity dependence. While shaded nodes represent observed (random) variables, clear nodes denote latent random variables.}
\label{F2.1}
\end{figure}

Even though these mean predictors account for a customer's risk characteristics, they typically still lead to ample residual heterogeneity. We can capture (part of) this individual heterogeneity by assuming independent prior distributions for the heterogeneity factors $U^{(j)}$ and $V^{(j)}$ in a Bayesian sense. As such, we consider a Gamma prior for $U^{(j)}$ and an Inverse-Gamma prior for $V^{(j)}$, with rate and scale parameters $\left(a^{(j)}_{U}, b^{(j)}_{U}\right)$ and $\left(a^{(j)}_{V}, b^{(j)}_{V}\right)$ for risk profile $j$, respectively. For the sake of generality we allow the parameters $b^{(j)}_{U}$ and $b^{(j)}_{V}$ to be different from $a^{(j)}_{U}$ and $a^{(j)}_{V} - 1$ for all $j$, respectively, but it is common to assume this in Bayesian experience rating to leave the prior risk premium unaffected. The resulting Bayesian HMM can be visually represented using plate notation by the graphical model in \Cref{F2.1}, which shows the dependencies between all latent and observed random variables and parameters. It depicts the time dependent structure of the latent risk profiles and how these influence the claim count and size distributions, and it demonstrates that at any given point in time $t$ it is sufficient to know only the latest latent risk profile $Z_{i, t}$ to have independence between the claim counts and sizes. Note that we do not assume prior distributions for the transition probabilities $\bm{w}_{h}$ of the latent risk profiles since we can only use information on past observables to price \textit{a posteriori}.

While these priors are common in Bayesian experience rating, they are typically motivated due to their conjugateness to the likelihoods. The posteriors are therefore available in closed-form as Gamma and Inverse-Gamma distributions with updated, posterior parameters $\Big(a^{(j)}_{U} + \sum_{\tau = 1}^{t - 1} N_{i, \tau}, b^{(j)}_{U} + \sum_{\tau = 1}^{t - 1} e_{i, \tau} \lambda^{(j)}_{i, \tau}\Big)$ and $\Big(a^{(j)}_{V} + \sum_{\tau = 1}^{t - 1} N_{i, \tau} \mu^{(j)}_{i, \tau}, b^{(j)}_{V} + \varphi^{(j)} \sum_{\tau = 1}^{t - 1} \sum_{n = 0}^{N_{i, \tau}} X_{i, \tau, n}\Big)$ for $U^{(j)}$ and $V^{(j)}$, respectively, given a claims history of $t - 1$ periods if the current latent risk profile is $Z_{i, t} = j$. Moreover, the marginal, or prior predictive, distributions for profile $j$ equal a NB distribution with $a^{(j)}_{U}$ successes and success probability $b^{(j)}_{U} / \left(b^{(j)}_{U} + e_{i, t} \lambda^{(j)}_{i, t}\right)$ for the claim counts and a GB2 distribution with shape parameters $\left(\mu^{(j)}_{i, t}, a^{(j)}_{V}, 1\right)$ and scale parameter $b^{(j)}_{V} / \varphi^{(j)}$ for the individual claim sizes. The posterior predictive distributions are obtained by simply replacing the prior parameters by their posterior analogue.

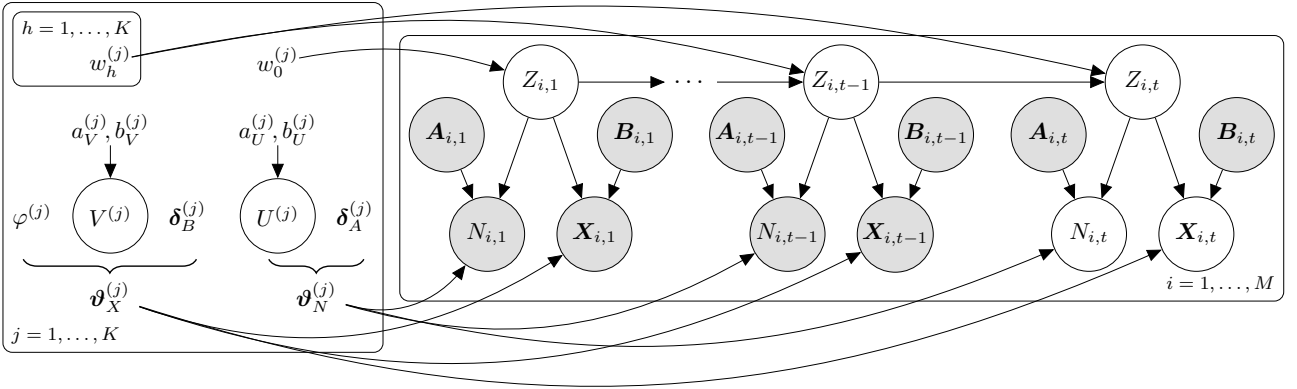
\begin{figure}[t!]
\centering
\if1\tikzpdf
    \vspace{1pt}
\fi
\if0\tikzpdf
    \vspace{-7pt}
\fi
\scalebox{0.873}{
\if1\tikzpdf
    \includegraphics[width=1.137\textwidth]{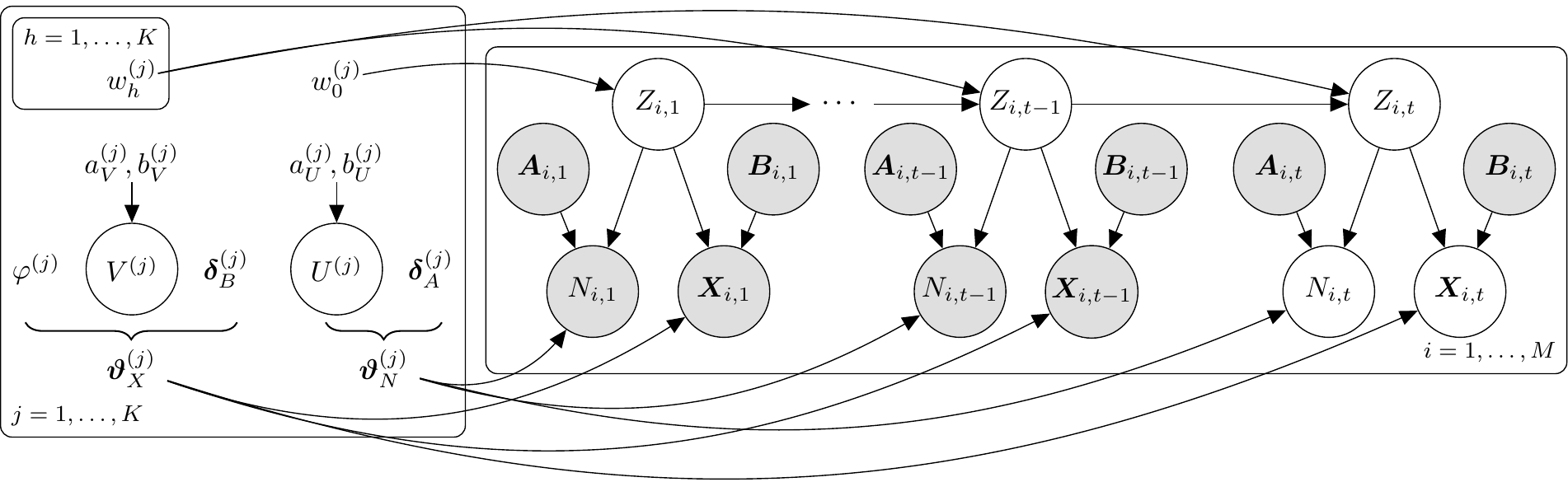}
\fi
\if0\tikzpdf
    \begin{tikzpicture}[decoration={brace, amplitude=6pt},latent/.append style={minimum size=11.3mm},obs/.append style={minimum size=11.3mm}]
    \node[latent] (Z1) {$Z_{i, 1}$};
    \node (Z2) [right =1.3cm of Z1] {$\cdots$};
    \node[latent] (Ztm1) [right =1.3cm of Z2] {$Z_{i, t - 1}$};
    \node[latent] (Zt) [right =3.4cm of Ztm1] {$Z_{i, t}$};
    \node[obs] (N1) [below left =1.5cm and 0cm of Z1] {$N_{i, 1}$};
    \node[obs] (A1) [above left =0.7cm and -0.2cm of N1] {$\bm{A}_{i, 1}$};
    \node[obs] (X1) [below right =1.5cm and 0cm of Z1] {$\bm{X}_{i, 1}$};
    \node[obs] (B1) [above right =0.7cm and -0.2cm of X1] {$\bm{B}_{i, 1}$};
    \node[obs] (Ntm1) [below left =1.5cm and 0cm of Ztm1] {$N_{i, t - 1}$};
    \node[obs] (Atm1) [above left =0.7cm and -0.2cm of Ntm1] {$\bm{A}_{i, t - 1}$};
    \node[obs] (Xtm1) [below right =1.5cm and 0cm of Ztm1] {$\bm{X}_{i, t - 1}$};
    \node[obs] (Btm1) [above right =0.7cm and -0.2cm of Xtm1] {$\bm{B}_{i, t - 1}$};
    \node[latent] (Nt) [below left =1.5cm and 0cm of Zt] {$N_{i, t}$};
    \node[obs] (At) [above left =0.7cm and -0.2cm of Nt] {$\bm{A}_{i, t}$};
    \node[latent] (Xt) [below right =1.5cm and 0cm of Zt] {$\bm{X}_{i, t}$};
    \node[obs] (Bt) [above right =0.7cm and -0.2cm of Xt] {$\bm{B}_{i, t}$};
    \edge {Z1} {N1, X1, Z2};
    \edge {A1} {N1};
    \edge {B1} {X1};
    \edge {Z2} {Ztm1};
    \edge {Ztm1} {Ntm1, Xtm1, Zt};
    \edge {Atm1} {Ntm1};
    \edge {Btm1} {Xtm1};
    \edge {Zt} {Nt, Xt};
    \edge {At} {Nt};
    \edge {Bt} {Xt};
    {\tikzset{plate caption/.append style={below left=1pt and 0pt of #1.south east}}
    \plate {i} {(Z1)(A1)(N1)(Ztm1)(Ntm1)(Xtm1)(Zt)(Bt)(Xt)} {\scriptsize$i = 1, \dots, M$};}
    
    \node[const] (dA) [above left =-0.37cm and 1.3cm of N1] {\small$\bm{\delta}_{A}^{(j)}$} coordinate[below=0.3cm of dA.south] (theta_N);
    \node[latent] (U) [left =0.3cm of dA] {\small$U^{(j)}$};
    \node[const] (abU) [above =0.5cm of U] {\small$a_{U}^{(j)}\!, b_{U}^{(j)}$};
    \node [fit=(dA.south) (U.south),yshift = 0.05cm] (theta_N_j) {};              
    \draw [decorate,line width=0.6pt] (theta_N_j.south east) -- (theta_N_j.south west) node [midway, below=0.2cm] (theta_N) {\small$\bm{\vartheta}^{(j)}_{N}$};
    \node[const] (dB) [left =0.5cm of U] {\small$\bm{\delta}_{B}^{(j)}$} coordinate[below=0.3cm of dB.south] (theta_X);
    \node[latent] (V) [left =0.3cm of dB] {\small$V^{(j)}$};
    \node[const] (abV) [above =0.5cm of V] {\small$a_{V}^{(j)}\!, b_{V}^{(j)}$};
    \node[const] (varphi) [left =0.3cm of V] {\small$\varphi^{(j)}$};
    \node [fit=(dB.south) (V.south) (varphi.south),yshift = 0.05cm] (theta_X_j) {};
    \draw [decorate,line width=0.6pt] (theta_X_j.south east) -- (theta_X_j.south west) node [midway, below=0.2cm] (theta_X) {\small$\bm{\vartheta}^{(j)}_{X}$};
    \node[const] (w0) [above =0.55cm of abU] {\small$w_{0}^{(j)}$};
    \node[const] (W) [above =0.55cm of abV] {\small$w_{h}^{(j)}$};
    \path (w0) edge [bend left=14,->,>={triangle 45}] (Z1);
    \path (W) edge [bend left=13,->,>={triangle 45}] (Ztm1);
    \path (W) edge [bend left=12,->,>={triangle 45}] (Zt);
    \edge {abU} {U};
    \path (theta_N) edge [bend right=35,->,>={triangle 45}] (N1);
    \path (theta_N) edge [bend right=23,->,>={triangle 45}] (Ntm1);
    \path (theta_N) edge [bend right=19,->,>={triangle 45}] (Nt);
    \edge {abV} {V};
    \path (theta_X) edge [bend right=26,->,>={triangle 45}] (X1);
    \path (theta_X) edge [bend right=23,->,>={triangle 45}] (Xtm1);
    \path (theta_X) edge [bend right=21,->,>={triangle 45}] (Xt);
    {\tikzset{plate caption/.append style={above left=3pt and 0pt of #1.north east}} \plate {h} {(W)} {\scriptsize$h = 1, \dots, K$}}
    {\tikzset{plate caption/.append style={below right=2pt and 0pt of #1.south west}} \plate {theta_left} {(w0)(W)(U)(abU)(dA)(V)(abV)(dB)(varphi)(h)(theta_N)(theta_X)} {\scriptsize$j = 1, \dots, K$}}
\end{tikzpicture}
\fi
}
\if1\tikzpdf
    \vspace{-22pt}
\fi
\if0\tikzpdf
    \vspace{-37pt}
\fi
\caption{A graphical representation of the (past) information affecting the posterior risk premium in period $t > 1$ in the proposed Bayesian HMM. While shaded nodes represent observed (random) variables, clear nodes denote latent random variables.}
\label{F2.2}
\end{figure}

Given these posterior (predictive) distributions, we are able to derive a customer's individual risk premium. The collective, prior risk premium is, for instance, given by
\begin{equation*}
    \pi_{i, t} = \sum_{j = 1}^{K} \mathbb{P}\left[Z_{i, t} = j\right] \pi_{i, t}^{(j)} = \sum_{j = 1}^{K} \left( \bm{w}_{0} \bm{W}^{t - 1}\right)_{j} \exp\!\left(\bm{A}_{i, t}^{\prime} \bm{\delta}^{(j)}_{A} + \bm{B}_{i, t}^{\prime} \bm{\delta}^{(j)}_{B}\right) \frac{a^{(j)}_{U}}{b^{(j)}_{U}} \frac{b^{(j)}_{V}}{a^{(j)}_{V} - 1},
\end{equation*}
where $\left( \bm{w}_{0} \bm{W}^{t - 1}\right)_{j}$ reduces to $w^{(j)}_{0}$ in case of the initial period with $t = 1$. The Bayesian, posterior risk premium in period $t > 1$, on the other hand, is defined as
\begin{align*}
    \pi_{i, t}^{*} &= \sum_{j = 1}^{K} \mathbb{P}^{*}\!\left[Z_{i, t} = j\right] {\pi_{i, t}^{(j)}}^{*} = \sum_{j = 1}^{K} \mathbb{P}^{*}\!\left[Z_{i, t} = j\right] \pi_{i, t}^{(j)} \frac{\mathbb{E}^{*}\!\left[U^{(j)}\right]}{\mathbb{E}\left[U^{(j)}\right]} \frac{\mathbb{E}^{*}\!\left[V^{(j)}\right]}{\mathbb{E}\left[V^{(j)}\right]} \\
    &= \sum_{j = 1}^{K} \mathbb{P}^{*}\!\left[Z_{i, t} = j\right] \exp\!\left(\bm{A}_{i, t}^{\prime} \bm{\delta}^{(j)}_{A} + \bm{B}_{i, t}^{\prime} \bm{\delta}^{(j)}_{B}\right) \frac{a^{(j)}_{U} + \sum_{\tau = 1}^{t - 1} N_{i, \tau}}{b^{(j)}_{U} + \sum_{\tau = 1}^{t - 1} e_{i, \tau} \lambda^{(j)}_{i, \tau}} \frac{b^{(j)}_{V} + \varphi^{(j)} \sum_{\tau = 1}^{t - 1} \sum_{n = 0}^{N_{i, \tau}} X_{i, \tau, n}}{a^{(j)}_{V} + \sum_{\tau = 1}^{t - 1} N_{i, \tau} \mu^{(j)}_{i, \tau} - 1},
\end{align*}
where the asterisk indicates that we condition on the claims history observed before period $t$, namely on $\bm{N}_{i, 1:(t - 1)}$ and $\bm{X}_{i, 1:(t - 1)}$ with $\bm{X}_{i, \tau} = \left(X_{i, \tau, 1}, \dots, X_{i, \tau, N_{i, \tau}}\right)$ as shown in \Cref{F2.2}. More specifically, \Cref{F2.2} indicates the information that is observed and can be used to predict the (future) posterior risk premium in period $t$, whereas \Cref{F2.1} displays the (historical) information that is available to estimate the model parameters. Recall that we only condition on past values of $N_{i, \tau}$ and $X_{i, \tau, n}$ since the risk profiles $Z_{i, \tau}$ are latent by construction and that we must regard all possible sequences ending in $Z_{i, t}$ to determine the assignment probabilities. However, since these sequences now also include the observed claims history, the latent risk profiles are no longer time-homogeneous \textit{a posteriori}. The posterior assignment probabilities\vspace{-5pt}
\begin{equation*}
    \mathbb{P}^{*}\!\left[Z_{i, t} = j\right] = \frac{\sum_{h = 1}^{K} \alpha^{(h)}_{i, t - 1} \, \mathbb{P}\left[Z_{i, t} = j \big| Z_{i, t - 1} = h\right]}{\sum_{h = 1}^{K} \alpha^{(h)}_{i, t - 1}}
\end{equation*}
are therefore computationally easier to determine through the recursive forward probabilities
\begin{align*}
    \alpha^{(h)}_{i, t - 1} &= \mathbb{P}\left[ Z_{i, t - 1} = h, \bm{N}_{i, 1:(t - 1)}, \bm{X}_{i, 1:(t - 1)}\right] \\
    &= \left[\mathbb{P}^{(h)}\!\left(N_{i, t - 1}\right) \prod_{n = 0}^{N_{i, t - 1}} \mathbb{P}^{(h)}\!\left(X_{i, t - 1, n}\right)\right] \sum_{k = 1}^{K} \alpha^{(k)}_{i, t - 2} \, \mathbb{P}\left[Z_{i, t - 1} = h \big| Z_{i, t - 2} = k\right]
\end{align*}
with initial condition $\alpha^{(h)}_{i, 1} = \mathbb{P}\left[ Z_{i, 1} = h\right] \mathbb{P}^{(h)}\!\left[N_{i, 1}\right] \prod_{n = 0}^{N_{i, 1}} \mathbb{P}^{(h)}\!\left[X_{i, 1, n}\right]$,
transition probabilities $\mathbb{P}\left[Z_{i, 1} = h\right] \linebreak= w^{(h)}_{0}$ and $\mathbb{P}\left[Z_{i, t} = j \big| Z_{i, t - 1} = h\right] = w^{(j)}_{h}$ for $t > 1$ that are independent of a customer's claims history \textit{a priori}, and marginal probabilities
\begin{align*}
    \mathbb{P}^{(h)}\!\left[N_{i, t} = n_{i, t}\right] &= \frac{\Gamma\left(n_{i, t} + a^{(h)}_{U}\right)}{n_{i, t}! \, \Gamma\left(a^{(h)}_{U}\right)} \left(\frac{e_{i, t} \lambda^{(h)}_{i, t}}{b^{(h)}_{U} + e_{i, t} \lambda^{(h)}_{i, t}}\right)^{n_{i, t}} \left(\frac{b^{(h)}_{U}}{b^{(h)}_{U} + e_{i, t} \lambda^{(h)}_{i, t}}\right)^{a^{(h)}_{U}}, \\
    \mathbb{P}^{(h)}\!\left[X_{i, t, 1} = x_{i, t, 1}\right] &= \frac{\Gamma\left(\mu^{(h)}_{i, t} + a^{(h)}_{V}\right)}{\Gamma\left(\mu^{(h)}_{i, t}\right) \Gamma\left(a^{(h)}_{V}\right)} \frac{{b^{(h)}_{V}}^{a^{(h)}_{V}} {\varphi^{(h)}}^{\mu^{(h)}_{i, t}} x_{i, t, 1}^{\mu^{(h)}_{i, t} - 1}}{\left(b^{(h)}_{V} + \varphi^{(h)} x_{i, t, 1}\right)^{\mu^{(h)}_{i, t} + a^{(h)}_{V}}},
\end{align*}
where $\mathbb{P}^{(h)}\!\left[X_{i, t, 0}\right] = 1$ for every $t$. This implies that the assignment probabilities vary with a customer's risk characteristics and can actually account for a customer's claims experience \textit{a posteriori}.

The Bayesian premium thus updates the prior beliefs about a customer's risk profile based on the claims history and adds a Bonus-Malus correction for each profile. The transition rules of this Bayesian premium become more intuitive when we rewrite them in terms of classical credibility expressions as
\begin{align*}
    \pi_{i, t}^{*} &= \sum_{j = 1}^{K} \mathbb{P}^{*}\!\left[Z_{i, t} = j\right] \pi_{i, t}^{(j)} \left[\kappa^{(j)}_{U, t - 1} + \left(1 - \kappa^{(j)}_{U, t - 1}\right) \frac{\sum_{\tau = 1}^{t - 1} N_{i, \tau}}{\mathbb{E}^{(j)}\!\left(\sum_{\tau = 1}^{t - 1} N_{i, \tau}\right)}\right] \\
    &\quad \times \left[\kappa^{(j)}_{V, t - 1} + \left(1 - \kappa^{(j)}_{V, t - 1}\right) \frac{\sum_{\tau = 1}^{t - 1} \sum_{n = 0}^{N_{i, \tau}} X_{i, \tau, n}}{\mathbb{E}^{(j)}\!\left(\sum_{\tau = 1}^{t - 1} \sum_{n = 0}^{N_{i, \tau}} X_{i, \tau, n}\right)} \right],
\end{align*}
by recognizing that for every $j = 1, \dots, K$, with $\mathbb{V}^{(j)}\!\left[\cdot\right]$ the variance conditional on $Z_{i, t} = j$,

\vspace{-15pt}{\small\begin{align*}
    \kappa^{(j)}_{U, t - 1} &= \frac{b^{(j)}_{U}}{b^{(j)}_{U} + \sum_{\tau = 1}^{t - 1} e_{i, \tau} \lambda^{(j)}_{i, \tau}} = \frac{\mathbb{E}\left[\mathbb{V}^{(j)}\!\left(\sum_{\tau = 1}^{t - 1} N_{i, \tau} \big| U^{(j)}\right) \right]}{\mathbb{E}\left[\mathbb{V}^{(j)}\!\left(\sum_{\tau = 1}^{t - 1} N_{i, \tau} \big| U^{(j)}\right) \right] + \mathbb{V}\left[\mathbb{E}^{(j)}\!\left(\sum_{\tau = 1}^{t - 1} N_{i, \tau} \big| U^{(j)}\right) \right]}, \\
    \kappa^{(j)}_{V, t - 1} &= \frac{a^{(j)}_{V} - 1}{a^{(j)}_{V} - 1 + \sum_{\tau = 1}^{t - 1} N_{i, \tau} \mu^{(j)}_{i, \tau}} = \frac{\mathbb{E}\left[\mathbb{V}^{(j)}\!\left(\sum_{\tau = 1}^{t - 1} \sum_{n = 0}^{N_{i, \tau}} X_{i, \tau, n} \big| V^{(j)}\right)\right]}{\mathbb{E}\left[\mathbb{V}^{(j)}\!\left(\sum_{\tau = 1}^{t - 1} \sum_{n = 0}^{N_{i, \tau}} X_{i, \tau, n} \big| V^{(j)}\right)\right] + \mathbb{V}\left[\mathbb{E}^{(j)}\!\left(\sum_{\tau = 1}^{t - 1} \sum_{n = 0}^{N_{i, \tau}} X_{i, \tau, n} \big| V^{(j)}\right)\right]},
\end{align*}}\hspace{-4.1pt}
conditional on $\bm{N}_{i, 1:(t - 1)}$ for the claim sizes for the sake of brevity \citep{buhlmann2005,ohlsson2010}. As expected, the credibility weights $\kappa^{(j)}_{U, t - 1}$ and $\kappa^{(j)}_{V, t - 1}$ equal one if no prior claims experience is available, i.e., when $t = 1$. Once we observe more claims history, and hence $t$ and $\sum_{\tau = 1}^{t - 1} e_{i, \tau}$ increase, $\kappa^{(j)}_{U, t - 1}$ approaches zero and more weight is given to the observed individual number of claims relative to the prior expectation. We find that $\kappa^{(j)}_{V, t - 1}$ only starts to approach zero and to give more weight to the total claim amount of a customer once we observe more non-zero claims, since we can only learn better how much a customer will claim if the number of claims is non-zero and increases. The posterior updating of the risk premium is thus consistent with the usual transition rules in BMSs for claim frequencies and severities, although the effect of a transition may be amplified since our beliefs about a customer's risk profile, or $\mathbb{P}^{*}\!\left[Z_{i, t} = j\right]$, are updated as well (see, e.g., \citet{denuit2007,kaas2008,gomezdeniz2016,oh2020,verschuren2021}). Hence, the Bayesian risk premium constitutes a Bonus-Malus corrected version of the prior risk premium based on a customer's posterior observed number and size of claims.

Using these credibility expressions we can also investigate what effect the assumptions imposed in this section have on the model's dependence structure. Consider the example from the previous section, for instance, where we considered only $K = 2$ risk profiles and the covariance can be expressed similar to \Cref{eq2.1}. When denoting the expected claim frequency and severity for the first profile as $\lambda_{i, t}$ and $\mu_{i, t}$, respectively, the expectations for the second profile can again be expressed as a scaled version of the ones for the first profile, namely as $r_{i, t} \lambda_{i, t}$ and $s_{i, t} \mu_{i, t}$. However, the model structure introduced in this section allows us to determine how these scaling factors are specified. More specifically, \textit{a priori} these factors equal
\begin{align*}
    r_{i, t} &= \exp\!\left[\bm{A}_{i, t}^{\prime} \left(\bm{\delta}^{(2)}_{A} - \bm{\delta}^{(1)}_{A} \right)\right] \frac{a^{(2)}_{U} \big/\, b^{(2)}_{U}}{a^{(1)}_{U} \big/\, b^{(1)}_{U}}, \\
    s_{i, t} &= \exp\!\left[\bm{B}_{i, t}^{\prime} \left(\bm{\delta}^{(2)}_{B} - \bm{\delta}^{(1)}_{B} \right)\right] \frac{b^{(2)}_{V} \Big/ \left(a^{(2)}_{V} - 1\right)}{b^{(1)}_{V} \Big/ \left(a^{(1)}_{V} - 1\right)},
\end{align*}
whereas \textit{a posteriori} they are given by
\begingroup
\allowdisplaybreaks
\begin{align*}
    r^{*}_{i, t} &= \exp\!\left[\bm{A}_{i, t}^{\prime} \left(\bm{\delta}^{(2)}_{A} - \bm{\delta}^{(1)}_{A} \right)\right] \frac{\left(a^{(2)}_{U} + \sum_{\tau = 1}^{t - 1} N_{i, \tau}\right) \Big/ \left(b^{(2)}_{U} + \sum_{\tau = 1}^{t - 1}e_{i, \tau} \lambda_{i, \tau}^{(2)}\right)}{\left(a^{(1)}_{U} + \sum_{\tau = 1}^{t - 1} N_{i, \tau}\right) \Big/ \left(b^{(1)}_{U} + \sum_{\tau = 1}^{t - 1}e_{i, \tau} \lambda_{i, \tau}^{(1)}\right)} \\
    &= r_{i, t} \, \frac{\kappa^{(2)}_{U, t - 1} + \left(1 - \kappa^{(2)}_{U, t - 1}\right) \sum_{\tau = 1}^{t - 1} N_{i, \tau} \Big/ \mathbb{E}^{(2)}\!\left(\sum_{\tau = 1}^{t - 1} N_{i, \tau}\right)}{\kappa^{(1)}_{U, t - 1} + \left(1 - \kappa^{(1)}_{U, t - 1}\right) \sum_{\tau = 1}^{t - 1} N_{i, \tau} \Big/ \mathbb{E}^{(1)}\!\left(\sum_{\tau = 1}^{t - 1} N_{i, \tau}\right)}, \\
    s^{*}_{i, t} &= \exp\!\left[\bm{B}_{i, t}^{\prime} \left(\bm{\delta}^{(2)}_{B} - \bm{\delta}^{(1)}_{B} \right)\right] \frac{\left(b^{(2)}_{V} + \varphi^{(2)} \sum_{\tau = 1}^{t - 1} \sum_{n = 0}^{N_{i, \tau}} X_{i, \tau, n}\right) \Big/ \left(a^{(2)}_{V} + \sum_{\tau = 1}^{t - 1} N_{i, \tau} \mu^{(2)}_{i, \tau} - 1\right)}{\left(b^{(1)}_{V} + \varphi^{(1)} \sum_{\tau = 1}^{t - 1} \sum_{n = 0}^{N_{i, \tau}} X_{i, \tau, n}\right) \Big/ \left(a^{(1)}_{V} + \sum_{\tau = 1}^{t - 1} N_{i, \tau} \mu^{(1)}_{i, \tau} - 1\right)} \\
    &= s_{i, t} \, \frac{\kappa^{(2)}_{V, t - 1} + \left(1 - \kappa^{(2)}_{V, t - 1}\right) \sum_{\tau = 1}^{t - 1} \sum_{n = 0}^{N_{i, \tau}} X_{i, \tau, n} \Big/ \mathbb{E}^{(2)}\!\left(\sum_{\tau = 1}^{t - 1} \sum_{n = 0}^{N_{i, \tau}} X_{i, \tau, n}\right)}{\kappa^{(1)}_{V, t - 1} + \left(1 - \kappa^{(1)}_{V, t - 1}\right) \sum_{\tau = 1}^{t - 1} \sum_{n = 0}^{N_{i, \tau}} X_{i, \tau, n} \Big/ \mathbb{E}^{(1)}\!\left(\sum_{\tau = 1}^{t - 1} \sum_{n = 0}^{N_{i, \tau}} X_{i, \tau, n}\right)}.
\end{align*}
\endgroup
From these expressions we can see that the posterior scaling factors represent a Bonus-Malus corrected version of their prior values as well and that the model structure does not impose any restrictions on their magnitude. Moreover, since the covariance can be expressed similar to \Cref{eq2.1}, the assignment probabilities $w_{i, t} = \mathbb{P}\left[Z_{i, t} = j\right]$ and $w^{*}_{i, t} = \mathbb{P}^{*}\!\left[Z_{i, t} = j\right]$ again affect only the degree of correlation and not the type of dependence, although their effect on the degree of correlation may differ \textit{a priori} from \textit{a posteriori}. In other words, the model structure can still allow for a positive or negative correlation between the claim counts and sizes, as the insurance application in this paper will show and similar to the Spearman's rho and Kendall's tau.

\subsection{Empirical Bayes Expectation-Maximization} \label{Section2.3}

While the previous section elaborated on the Bayesian frequency-severity experience rating framework, we did not yet specify how to estimate the latent Markovian risk profiles in this approach. Ordinarily, we would maximize the observed log-likelihood of the model in a frequentist setting to obtain estimates of the parameters $\bm{\vartheta} = \left(\bm{\vartheta}_{N}, \bm{\vartheta}_{X}, \bm{\vartheta}_{Z}\right) = \left[\left(\bm{\delta}_{A}, \bm{a}_{U}, \bm{b}_{U}\right), \left(\bm{\varphi}, \bm{\delta}_{B}, \bm{a}_{V}, \bm{b}_{V}\right), \left(\bm{w}_{0}, \bm{W}\right)\right]$. A full Bayesian analysis, on the other hand, does not solely rely on point estimates but generates the entire posterior distribution of the parameters. The empirical Bayes approach serves as a compromise between the standard frequentist setting, in which we can perform Maximum Likelihood Estimation (MLE) through an EM algorithm without any prior distributions, and the full Bayesian analysis based on posterior sampling \citep{robbins1964,norberg1980,casella1985,mashayekhi2002,denuit2007}. Even though the full Bayesian methodology may provide a more flexible structure of our estimation problem, the empirical Bayes approach is more consistent with standard practices in, and hence easier to implement by, the insurance industry and will therefore be used in this paper. This empirical Bayes approach considers the observed log-likelihood after marginalizing over $\bm{U}$ and $\bm{V}$ and estimates the prior parameters of these heterogeneity factors by standard MLE \citep{buhlmann2005,gomezdeniz2016,gomezdeniz2018,klugman2019}. However, the marginal observed, or incomplete, log-likelihood is given in terms of the observed data $\bm{N} = \left(\bm{N}_{1, 1:T_{1}}, \dots, \bm{N}_{M, 1:T_{M}}\right)$ and $\bm{X} = \left(\bm{X}_{1, 1:T_{1}}, \dots, \bm{X}_{M, 1:T_{M}}\right)$ by
\begin{equation*}
    \ell\left(\bm{\vartheta}\big| \bm{N}, \bm{X}\right) = \sum_{i = 1}^{M} \sum_{t = 1}^{T_{i}} \ln\left[ \sum_{j = 1}^{K} \mathbb{P}\left(Z_{i, t} = j \big| \bm{\vartheta}_{Z}\right) \mathbb{P}^{(j)}\!\left(N_{i, t} | \bm{\vartheta}_{N}\right) \prod_{n = 0}^{N_{i, t}} \mathbb{P}^{(j)}\!\left(X_{i, t, n} | \bm{\vartheta}_{X}\right)\right],
\end{equation*}
which includes a summation over all possible latent risk profiles $j$ in the logarithmic transformation.

Since the incomplete log-likelihood can be challenging to optimize directly, it is customary in mixture models to optimize this indirectly instead through the complete log-likelihood (see, e.g., \citet{miljkovic2016,blostein2019,pocuca2020}). The corresponding marginal complete log-likelihood is based on both the observed and unobserved data and is defined as
\begin{align*}
    \ell\left(\bm{\vartheta}\big| \bm{N}, \bm{X}, \bm{Z}\right) &= \sum_{i = 1}^{M} \left\{\sum_{j = 1}^{K} \mathbbm{1}\left[Z_{i, 1} = j\right] \ln\left[\mathbb{P}\left(Z_{i, 1} = j \big| \bm{w}_{0}\right)\right]\right. \\ 
    &\quad + \sum_{t = 2}^{T_{i}} \sum_{h = 1}^{K} \sum_{j = 1}^{K} \mathbbm{1}\left[Z_{i, t - 1} = h, Z_{i, t} = j\right] \ln\left[\mathbb{P}\left(Z_{i, t} = j \big| Z_{i, t - 1} = h, \bm{W}\right)\right] \\ 
    &\quad + \left.\sum_{t = 1}^{T_{i}} \sum_{j = 1}^{K} \mathbbm{1}\left[Z_{i, t} = j\right] \left[\ln\left( \mathbb{P}^{(j)}\!\left[N_{i, t} \big| \bm{\vartheta}_{N}\right] \right) + \sum_{n = 0}^{N_{i, t}} \ln\left( \mathbb{P}^{(j)}\!\left[X_{i, t, n} \big| \bm{\vartheta}_{X}\right] \right)\right] \right\}
\end{align*}
with respect to the marginal probabilities presented earlier, where we, initially, pretend to know the latent risk profiles $\bm{Z} = \left(\bm{Z}_{1, 1:T_{1}}, \dots, \bm{Z}_{M, 1:T_{M}}\right)$.
To maximize this complete log-likelihood, we usually employ the Baum-Welch algorithm of \citet{baum1970}, also known as the Forward-Backward algorithm and a special case of the EM algorithm of \citet{dempster1977}. This algorithm is known to lead to consistent and asymptotically Normal estimators under mild regulatory conditions (see, e.g., \citet{leroux1992,bickel1996,bickel1998,cappe2004}). It first replaces the latent profile assignments $\mathbbm{1}\left[Z_{i, t} = j\right]$ and $\mathbbm{1}\left[Z_{i, t - 1} = h, Z_{i, t} = j\right]$ by their posterior expectations
\begin{align*}
    \gamma^{(j)}_{i, t} &= \frac{\alpha_{i, t}^{(j)} \beta_{i, t}^{(j)}}{\sum_{k = 1}^{K} \alpha_{i, t}^{(k)} \beta_{i, t}^{(k)}}, \\
    \xi^{(h, j)}_{i, t} &= \frac{\alpha_{i, t - 1}^{(h)} \, \mathbb{P}\left[Z_{i, t} = j \big| Z_{i, t - 1} = h, \bm{\vartheta}\right] \beta_{i, t}^{(j)} \, \mathbb{P}^{(j)}\!\left[N_{i, t} \big| \bm{\vartheta}\right] \prod_{n = 0}^{N_{i, t}} \mathbb{P}^{(j)}\!\left[X_{i, t, n} \big| \bm{\vartheta}\right]}{\sum_{m = 1}^{K} \sum_{k = 1}^{K} \alpha_{i, t - 1}^{(m)} \, \mathbb{P}\left[Z_{i, t} = k \big| Z_{i, t - 1} = m, \bm{\vartheta}\right] \beta_{i, t}^{(k)} \, \mathbb{P}^{(k)}\!\left[N_{i, t} \big| \bm{\vartheta}\right] \prod_{n = 0}^{N_{i, t}} \mathbb{P}^{(k)}\!\left[X_{i, t, n} \big| \bm{\vartheta}\right]},
\end{align*}
respectively, given customer $i$'s entire observed claims history $\bm{N}_{i, 1:T_{i}}$ and $\bm{X}_{i, 1:T_{i}}$ to overcome the fact that the risk profiles are unobserved in reality. The expectations, or responsibilities in this first step, the E-step, rely on the probabilities introduced earlier and the recursive backward probabilities 
\begin{align*}
    \beta_{i, t}^{(j)} &= \mathbb{P}\left[\bm{N}_{i, (t + 1):T_{i}}, \bm{X}_{i, (t + 1):T_{i}} \big| Z_{i, t} = j, \bm{\vartheta}\right] \\
    &= \sum_{h = 1}^{K} \beta_{i, t + 1}^{(h)} \, \mathbb{P}\left[Z_{i, t + 1} = h \big| Z_{i, t} = j, \bm{\vartheta}\right] \mathbb{P}^{(h)}\!\left[N_{i, t + 1} \big| \bm{\vartheta}\right] \prod_{n = 0}^{N_{i, t + 1}} \mathbb{P}^{(h)}\!\left[X_{i, t + 1, n} \big| \bm{\vartheta}\right],
\end{align*}
with terminal condition $\beta_{i, T_{i}}^{(j)} = 1$. The forward process $\alpha_{i, t}^{(j)}$ can effectively be interpreted as a filter on the observational evidence by incorporating only past history, similar to the posterior updating of the (future) risk premium in \Cref{F2.2}. The backward process $\beta_{i, t}^{(j)}$, in turn, smoothens the parameter estimates by accounting for future history as well and thus only affects the (historical) estimation procedure in \Cref{F2.1} to account for the remaining observed claims history.

Given these posterior expectations, we maximize the complete log-likelihood in the second step, the M-step, of the Baum-Welch algorithm. More specifically, the expected marginal complete log-likelihood
\begin{align*}
    Q\left(\bm{\vartheta} \big| \tensor*[^{(r)}]{\!\bm{\vartheta}}{}\right) &= \sum_{i = 1}^{M} \left\{\sum_{j = 1}^{K} \tensor*[^{(r)}]{\!\gamma}{^{(j)}_{i, 1}} \ln\left[\mathbb{P}\left(Z_{i, 1} = j \big| \bm{w}_{0}\right)\right] + \sum_{t = 2}^{T_{i}} \sum_{h = 1}^{K} \sum_{j = 1}^{K} \tensor*[^{(r)}]{\!\xi}{^{(h, j)}_{i, t}} \ln\left[\mathbb{P}\left(Z_{i, t} = j \big| Z_{i, t - 1} = h, \bm{W}\right)\right]\right. \\ 
    &\quad + \left.\sum_{t = 1}^{T_{i}} \sum_{j = 1}^{K} \tensor*[^{(r)}]{\!\gamma}{^{(j)}_{i, t}} \left[\ln\left( \mathbb{P}^{(j)}\!\left[N_{i, t} \big| \bm{\vartheta}_{N}\right] \right) + \sum_{n = 0}^{N_{i, t}} \ln\left( \mathbb{P}^{(j)}\!\left[X_{i, t, n} \big| \bm{\vartheta}_{X}\right] \right)\right]\right\}
\end{align*}
is maximized with respect to the parameters $\bm{\vartheta}$, where the expectations $\tensor*[^{(r)}]{\!\gamma}{^{(j)}_{i, t}}$ and $\tensor*[^{(r)}]{\!\xi}{^{(h, j)}_{i, t}}$ are evaluated at the estimates $\!\tensor*[^{(r)}]{\!\bm{\vartheta}}{}$ from the previous iteration $r$. The parameters $\bm{w}_{0}$ and $\bm{W}$ can be estimated analytically while using their constraints $\sum_{j = 1}^{K} w^{(j)}_{h} = 1$ for every $h = 0, \dots, K$ as
\begin{equation*}
    \tensor*[^{(r + 1)}]{\!w}{^{(j)}_{0}} = \frac{\sum_{i = 1}^{M} \tensor*[^{(r)}]{\!\gamma}{^{(j)}_{i, 1}}}{M} \quad\quad\quad \textrm{and} \quad\quad\quad \tensor*[^{(r + 1)}]{\!w}{^{(j)}_{k}} = \frac{\sum_{i = 1}^{M} \sum_{t = 2}^{T_{i}} \tensor*[^{(r)}]{\!\xi}{^{(h, j)}_{i, t}}}{\sum_{i = 1}^{M} \sum_{t = 2}^{T_{i}} \tensor*[^{(r)}]{\!\gamma}{^{(h)}_{i, t - 1}}},
\end{equation*}
respectively, since $Q\left(\bm{\vartheta} \big| \tensor*[^{(r)}]{\!\bm{\vartheta}}{}\right)$ involves four independent sums over $\bm{w}_{0}$, $\bm{W}$, $\bm{\vartheta}_{N}$ and $\bm{\vartheta}_{X}$. On the other hand, the expected complete log-likelihood cannot be solved in closed-form for the remaining parameters and we therefore need to estimate $\bm{\vartheta}_{N}$ and $\bm{\vartheta}_{X}$ numerically using, for instance, the iterative Newton-Raphson method (see \ref{AppendixB}). By iterating over this E- and M-step, we maximize the more complicated incomplete log-likelihood as well and are able to estimate the parameters for the joint, frequency-severity experience rating approach proposed in this paper. 
\section{Data and empirical considerations} \label{Section3}

\subsection{Motor Third Party Liability insurance} \label{Section3.1}

To illustrate the implications of the proposed frequency-severity experience rating approach, we apply the Bayesian HMM to a Dutch automobile insurance portfolio. This portfolio consists of Motor Third Party Liability (MTPL) insurance policies from 2011 up to and including 2019 and contains data on the level of individual policyholders. It involves $193{,}744$ policies with a total exposure to risk of $108{,}088$ years and $5{,}662$ claims ranging from $\textrm{\texteuro}4.39$ to $\textrm{\texteuro}569{,}584.57$. These claims have an overall mean (median) of $\textrm{\texteuro} 3{,}445.43$ ($\textrm{\texteuro} 1{,}100.00$) and their sizes are thus highly skewed to the right. Furthermore, each policyholder has insured only one vehicle and the portfolio includes various customer- and vehicle-specific risk characteristics that are commonly adopted for pricing in the Dutch insurance market. For more details on the risk factors used in this paper, see \Cref{TA.1} in \ref{AppendixA}.

\begin{figure}[t!]
    \centering
    \begin{tabular}{c c}
        \centering
        \includegraphics[width=0.45\textwidth]{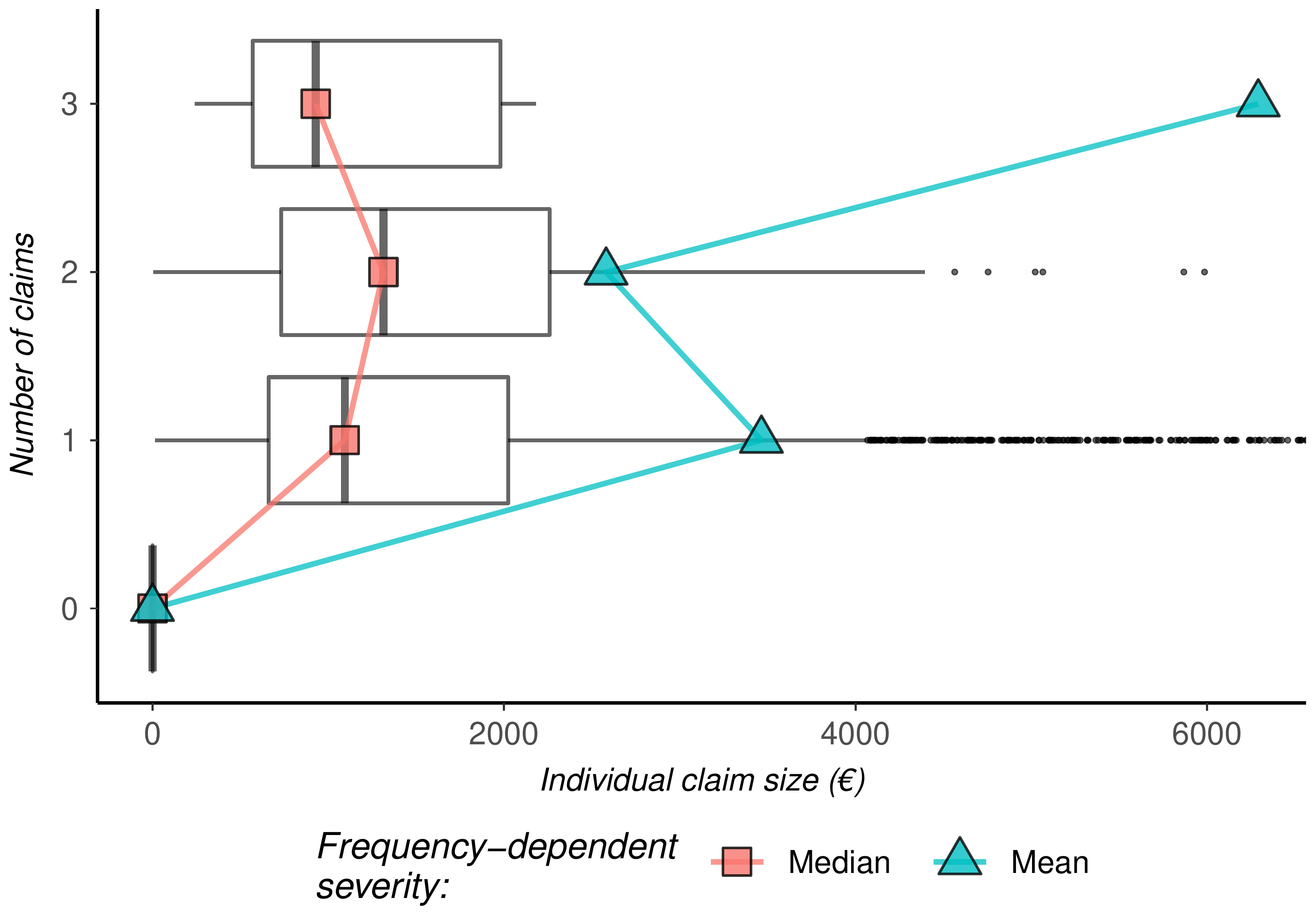}&
        \includegraphics[width=0.45\textwidth]{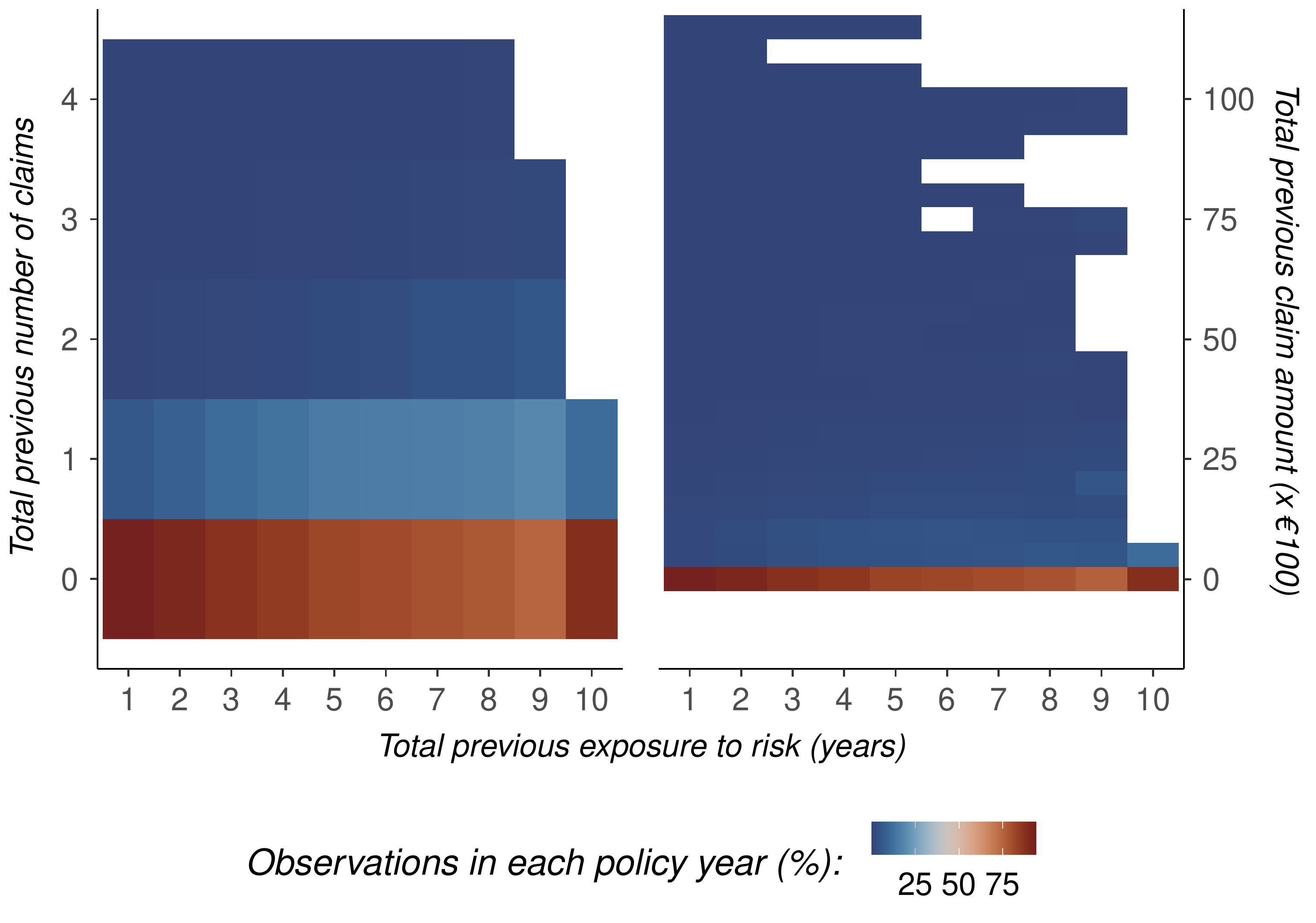}
    \end{tabular}\vspace{-6pt}
    \caption{Frequency-dependent severity distribution (left) and distribution of claims experience based on total previous number of claims and claim amount (right) in the observed MTPL insurance portfolio.}
	\label{F3.1}
\end{figure}

While insurers traditionally assume independence between the claim counts and sizes, \Cref{F3.1} demonstrates that this usually does not hold in practice. More specifically, \Cref{F3.1} (left) depicts the individual claim sizes that are observed for customers who have had from zero up to three claims in a single policy year, whereas \Cref{F3.1} (right) displays what percentage of customers in each policy year has obtained how many, and what amount of, claims. \Cref{F3.1} (left), for instance, shows that the overall distribution and median of the individual claim sizes vary non-linearly with the number of claims reported on a policy. Moreover, this effect appears to be particularly large when we consider the mean of the claim sizes, whose range has been truncated for illustrative purposes. A constant, linear dependence that is often assumed in the literature on frequency-severity modeling may therefore not be flexible enough to capture the non-linear association actually observed between the claim frequencies and severities. \Cref{F3.1} (right) additionally highlights that the claims experience of most policyholders does not contain any claims, or at most one, as they gain more claims history.

\subsection{Optimization methodology} \label{Section3.2}

Using the Dutch automobile insurance portfolio, we estimate the parameters for the frequency-severity experience rating approach developed in this paper. We consider a full and sparse representation of the Bayesian HMM where $\left(b^{(j)}_{U}, b^{(j)}_{V}\right) = \left(a^{(j)}_{U}, a^{(j)}_{V} - 1\right)$ to follow the Bayesian experience rating convention and $\left(\bm{\delta}^{(j)}_{A}, \varphi^{(j)}, \bm{\delta}^{(j)}_{B}\right) = \left(\bm{\delta}_{A}, \varphi, \bm{\delta}_{B}\right)$ to reduce the model complexity, respectively. The two specifications are numerically optimized in \texttt{R} using the BFGS method from the \texttt{maxLik} package developed by \citet{henningsen2011} which make use of the gradient vectors and Hessian matrices derived in \ref{AppendixB} for a varying number of latent risk profiles $K$. Moreover, as the number of risk profiles $K$ increases, it may become harder to distinguish between the different profiles. We therefore include a small (Ridge) penalty of $4 \cdot 10^{-6}$ on the squared $L_{2}$ norm of the parameters $\bm{\vartheta}_{N}$ and $\bm{\vartheta}_{X}$ to limit the size of their effects if they are not significant enough and to thereby avoid estimating spurious relationships \citep{hoerl1970,hastie2009}. This, in turn, enables us to estimate the Bayesian HMMs through the adjusted, empirical Bayes version of the Baum-Welch algorithm.

However, since the Baum-Welch algorithm is in fact a special case of EM, its initialization may determine whether it converges to a local, rather than a global, optimum. We therefore extensively search the parameter space before running the algorithm until convergence, similar to \citet{biernacki2003} and \citet{blostein2019}, and adopt the parameter estimates for the previous value of $K$ as a starting point for new searches where we increase $K$. Note that we start from the model with only a single profile, or $K = 1$, which corresponds to the classical experience rating approach under independence and will serve as a benchmark. Once the HMMs are optimized for various values of $K$, we compare them and look for the lowest Akaike Information Criterion (AIC) of \citet{akaike1974} and Bayesian Information Criterion (BIC) of \citet{schwarz1978}, as these criteria are typically used for model selection in mixture modeling (see, e.g., \citet{tzougas2014,tzougas2018,miljkovic2018,blostein2019,pocuca2020}). They are defined as $\textrm{AIC}\left(\bm{\vartheta}\right) = 2 P - 2 \ln\left[\ell\left(\bm{\vartheta} \big| \bm{N}, \bm{X}\right)\right]$ and $\textrm{BIC}\left(\bm{\vartheta}\right) = P \ln\left(\sum_{i = 1}^{M} T_{i}\right) - 2 \ln\left[\ell\left(\bm{\vartheta} \big| \bm{N}, \bm{X}\right)\right]$, with $P = 82K + \mathbbm{1}\left[K > 1\right] \left(K + 1\right) K$ and $P = 78 + 4K + \mathbbm{1}\left[K > 1\right] \left(K + 1\right) K$ in the full and sparse HMM, respectively, and where the BIC penalizes the number of parameters $P$ more severely. Besides that, we also report the optimal log-likelihood value of each HMM directly, assess the correlations, Spearman's rho's, and Kendall's tau's implied by the models, and consider the prior/posterior expected loss ratio $\sum_{i = 1}^{M} \sum_{\tau = 1}^{T_{i}} L_{i, \tau} \big/ \sum_{i = 1}^{M} \sum_{\tau = 1}^{T_{i}} e_{i, \tau} \pi_{i, \tau}$ of the aggregate claim amount in the portfolio for a more practical performance measure (see, e.g., \citet{ohlsson2010}). As a result, we can determine whether and to what extent we can improve our premium estimates on an aggregate level by allowing for a frequency-severity dependence as observed in practice. 
\section{Applications in Motor Third Party Liability insurance} \label{Section4}

\subsection{Inferring the number of mixture components} \label{Section4.1}

\begin{figure}[t!]
    \centering
    \begin{tabular}{c c}
        \centering
        \includegraphics[width=0.45\textwidth]{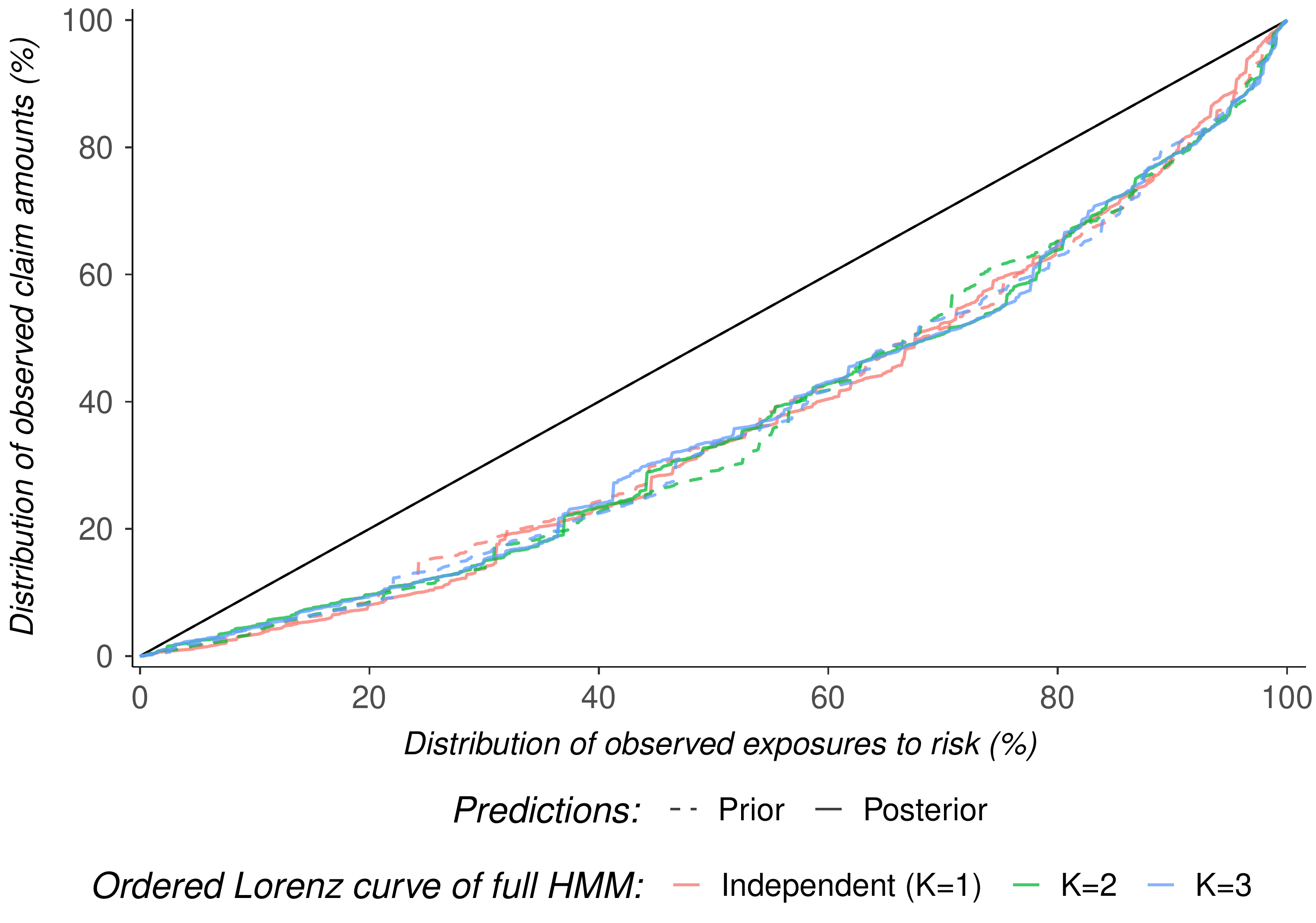}&
        \includegraphics[width=0.45\textwidth]{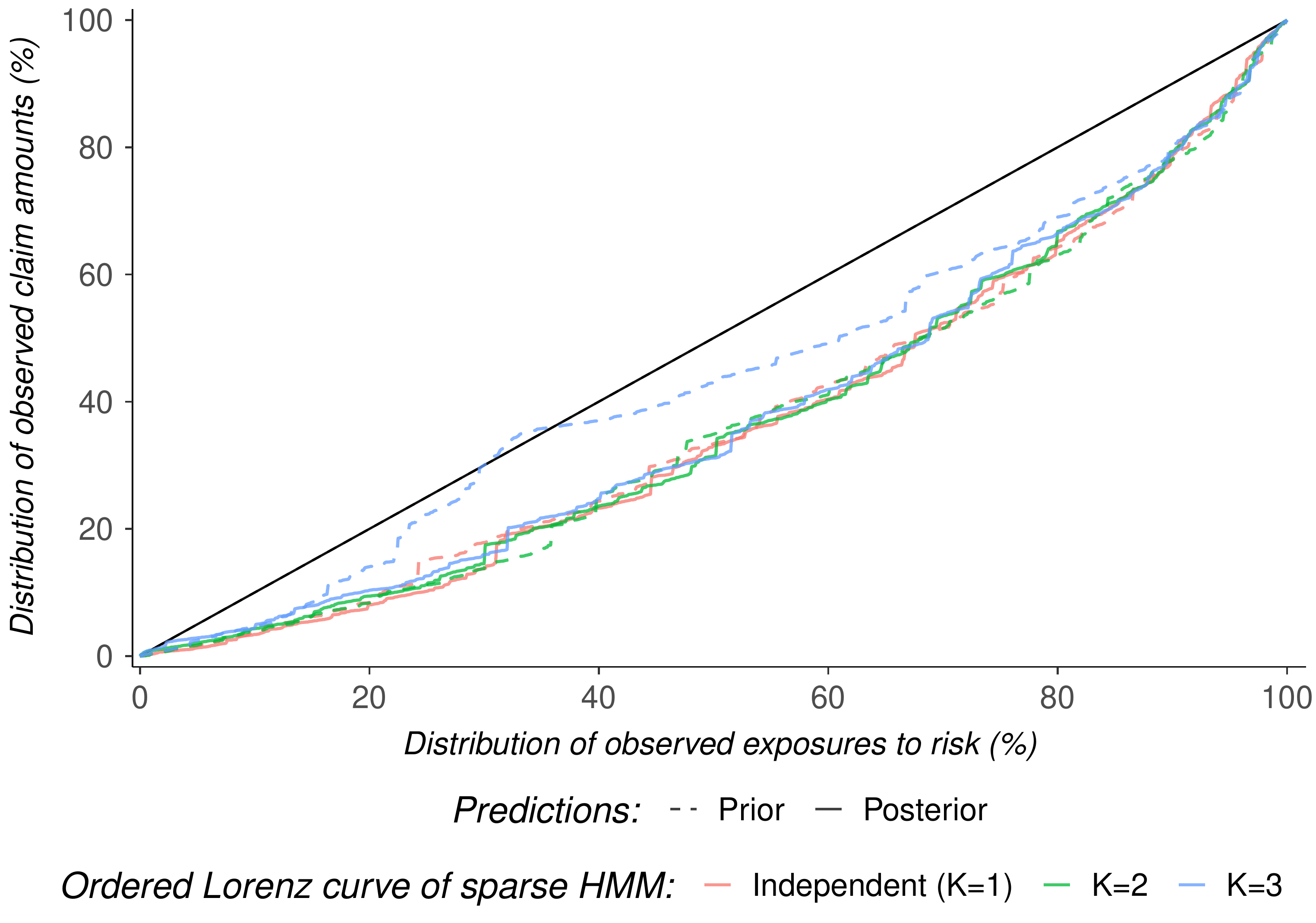}
    \end{tabular}\vspace{-6pt}
    \caption{Ordered Lorenz curve based on the prior/posterior predicted risk premia for $K \leq 3$ for each full (left) and sparse (right) MTPL insurance HMM.}
	\label{F4.2}
\end{figure}

Based on the methodology described earlier, we explore how well the proposed joint experience rating approach can explain the observed MTPL insurance claims. We consider both the full and sparse representation of the Bayesian HMM and compare them in terms of their log-likelihood value, AIC/BIC value, and prior/posterior loss ratio in \Cref{F4.1} and \Cref{TC.1} in \ref{AppendixC}, to infer the optimal number of mixture components. The discriminatory power of the predicted premia is investigated by considering the ordered Lorenz curves in \Cref{F4.2}. These Lorenz curves display what fraction of least risky policies according to each model incurs what proportion of observed claim amounts, and they therefore quantify how well we can identify and distinguish between risky and safe customers \citep{frees2011,frees2014,verschuren2021,henckaerts2021}. However, note that these Lorenz curves depict the total degree of risk discrimination of each model, i.e., compared to a model without any risk discrimination, while in practice we are usually more interested in how the risk classification of a certain model compares to that from a benchmark model. By ordering the predicted premia relative to those from a benchmark model rather than to a constant, we can calculate the ratio Gini coefficients in \Cref{T4.1}. These coefficients indicate how vulnerable the benchmark model is to alternative models due to the different ordering of risks and how much more profitable it is, on average, to use the alternative rate structure instead of the benchmark structure \citep{frees2014,verschuren2021,henckaerts2021}. According to the mini-max strategy of \citet{frees2014}, the benchmark model that minimizes the maximal coefficient is considered the least vulnerable to alternative specifications and hence the most profitable model to implement. Finally, we report the estimated prior parameters and transition probabilities for each HMM in \Cref{TC.2,TC.3}, respectively, whereas the estimated effects of all the risk characteristics are presented in \Cref{TC.9,TC.10,TC.11}.

\begin{figure}[t!]
    \centering
    \includegraphics[width=0.925\textwidth]{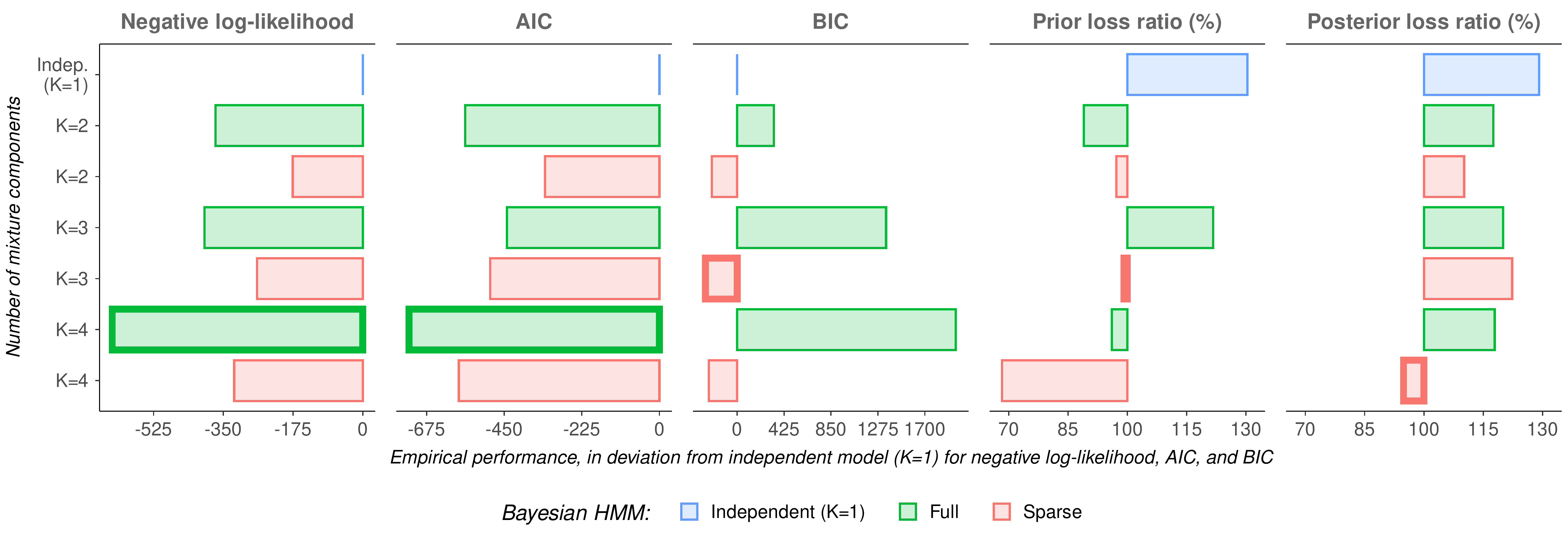}
    \vspace{-3pt}
    \caption{Empirical performance, in deviation from HMM with $K = 1$ for negative log-likelihood, AIC, and BIC, of the MTPL insurance HMMs. Recall that the lowest/closest to $100\%$ value (bold) indicates the least prediction error and the most statistical/empirical improvement. For more details on these values, see \Cref{TC.1}.}
	\label{F4.1}
\end{figure}

\begin{table}[t!]
    \caption{Ratio Gini coefficients in percentages for independent model ($K = 1$) and both full (left) and sparse (right) HMM representation, using either the \textit{a priori} (top) or \textit{a posteriori} (bottom) risk premium predictions. While the rows of this table denote the benchmark model, the columns represent the alternative model.}
    \label{T4.1}\vspace{-3pt}\vspace{-3pt}
    \centerline{\scalebox{0.8505}{\begin{tabularx}{1.18\textwidth}{l p{0.5em} rrr p{0.5em} rrr}
		\toprule \addlinespace[1ex] \vspace{1pt}
		\textbf{\textit{A priori}} && \multicolumn{3}{c}{\textbf{Full representation (alternative)}} && \multicolumn{3}{c}{\textbf{Sparse representation (alternative)}} \\ \cline{3-5} \cline{7-9} \addlinespace[0.4ex]
		\textbf{\textit{(benchmark)}} && \textbf{Independent ($\mathbf{K = 1}$)} & $\mathbf{K = 2}$ & $\mathbf{K = 3}$ && \textbf{Independent ($\mathbf{K = 1}$)} & $\mathbf{K = 2}$ & $\mathbf{K = 3}$ \\ \hline \addlinespace[0.4ex]
		\textit{Independent}\hspace{-7pt} &&  & \textbf{7.4913} & 1.0682 &&  & \textbf{11.6099} & 9.2879 \\
		\quad \textit{($\mathit{K = 1}$)} \\
		$\mathit{K = 2}$ && \textbf{0.3656} &  &  && \textbf{10.0734} &  &  \\
		$\mathit{K = 3}$ && \textbf{34.7476} &  &  && \textbf{4.5966} &  &  \\ \hline \addlinespace[0.4ex]
		\textbf{\textit{A posteriori}} && \multicolumn{3}{c}{\textbf{Full representation (alternative)}} && \multicolumn{3}{c}{\textbf{Sparse representation (alternative)}} \\ \cline{3-5} \cline{7-9} \addlinespace[0.4ex]
		\textbf{\textit{(benchmark)}} && \textbf{Independent ($\mathbf{K = 1}$)} & $\mathbf{K = 2}$ & $\mathbf{K = 3}$ && \textbf{Independent ($\mathbf{K = 1}$)} & $\mathbf{K = 2}$ & $\mathbf{K = 3}$ \\ \hline \addlinespace[0.4ex]
		\textit{Independent}\hspace{-7pt} &&  & 5.7824 & \textbf{8.5380} &&  & 16.9356 & \textbf{18.1519} \\
		\quad \textit{($\mathit{K = 1}$)} \\
		$\mathit{K = 2}$ && \textbf{0.5465} &  &  && \textbf{0.0603} &  &  \\
		$\mathit{K = 3}$ && \textbf{0.1472} &  &  && \textbf{-0.0600} &  &  \\ \bottomrule
	\end{tabularx}}}
\end{table}

From the empirical performance measures in \Cref{F4.1} and \Cref{TC.1}, we observe that the Bayesian HMMs tend to perform better as we consider more mixture components. Both model representations, for instance, improve the (negative) log-likelihood and AIC when more mixture components are included and lead to a prior/posterior loss ratio closer to $100\%$. However, we find that the full model representation actually performs worse as $K$ increases when we take the model complexity into account through the BIC and that its posterior loss ratio is always further away from $100\%$ than its prior loss ratio. The sparse model representation, on the other hand, leads to a substantially lower, and hence better, BIC value and seems to improve its prior/posterior loss ratio in a more robust manner as $K$ increases. Nonetheless, the gain in the log-likelihood decreases in the number of mixture components $K$ and no longer justifies the increase in model complexity for $K > 3$ in the sparse model. This, in turn, implies that it is sufficient to consider (at most) $K = 3$ latent risk profiles and that the Bayesian HMMs start to overfit once we allow for more mixture components. We therefore focus on the results for $K \leq 3$ in the remainder of this analysis.

Compared to the classical experience rating approach under independence, the proposed Bayesian HMMs lead to much more accurate aggregate claim amount predictions. The ordered Lorenz curves in \Cref{F4.2} additionally indicate that we can distinguish considerably better between customer risks \textit{a posteriori}, in particular in the sparse HMMs. Nonetheless, the different (posterior) orderings of risks appear to yield relatively similar discriminatory power for both approaches. The discriminatory power of the independent model ($K = 1$) on the one hand and the proposed Bayesian HMMs on the other hand is therefore quantified relative to one another by the ratio Gini coefficients in \Cref{T4.1}, \textit{a priori} as well as \textit{a posteriori} and for both the full and sparse HMMs. From this table we find, for instance, that the independent model ($K = 1$) leads to a maximal ratio Gini coefficient of $7.4913$ or $11.6099$ \textit{a priori} when considering the full or sparse HMMs as alternative models, respectively. The HMMs themselves lead to a somewhat lower maximal coefficient of $0.3656$ or $4.5966$ when considering the independent model ($K = 1$) as the alternative model. This difference persists \textit{a posteriori}, where especially the sparse HMMs lead to a substantially lower maximal ratio Gini coefficient of $-0.0600$ compared to a maximal coefficient of $18.1519$ for the independent model ($K = 1$). The interpretation of these results is that the Bayesian HMMs are considerably less vulnerable to alternative specifications than the independent model ($K = 1$) and that, on average, they are much more profitable to use than the independent model ($K = 1$) since they can better distinguish between customer risks. The proposed HMMs thus not only improve our premium estimates on an aggregate level but also on an individual level by better aligning these estimates to the customer risks actually observed in the portfolio.

\subsection{Identification of underlying risk profiles} \label{Section4.2}

\begin{figure}[t!]
    \centering
    \begin{subfigure}{\textwidth}
        \centering
        \includegraphics[width=0.925\textwidth]{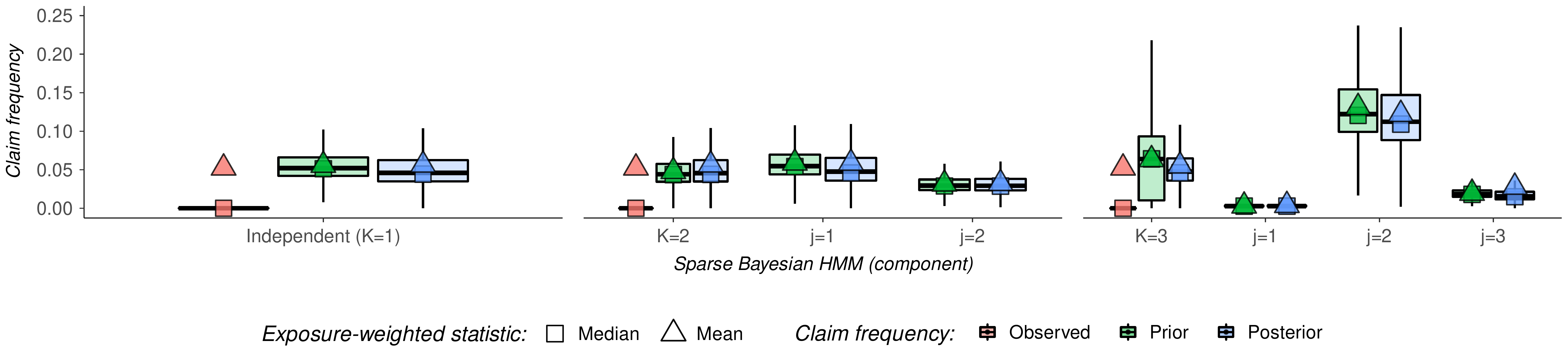}
        \vspace{-3pt}
        \caption{Observed and predicted prior/posterior claim frequencies}
        \label{F4.3a}
    \end{subfigure}
    
    \vspace{3pt}
    
    \begin{subfigure}{\textwidth}
        \centering
        \includegraphics[width=0.925\textwidth]{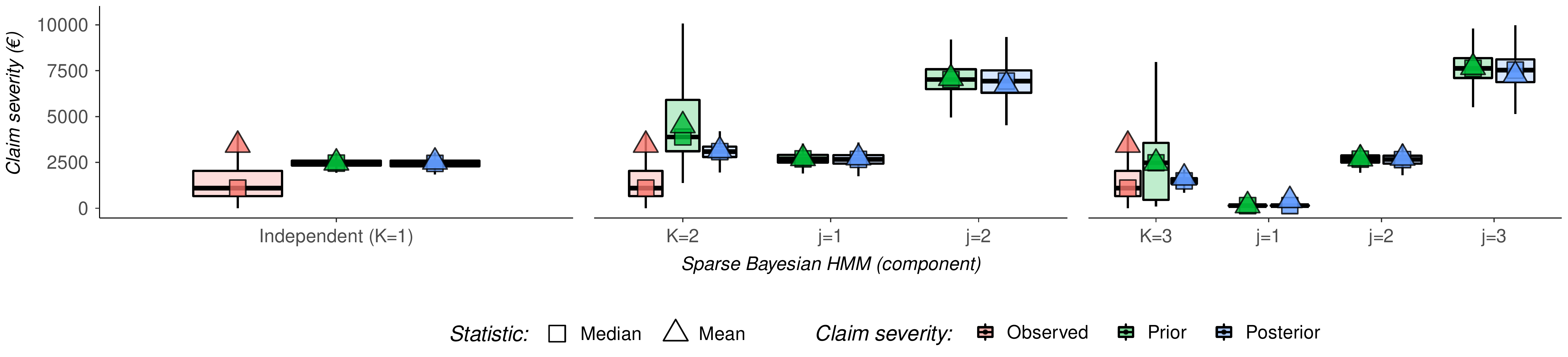}
        \vspace{-3pt}
        \caption{Observed and predicted prior/posterior claim severities}
        \label{F4.3b}
    \end{subfigure}
    
    \vspace{3pt}
    
    \begin{subfigure}{\textwidth}
        \centering
        \includegraphics[width=0.925\textwidth]{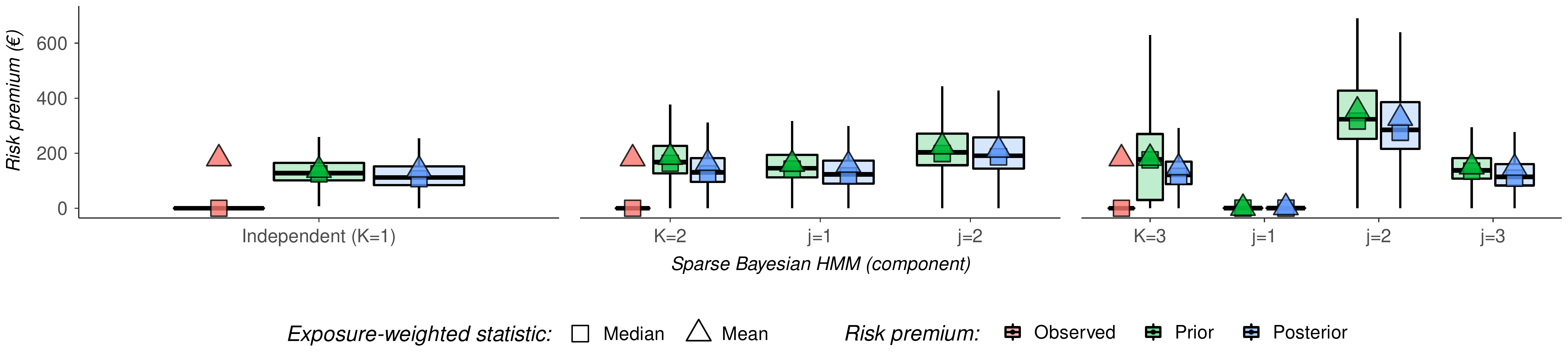}
        \vspace{-3pt}
        \caption{Observed and predicted prior/posterior risk premia}
        \label{F4.3c}
    \end{subfigure}\vspace{-3pt}
    \caption{Distribution of observed and predicted prior/posterior claim frequencies (panel (a)), claim severities (panel (b)), and risk premia (panel (c)) in total and for each latent risk profile of the sparse HMM representation in MTPL insurance. For more details on these distributions, see \Cref{TC.4}.}
	\label{F4.3}
\end{figure}

\begin{figure}[t!]
    \centering
    \includegraphics[width=0.925\textwidth]{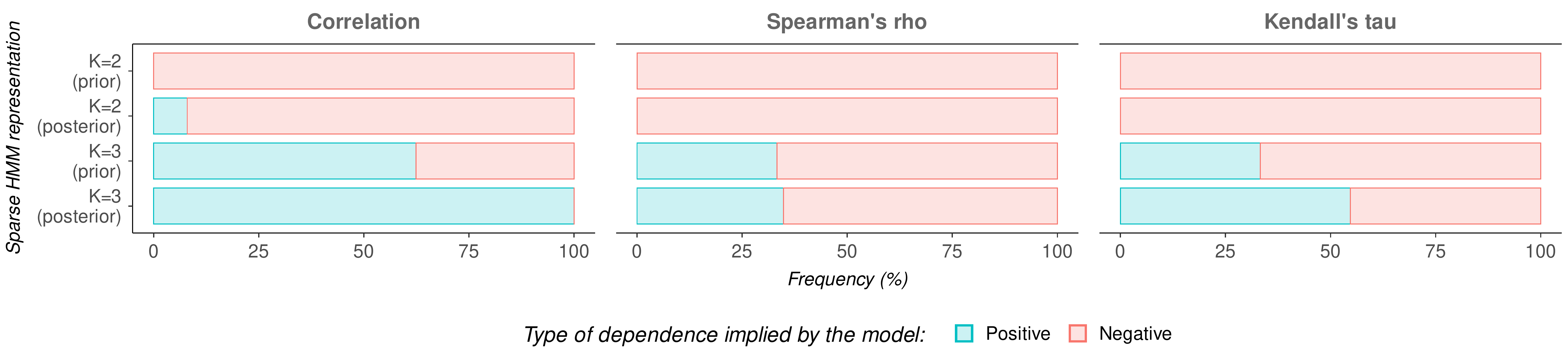}
    \vspace{-3pt}
    \caption{Frequency of the type of dependence, i.e., positive or negative correlation, Spearman's rho, and Kendall's tau, as implied by the estimated sparse HMM representation in MTPL insurance.}
	\label{F4.4}
\end{figure}

\begin{figure}[t!]
    \centering
    \begin{subfigure}{\textwidth}
        \centering
        \begin{tabular}{c c c}
            \centering
            \includegraphics[width=0.30\textwidth]{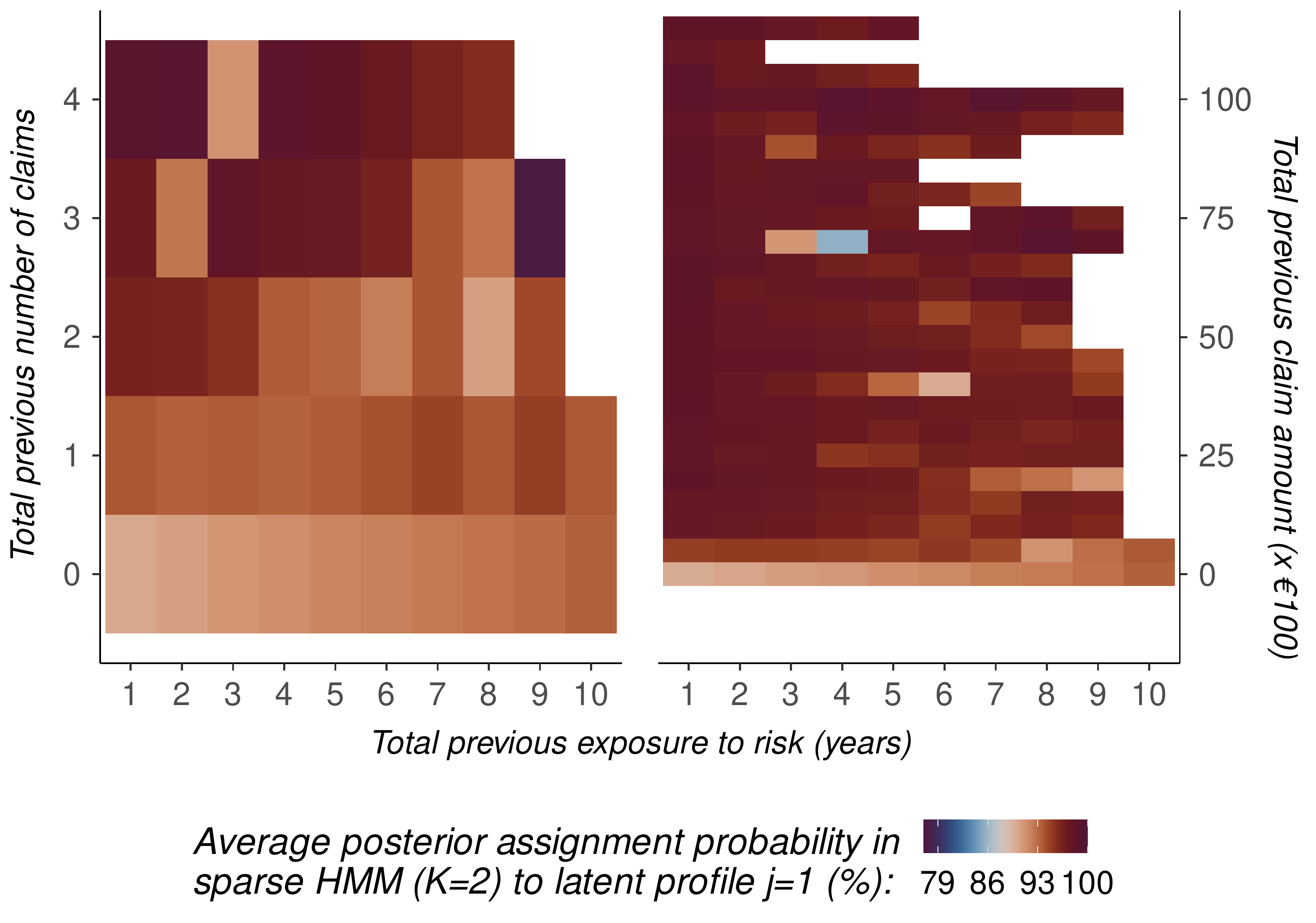}&
            \includegraphics[width=0.30\textwidth]{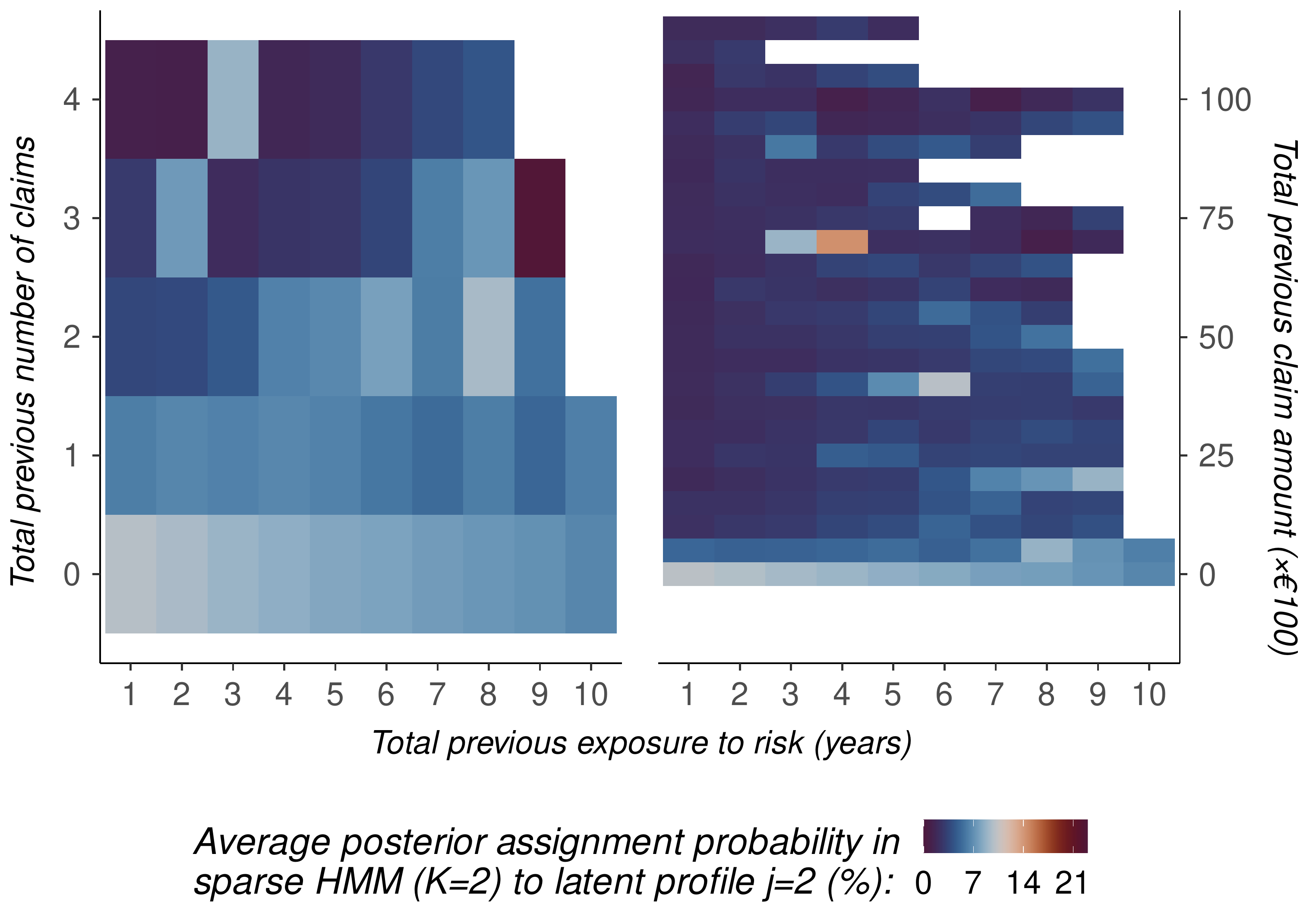}&
            \phantom{\includegraphics[width=0.30\textwidth]{Figures/Figure_8_a_middle.pdf}}
        \end{tabular}\vspace{-6pt}
        \caption{Based on $K = 2$ latent risk profiles}
        \label{F4.5a}
    \end{subfigure}
    
    \vspace{3pt}
    
    \begin{subfigure}{\textwidth}
        \centering
        \begin{tabular}{c c c}
            \centering
            \includegraphics[width=0.30\textwidth]{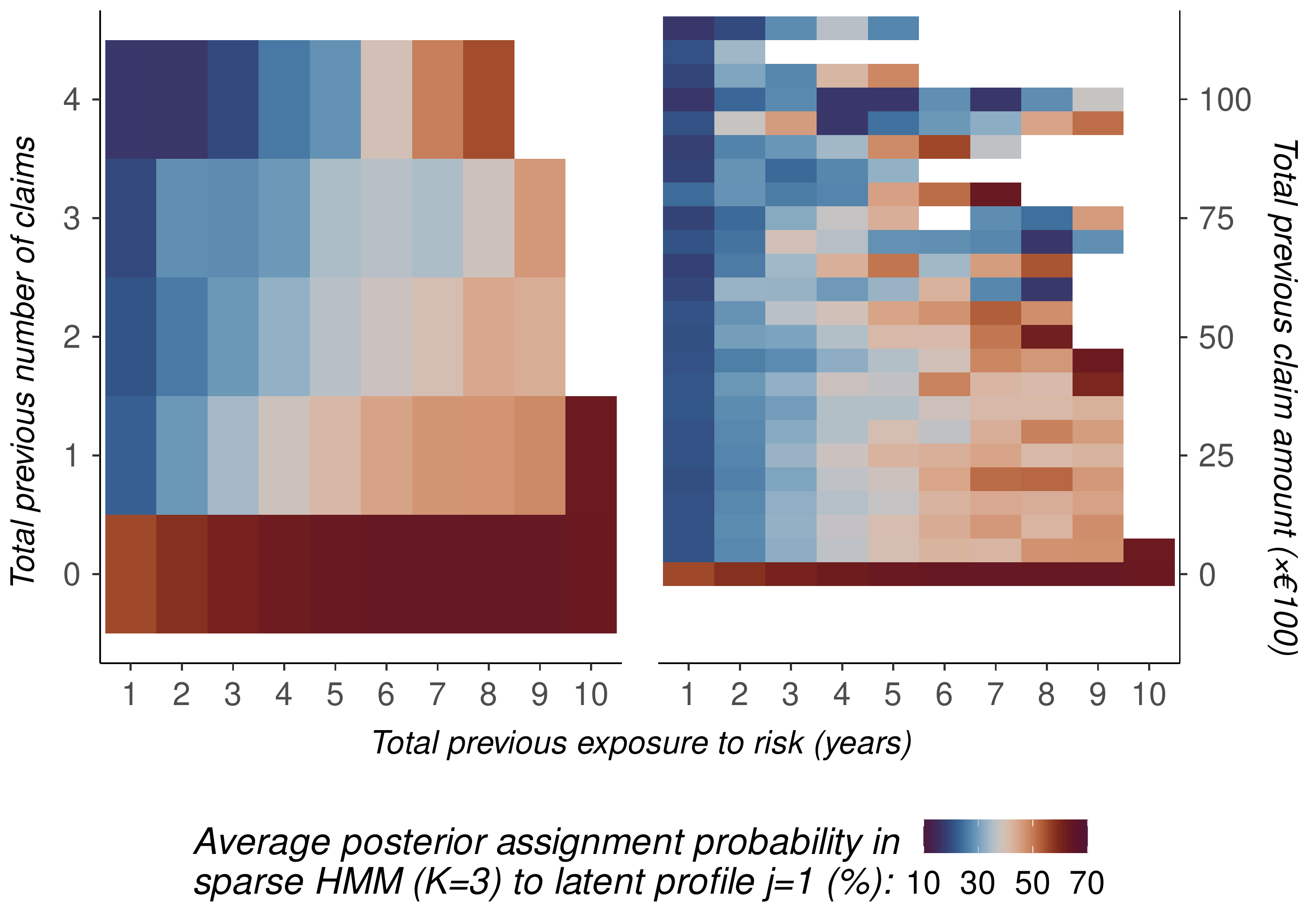}&
            \includegraphics[width=0.30\textwidth]{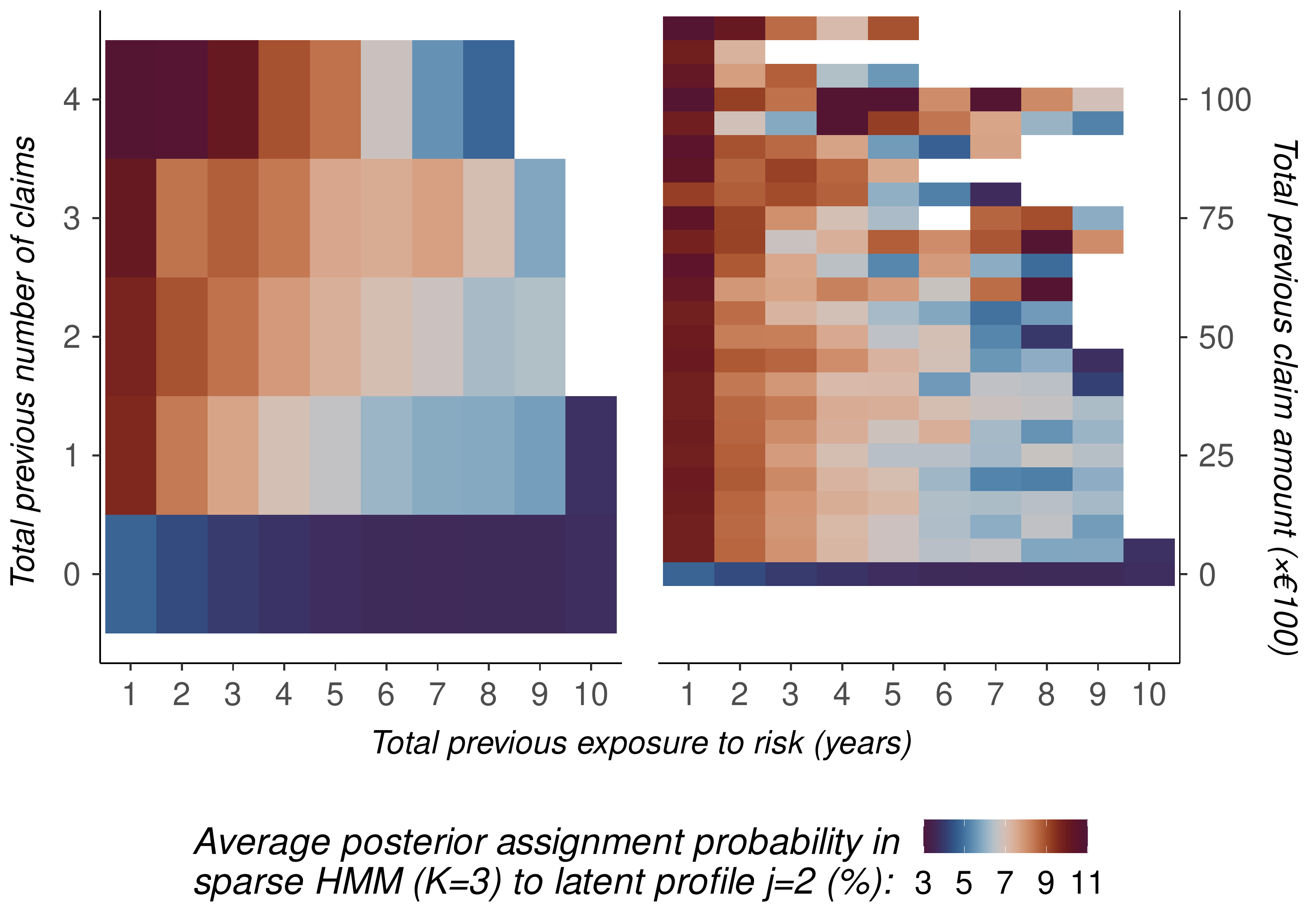}&
            \includegraphics[width=0.30\textwidth]{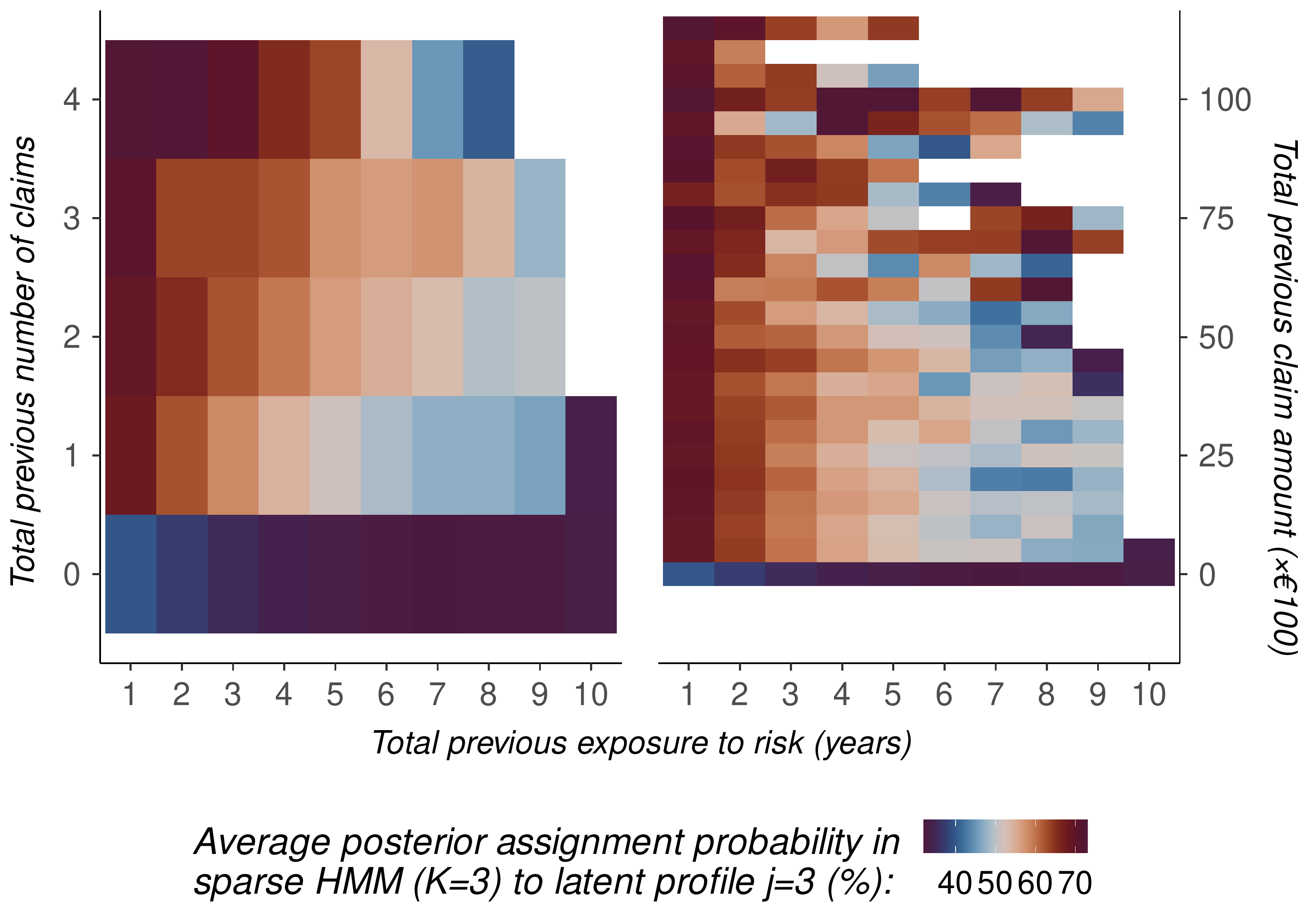}
        \end{tabular}\vspace{-6pt}
        \caption{Based on $K = 3$ latent risk profiles}
        \label{F4.5b}
    \end{subfigure}\vspace{-3pt}
    \caption{Distribution of average posterior assignment probabilities over total previous number of claims and claim amount for profile $j = 1$ (left), $j = 2$ (middle), and $j = 3$ (right) with $K = 2$ (panel (a)) and $K = 3$ (panel (b)) latent risk profiles for the sparse HMM representation in MTPL insurance.}
	\label{F4.5}
\end{figure}

\begin{figure}[t!]
    \centering
    \begin{tabular}{c c c}
        \centering
        \includegraphics[width=0.30\textwidth]{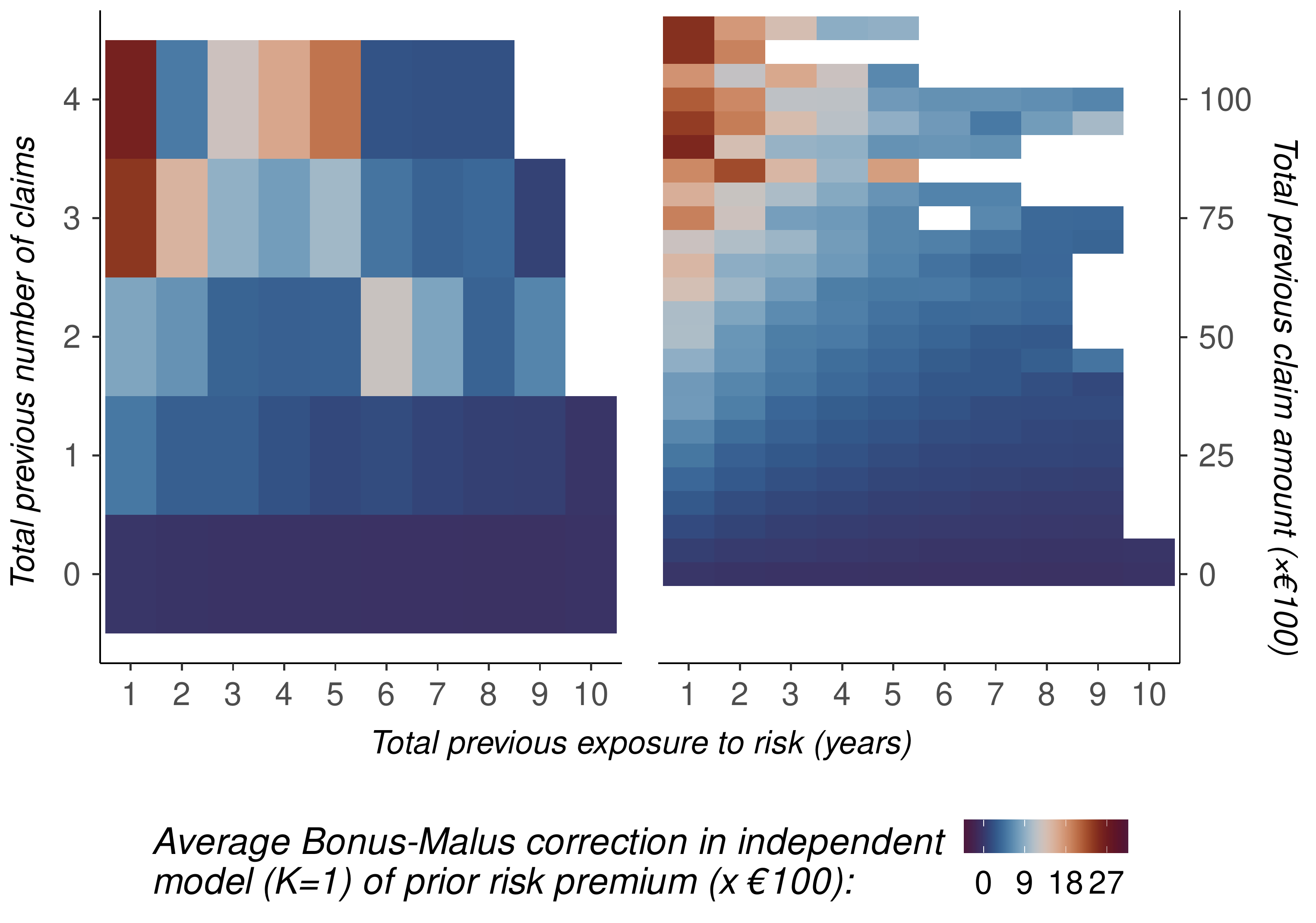}&
        \includegraphics[width=0.30\textwidth]{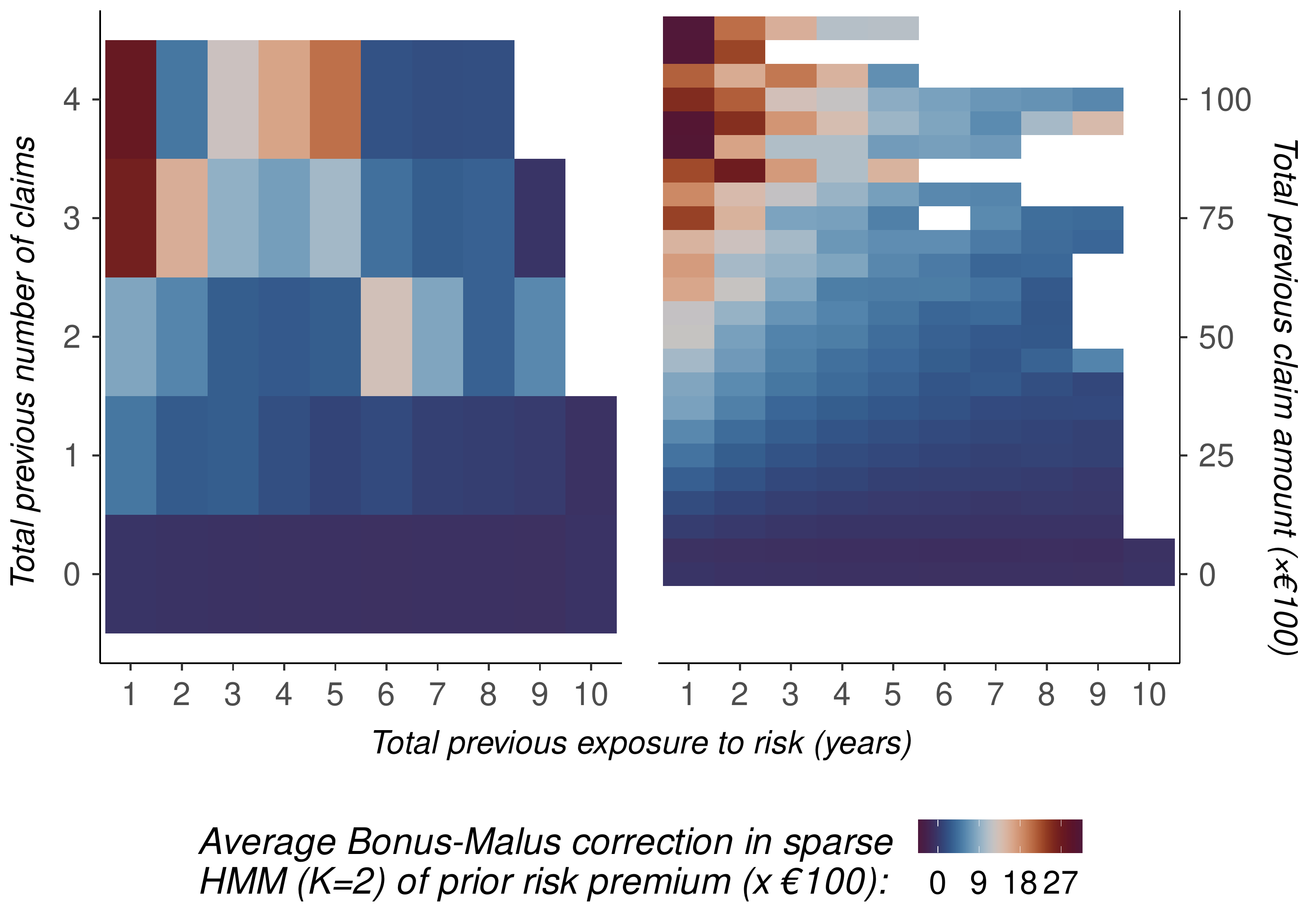}&
        \includegraphics[width=0.30\textwidth]{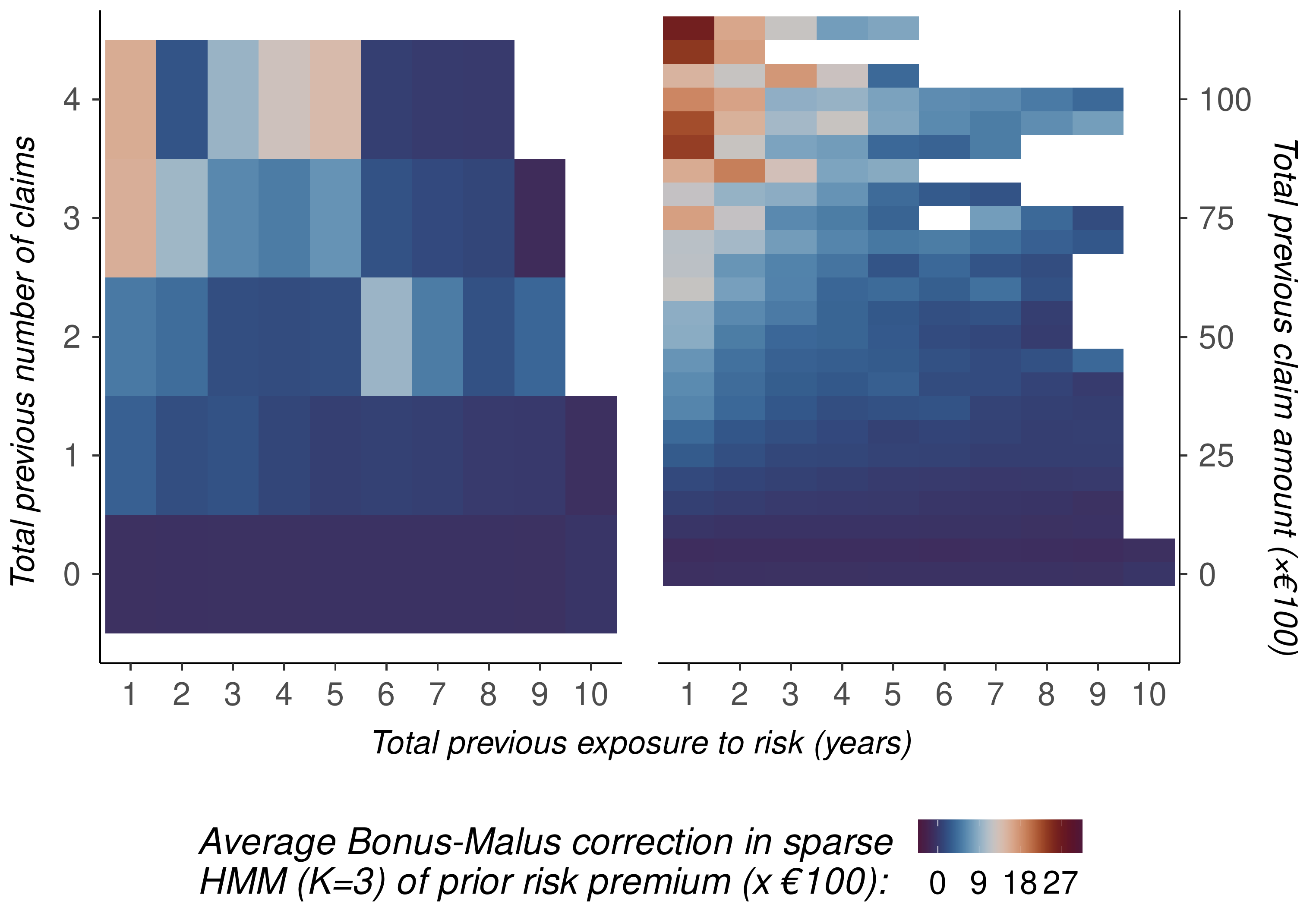}
        \end{tabular}\vspace{-6pt}
    \caption{Average Bonus-Malus correction of prior risk premium in terms of total previous number of claims and claim amount in independent model ($K = 1$, left) and with $K = 2$ (middle) and $K = 3$ (right) latent risk profiles for the sparse HMM representation in MTPL insurance.}
	\label{F4.6}
\end{figure}

Given the number of mixture components, we can identify the risk profiles underlying a customer's claiming behavior. We display the predicted prior/posterior claim frequencies, claim severities, and risk premia for each latent risk profile of the sparse model representation in \Cref{F4.3} and \Cref{TC.4} and compare their combined, total predictions to the observations in the portfolio. The type of dependence implied by the sparse HMMs, i.e., positive or negative correlation, Spearman's rho, and Kendall's tau, is presented for all policies in \Cref{F4.4}. Moreover, the effect of a customer's claims experience on the posterior assignment probabilities of all these risk profiles and on the Bonus-Malus corrections of the prior risk premia is shown in \Cref{F4.5,F4.6}, respectively. Note that for the sake of convenience, \Cref{F4.6} is also given in tabular form in \Cref{TC.5}. While we focus on the sparse model representation here, since it leads to the best BIC value and posterior loss ratio, the results for the full model representation are given in \Cref{TC.6,TC.7} and \Cref{FC.1,FC.2,FC.3}.

Based on the distributions of the prior and posterior predictions in \Cref{F4.3} and \Cref{TC.4}, we can clearly see that each of the latent risk profiles is associated with distinctive claiming behavior. The Bayesian HMM with $K = 2$ mixture components, for instance, identifies one below average risk profile ($j = 1$) where customers claim occasionally, and if so a moderate amount, and one above average risk profile ($j = 2$) where claims occur slightly less often but are much larger. The Bayesian HMM with $K = 3$ mixture components now introduces a third risk profile, which can be seen as a refinement of the below and above average risk profile, and adjusts the other two risk profiles accordingly. More specifically, it now identifies a good ($j = 1$) and medium ($j = 3$) risk profile where customers claim either (almost) never or occasionally, and if so very little or a lot, respectively, but also a bad risk profile ($j = 2$) where claims occur relatively often and are moderate in size. These profiles thus seem to suggest a negative dependence between the claim frequencies and severities and we find that this is in fact implied by the model for practically all policies with $K = 2$ profiles and for the majority of policies with $K = 3$ profiles in \Cref{F4.4}, both \textit{a priori} and \textit{a posteriori}. This is in line with the (marginally) decreasing pattern in the medians of the frequency-dependent severities in \Cref{F3.1} (left) and it is worth mentioning that qualitatively similar claiming behavior is expected when using the full model representation in \Cref{TC.6} and \Cref{FC.1}. The identified customer risk profiles and frequency-severity dependencies therefore appear to be robust with respect to the exact model specification. Note additionally that the means of the predicted total risk premia tend to slightly differ from their observed means due to some very large claim size outliers, but that \Cref{TC.8} shows that these large claims have only a limited impact on the GB2 distribution.

While a customer's claims experience leads to a posterior adjustment of the prior risk premia of the latent risk profiles, it also influences the weight that is \textit{a posteriori} assigned to each profile. We therefore depict how these posterior assignment probabilities evolve on average with respect to a customer's total previous exposure to risk and total previous number of claims or claim amount in \Cref{F4.5}. In case of $K = 2$ mixture components, we observe in \Cref{F4.5a} that the below average risk profile ($j = 1$) becomes more likely as a customer has claimed more often and a lot, whereas more importance is given to the above average risk profile ($j = 2$) when a customer displays more claims-free history. A similar but clearer pattern is shown in \Cref{F4.5b} for $K = 3$ mixture components, where the likelihood of the good ($j = 1$) and medium ($j = 3$) risk profile is now somewhat reduced to accommodate a third, bad risk profile ($j = 2$). This bad risk profile appears to be a similar but more extreme version of the medium risk profile since it receives more weight for customers who have had a lot of (large) claims in the past but requires more/larger claims for the weight to increase than the medium profile. The posterior weighting mechanism thus ensures that a customer's observed claims experience matches the claiming behavior associated with the most probable latent risk profile(s). From \Cref{F4.6} and \Cref{TC.5}, we additionally observe that the total posterior, Bonus-Malus correction of a customer's prior risk premium on average increases with the number of previous claims and past aggregate claim amount, as expected. However, this correction seems to slightly diminish for medium risk customers and to increasingly target bad risk customers as we allow for more mixture components $K$. The joint, frequency-severity experience rating approach proposed in this paper is therefore not only capable of identifying customer risk profiles with distinctive claiming behavior, but is also able to better distinguish bad and medium risk customers from good risk customers. 
\section{Conclusion} \label{Section5}

In this paper, we have presented and applied a novel joint experience rating approach to dynamically price insurance policies based on a customer's entire claims history. While Bonus-Malus Systems and alternative Bayesian forms of experience rating traditionally consider the number of claims irrespective of their sizes, we have allowed for either a positive or negative individual dependence between the claim frequencies and severities by adopting a Bayesian Hidden Markov Model for a customer's underlying risk profile. The resulting latent Markovian risk profiles evolve over time to account for any updates in a customer's claims experience and lead to a dynamic, claims experience-weighted mixture of standard credibility premia. As such, we have introduced a novel pricing framework based on dynamic mixture models to the insurance literature that strongly resembles the classical approach under independence but added the flexibility to incorporate both a customer's claim frequency and severity experience.

In our application of this joint experience rating approach, we considered an automobile insurance portfolio containing MTPL insurance policies from a large Dutch insurer. Using this portfolio, we compared the traditional experience rating approach based on independent GLMs to the proposed Bayesian HMM for a varying number of mixture components under both a full and sparse representation. While the full HMM performed best in terms of the log-likelihood and AIC, the sparse HMM led to the best performance when taking the model complexity into account in the BIC and in terms of the posterior loss ratio. Nonetheless, both HMM representations already substantially improved the classical approach under independence on an individual as well as aggregate level, with (at most) three mixture components and regardless of the increase in the model complexity. Moreover, the HMMs identified customer risk profiles with distinctive claiming behavior, that enabled us to better distinguish bad and medium risk customers from good risk customers and indicated a negative correlation, Spearman's rho, and Kendall's tau between the claim frequencies and severities for the majority of policies. The posterior weighting mechanism of these risk profiles additionally proved to be consistent with a customer's observed claims experience and led to total posterior, Bonus-Malus corrections of the prior risk premium that seemed to increasingly target bad risk customers as we allowed for more mixture components. The Bayesian HMMs introduced in this paper therefore pose a very flexible and promising experience rating approach for non-life insurers to incorporate the frequency-severity dependence in their dynamic pricing strategies.

While this paper has primarily considered a customer's risk characteristics in the claim frequency and severity components, we can also include these characteristics in the latent risk profiles. These risk factors may, for instance, enable us to further tailor the (initial) transition and assignment probabilities to an individual customer. However, since a customer's risk characteristics already affect the posterior assignment probabilities indirectly through the claim frequency and severity distributions, this will probably not yield very different results and mainly lead to a considerable increase in model complexity. A more interesting avenue for future research may therefore be to extend the proposed Bayesian HMMs for customers owning several insured vehicles and/or holding multiple insurance coverages. This extension may, in turn, lead to more specific risk profiles and allow us to identify profitable up- and/or cross-selling opportunities. Alternatively, we may consider a full Bayesian approach to estimate the HMMs and assume a prior for the number of mixture components to determine its optimal value, similar to, e.g., \citet{brown2015}. 

\section*{Acknowledgments}

The author gratefully acknowledges financial support from Nationale-Nederlanden. Any errors made or views expressed in this paper are the responsibility of the author alone. In addition, the author would like to thank Noud van Giersbergen, Peter Boswijk, Michel Vellekoop, and two anonymous referees for their valuable comments on this paper. 

\begingroup
    \setlength{\bibsep}{
    \if1\double
        7.5pt
    \else
        0.0pt
    \fi}
    \def\bibfont{\small}
    \bibliography{References_Paper} 
\endgroup

\appendix
{\renewcommand{\thesection}{Appendix \Alph{section}}
\renewcommand{\theequation}{\Alph{section}.\arabic{equation}}
\counterwithin*{equation}{section}
\renewcommand{\thefigure}{\Alph{section}.\arabic{figure}}
\counterwithin*{figure}{section}
\renewcommand{\thetable}{\Alph{section}.\arabic{table}}
\counterwithin*{table}{section}
\renewcommand{\thealgocf}{\Alph{section}.\arabic{algocf}}
\counterwithin*{algocf}{section}
\newpage
\section{Risk factors for automobile insurance} \label{AppendixA}

\begin{table}[ht!]
    \vspace{-7pt}
    \caption{Description of the key variables (top) and risk factors (bottom) used in both the frequency and severity components for MTPL insurance.}
    \label{TA.1}\vspace{-3pt}
	\centerline{\scalebox{0.80}{\begin{tabularx}{1.25\textwidth}{l l l }
		\toprule \addlinespace[1ex] \vspace{1pt}
		\textbf{Variable}\hspace{46pt}\hspace{10pt}\hspace{4pt} & \textbf{Values}\hspace{25pt}\hspace{10pt} & \textbf{Description} \\ \hline \addlinespace[0.4ex]
		\texttt{Count} & Integer & The number of claims filed by the policyholder. \\
		\texttt{Exposure} & Continuous & The exposure to risk in years. \\
		\texttt{Size} & Continuous & The size of the claim in euros.  \\ \hline \addlinespace[0.4ex]
		\texttt{Cust\_Age} & Continuous & Age of the policyholder in years. \\
		\texttt{Cust\_Residence} & $10$ categories & Residential area of the policyholder. \\
		\texttt{Veh\_Age} & Continuous & Age of the insured vehicle in years. \\
		\texttt{Veh\_BodyDoors} & $10$ categories & Bodywork of the insured vehicle, accounting for its number of doors. \\
		\texttt{Veh\_CatValue} & Continuous & Catalogue value of the insured vehicle in multiples of five hundred euro. \\
		\texttt{Veh\_FuelType} & $7$ categories & Type of fuel used by the insured vehicle. \\
		\texttt{Veh\_Mileage} & $3$ categories & Mileage of the insured vehicle. \\
		\texttt{Veh\_PowerWeight} & Continuous & Horsepower of the insured vehicle, accounting for its weight. \\
		\texttt{Veh\_Region} & $10$ categories & Geographical region in which the insured vehicle is used. \\
		\texttt{Veh\_Weight} & Continuous & Weight of the insured vehicle in multiples of fifty kilogram. \\
		\bottomrule
	\end{tabularx}}}
\end{table} 
\section{Numerical optimization of M-step} \label{AppendixB}

Given the posterior expectations from the E-step and the previous parameter estimates, we numerically maximize over the remaining parameters in the M-step of the Baum-Welch algorithm. Under the Bayesian experience rating convention of $\left(b^{(j)}_{U}, b^{(j)}_{V}\right) = \left(a^{(j)}_{U}, a^{(j)}_{V} - 1\right)$, the $r$-th M-step in the full representation entails for every $j$

\vspace{-15pt}{\small\begin{equation*}
    \arg\max_{\bm{\vartheta}^{(j)}_{N}}\sum_{i = 1}^{M} \sum_{t = 1}^{T_{i}} \tensor*[^{(r)}]{\!\gamma}{^{(j)}_{i, t}} \ln\left[ \mathbb{P}^{(j)}\!\left(N_{i, t} \big| \bm{\vartheta}^{(j)}_{N}\right) \right] \quad \textrm{and} \quad \arg\max_{\bm{\vartheta}^{(j)}_{X}}\sum_{i = 1}^{M} \sum_{t = 1}^{T_{i}} \tensor*[^{(r)}]{\!\gamma}{^{(j)}_{i, t}} \sum_{n = 0}^{N_{i, t}} \ln\left[ \mathbb{P}^{(j)}\!\left(X_{i, t, n} \big| \bm{\vartheta}^{(j)}_{X}\right) \right].
\end{equation*}}\hspace{-4.1pt}
In the sparse HMM where $\left(\bm{\delta}^{(j)}_{A}, \varphi^{(j)}, \bm{\delta}^{(j)}_{B}\right) = \left(\bm{\delta}_{A}, \varphi, \bm{\delta}_{B}\right)$ for every $j$, these $2K$ optimizations reduce to the $2$ problems

\vspace{-15pt}{\small\begin{equation*}
    \arg\max_{\bm{\vartheta}_{N}}\sum_{i = 1}^{M} \sum_{t = 1}^{T_{i}} \sum_{j = 1}^{K} \tensor*[^{(r)}]{\!\gamma}{^{(j)}_{i, t}} \ln\left[ \mathbb{P}^{(j)}\!\left(N_{i, t} \big| \bm{\vartheta}_{N}\right) \right] \quad \textrm{and} \quad \arg\max_{\bm{\vartheta}_{X}}\sum_{i = 1}^{M} \sum_{t = 1}^{T_{i}} \sum_{j = 1}^{K} \tensor*[^{(r)}]{\!\gamma}{^{(j)}_{i, t}} \sum_{n = 0}^{N_{i, t}} \ln\left[ \mathbb{P}^{(j)}\!\left(X_{i, t, n} \big| \bm{\vartheta}_{X}\right) \right].
\end{equation*}}\hspace{-4.1pt}
Regardless of the characterization, these optimization problems do not allow a full closed-form solution and need to be solved numerically using, for instance, the Newton-Raphson or Fisher scoring method.

Both of these optimization methods rely on the gradient vectors and Hessian matrices of the log-likelihoods. Based on the expected marginal complete log-likelihoods shown earlier, we find that the gradients are given by $\bm{g}^{(j)}_{N}\!\left(\bm{\vartheta}^{(j)}_{N}\right)$ and $\bm{g}^{(j)}_{X}\!\left(\bm{\vartheta}^{(j)}_{X}\right)$ or $\bm{g}_{N}\left(\bm{\vartheta}_{N}\right)$ and $\bm{g}_{X}\left(\bm{\vartheta}_{X}\right)$ with respective elements

\vspace{-15pt}{\small\begin{align*}
    \bm{g}^{(j)}_{N, 1}\left(\bm{\vartheta}^{(j)}_{N}\right) &= \sum_{i = 1}^{M} \sum_{t = 1}^{T_{i}} \gamma^{(j)}_{i, t} \frac{a^{(j)}_{U} \left(N_{i, t} - e_{i, t} \lambda^{(j)}_{i, t}\right)}{a^{(j)}_{U} + e_{i, t} \lambda^{(j)}_{i, t}} \bm{A}_{i, t}, \\
    g^{(j)}_{N, 2}\left(\bm{\vartheta}^{(j)}_{N}\right) &= \sum_{i = 1}^{M} \sum_{t = 1}^{T_{i}} \gamma^{(j)}_{i, t} \left[ \psi\left(N_{i, t} + a^{(j)}_{U}\right) - \psi\left(a^{(j)}_{U}\right) + \ln\left(a^{(j)}_{U}\right) - \ln\left(a^{(j)}_{U} + e_{i, t} \lambda^{(j)}_{i, t}\right) + \frac{e_{i, t} \lambda^{(j)}_{i, t} - N_{i, t}}{a^{(j)}_{U} + e_{i, t} \lambda^{(j)}_{i, t}} \right]
\end{align*}}\hspace{-4.1pt}
or

\vspace{-15pt}{\small\begin{align*}
    \bm{g}_{N, 1}\left(\bm{\vartheta}_{N}\right) &= \sum_{i = 1}^{M} \sum_{t = 1}^{T_{i}} \sum_{j = 1}^{K} \gamma^{(j)}_{i, t} \frac{N_{i, t} b^{(j)}_{U} - e_{i, t} \lambda_{i, t} a^{(j)}_{U}}{b^{(j)}_{U} + e_{i, t} \lambda_{i, t}} \bm{A}_{i, t}, \\
     g^{(j)}_{N, 2}\left(\bm{\vartheta}_{N}\right) &= \sum_{i = 1}^{M} \sum_{t = 1}^{T_{i}} \gamma^{(j)}_{i, t} \left[\psi\left(N_{i, t} + a^{(j)}_{U}\right) - \psi\left(a^{(j)}_{U}\right) + \ln\left(b^{(j)}_{U}\right) - \ln\left(b^{(j)}_{U} + e_{i, t} \lambda_{i, t}\right)\right], \\
     g^{(j)}_{N, 3}\left(\bm{\vartheta}_{N}\right) &= \sum_{i = 1}^{M} \sum_{t = 1}^{T_{i}} \gamma^{(j)}_{i, t} \frac{e_{i, t} \lambda_{i, t} a^{(j)}_{U} - N_{i, t} b^{(j)}_{U}}{b^{(j)}_{U} \left(b^{(j)}_{U} + e_{i, t} \lambda_{i, t}\right)}
\end{align*}}\hspace{-4.1pt}
for the number of claims and

\vspace{-15pt}{\small\begin{align*}
    g^{(j)}_{X, 1}\left(\bm{\vartheta}^{(j)}_{X}\right) &= \left. \sum_{i = 1}^{M} \sum_{t = 1}^{T_{i}} \gamma^{(j)}_{i, t} \left\{\frac{\mu^{(j)}_{i, t}}{\varphi^{(j)}} \right[ N_{i, t} \left( \psi\left[\mu^{(j)}_{i, t} + a^{(j)}_{V} \right] - \psi\left[\mu^{(j)}_{i, t}\right] + \ln\left[\varphi^{(j)}\right] + 1 \right) \right. \\
    &\quad + \left.\left. \sum_{n = 0}^{N_{i, t}} \left(\ln\left[X_{i, t, n}\right] - \ln\left[a^{(j)}_{V} - 1 + \varphi^{(j)} X_{i, t, n}\right] \right) \right] - \sum_{n = 0}^{N_{i, t}} \frac{\left[\mu^{(j)}_{i, t} + a^{(j)}_{V}\right] X_{i, t, n}}{a^{(j)}_{V} - 1 + \varphi^{(j)} X_{i, t, n}} \right\}, \\
    \bm{g}^{(j)}_{X, 2}\left(\bm{\vartheta}^{(j)}_{X}\right) &= \left. \sum_{i = 1}^{M} \sum_{t = 1}^{T_{i}} \gamma^{(j)}_{i, t} \mu^{(j)}_{i, t} \right\{ N_{i, t} \left[ \psi\left(\mu^{(j)}_{i, t} + a^{(j)}_{V}\right) - \psi\left(\mu^{(j)}_{i, t}\right) + \ln\left(\varphi^{(j)}\right) \right] \\
    &\quad + \left. \sum_{n = 0}^{N_{i, t}} \left[\ln\left(X_{i, t, n}\right) - \ln\left(a^{(j)}_{V} - 1 + \varphi^{(j)} X_{i, t, n}\right) \right] \right\} \bm{B}_{i, t}, \\
    g^{(j)}_{X, 3}\left(\bm{\vartheta}^{(j)}_{X}\right) &= \left. \sum_{i = 1}^{M} \sum_{t = 1}^{T_{i}} \gamma^{(j)}_{i, t} \right\{ N_{i, t} \left[\psi\left(\mu^{(j)}_{i, t} + a^{(j)}_{V}\right) - \psi\left(a^{(j)}_{V}\right) + \ln\left(a^{(j)}_{V} - 1\right) + \frac{a^{(j)}_{V}}{a^{(j)}_{V} - 1}\right] \\
    &\quad - \left. \sum_{n = 0}^{N_{i, t}} \left[\ln\left(a^{(j)}_{V} - 1 + \varphi^{(j)} X_{i, t, n}\right) + \frac{\mu^{(j)}_{i, t} + a^{(j)}_{V}}{a^{(j)}_{V} - 1 + \varphi^{(j)} X_{i, t, n}} \right] \right\}
\end{align*}}\hspace{-4.1pt}
or

\vspace{-15pt}{\small\begin{align*}
    g_{X, 1}\left(\bm{\vartheta}_{X}\right) &= \left.\left. \sum_{i = 1}^{M} \sum_{t = 1}^{T_{i}} \sum_{j = 1}^{K} \gamma^{(j)}_{i, t} \right\{\frac{\mu_{i, t}}{\varphi} \right[ N_{i, t} \left( \psi\left[\mu_{i, t} + a^{(j)}_{V} \right] - \psi\left[\mu_{i, t}\right] + \ln\left[\varphi\right] + 1 \right) \\
    &\quad + \left.\left. \sum_{n = 0}^{N_{i, t}} \left(\ln\left[X_{i, t, n}\right] - \ln\left[b^{(j)}_{V} + \varphi X_{i, t, n}\right] \right) \right] - \sum_{n = 0}^{N_{i, t}} \frac{\left[\mu_{i, t} + a^{(j)}_{V}\right] X_{i, t, n}}{b^{(j)}_{V} + \varphi X_{i, t, n}} \right\}, \\
    \bm{g}_{X, 2}\left(\bm{\vartheta}_{X}\right) &= \left. \sum_{i = 1}^{M} \sum_{t = 1}^{T_{i}} \sum_{j = 1}^{K} \gamma^{(j)}_{i, t} \mu_{i, t} \right\{ N_{i, t} \left[ \psi\left(\mu_{i, t} + a^{(j)}_{V}\right) - \psi\left(\mu_{i, t}\right) + \ln\left(\varphi\right) \right] \\
    &\quad + \left. \sum_{n = 0}^{N_{i, t}} \left[\ln\left(X_{i, t, n}\right) - \ln\left(b^{(j)}_{V} + \varphi X_{i, t, n}\right) \right] \right\} \bm{B}_{i, t}, \\
    g^{(j)}_{X, 3}\left(\bm{\vartheta}_{X}\right) &= \sum_{i = 1}^{M} \sum_{t = 1}^{T_{i}} \gamma^{(j)}_{i, t} \left\{N_{i, t} \left[\psi\left(\mu_{i, t} + a^{(j)}_{V}\right) - \psi\left(a^{(j)}_{V}\right) + \ln\left(b^{(j)}_{V}\right)\right] - \sum_{n = 0}^{N_{i, t}} \ln\left[b^{(j)}_{V} + \varphi X_{i, t, n}\right]\right\}, \\
    g^{(j)}_{X, 4}\left(\bm{\vartheta}_{X}\right) &= \sum_{i = 1}^{M} \sum_{t = 1}^{T_{i}} \gamma^{(j)}_{i, t} \sum_{n = 0}^{N_{i, t}} \frac{\varphi X_{i, t, n} a^{(j)}_{V} - \mu_{i, t} b^{(j)}_{V}}{b^{(j)}_{V} \left(b^{(j)}_{V} + \varphi X_{i, t, n}\right)}
\end{align*}}\hspace{-4.1pt}
for the claim sizes, for all $j \in \{1, \dots, K\}$ and where $\psi(x) = \frac{\mathrm{d}}{\mathrm{d}x} \ln\left(\Gamma(x)\right)$ denotes the digamma function. The respective Hessians $\bm{H}^{(j)}_{N}\left(\bm{\vartheta}^{(j)}_{N}\right)$ and $\bm{H}^{(j)}_{X}\left(\bm{\vartheta}^{(j)}_{X}\right)$ or $\bm{H}_{N}\left(\bm{\vartheta}_{N}\right)$ and $\bm{H}_{X}\left(\bm{\vartheta}_{X}\right)$ are given by

\vspace{-15pt}{\small\begin{align*}
    \bm{H}^{(j)}_{N, 1, 1}\left(\bm{\vartheta}^{(j)}_{N}\right) &= - \sum_{i = 1}^{M} \sum_{t = 1}^{T_{i}} \gamma^{(j)}_{i, t} \frac{e_{i, t} \lambda^{(j)}_{i, t} a^{(j)}_{U} \left(a^{(j)}_{U} + N_{i, t}\right)}{\left(a^{(j)}_{U} + e_{i, t} \lambda^{(j)}_{i, t}\right)^2} \bm{A}_{i, t} \bm{A}_{i, t}^{\prime}, \\
    \bm{H}^{(j)}_{N, 1, 2}\left(\bm{\vartheta}^{(j)}_{N}\right) &= \sum_{i = 1}^{M} \sum_{t = 1}^{T_{i}} \gamma^{(j)}_{i, t} \frac{e_{i, t} \lambda^{(j)}_{i, t} \left(N_{i, t} - e_{i, t} \lambda^{(j)}_{i, t}\right)}{\left(a^{(j)}_{U} + e_{i, t} \lambda^{(j)}_{i, t}\right)^2} \bm{A}_{i, t}^{\prime}, \\
    H^{(j)}_{N, 2, 2}\left(\bm{\vartheta}^{(j)}_{N}\right) &= \sum_{i = 1}^{M} \sum_{t = 1}^{T_{i}} \gamma^{(j)}_{i, t} \left[\psi_{1}\left(N_{i, t} + a^{(j)}_{U}\right) - \psi_{1}\left(a^{(j)}_{U}\right) + \frac{1}{a^{(j)}_{U}} - \frac{a^{(j)}_{U} + 2 e_{i, t} \lambda^{(j)}_{i, t} - N_{i, t}}{\left(a^{(j)}_{U} + e_{i, t} \lambda^{(j)}_{i, t}\right)^2}\right]
\end{align*}}\hspace{-4.1pt}
or

\vspace{-15pt}{\small\begin{align*}
    \bm{H}_{N, 1, 1}\left(\bm{\vartheta}_{N}\right) &= - \sum_{i = 1}^{M} \sum_{t = 1}^{T_{i}} \sum_{j = 1}^{K} \gamma^{(j)}_{i, t} \frac{e_{i, t} \lambda_{i, t} b^{(j)}_{U} \left(a^{(j)}_{U} + N_{i, t}\right)}{\left(b^{(j)}_{U} + e_{i, t} \lambda_{i, t}\right)^2} \bm{A}_{i, t} \bm{A}_{i, t}^{\prime}, \\
    \bm{H}^{(j)}_{N, 1, 2}\left(\bm{\vartheta}_{N}\right) &= - \sum_{i = 1}^{M} \sum_{t = 1}^{T_{i}} \gamma^{(j)}_{i, t} \frac{e_{i, t} \lambda_{i, t}}{b^{(j)}_{U} + e_{i, t} \lambda_{i, t}} \bm{A}_{i, t}^{\prime}, \\
    \bm{H}^{(j)}_{N, 1, 3}\left(\bm{\vartheta}_{N}\right) &= \sum_{i = 1}^{M} \sum_{t = 1}^{T_{i}} \gamma^{(j)}_{i, t} \frac{e_{i, t} \lambda_{i, t} \left(a^{(j)}_{U} + N_{i, t}\right)}{\left(b^{(j)}_{U} + e_{i, t} \lambda_{i, t}\right)^2} \bm{A}_{i, t}^{\prime}, \\
    H^{(h, j)}_{N, 2, 2}\left(\bm{\vartheta}_{N}\right) &= \mathbbm{1}\left[h = j\right] \sum_{i = 1}^{M} \sum_{t = 1}^{T_{i}} \gamma^{(j)}_{i, t} \left[\psi_{1}\left(N_{i, t} + a^{(j)}_{U}\right) - \psi_{1}\left(a^{(j)}_{U}\right)\right], \\
    H^{(h, j)}_{N, 2, 3}\left(\bm{\vartheta}_{N}\right) &= \mathbbm{1}\left[h = j\right] \sum_{i = 1}^{M} \sum_{t = 1}^{T_{i}} \gamma^{(j)}_{i, t} \frac{e_{i, t} \lambda_{i, t}}{b^{(j)}_{U} \left(b^{(j)}_{U} + e_{i, t} \lambda_{i, t}\right)}, \\
    H^{(h, j)}_{N, 3, 3}\left(\bm{\vartheta}_{N}\right) &= - \mathbbm{1}\left[h = j\right] \sum_{i = 1}^{M} \sum_{t = 1}^{T_{i}} \gamma^{(j)}_{i, t} \frac{e_{i, t} \lambda_{i, t} a^{(j)}_{U} \left(e_{i, t} \lambda_{i, t} + 2 b^{(j)}_{U}\right) - N_{i, t} {b^{(j)}_{U}}^2}{{b^{(j)}_{U}}^2 \left(b^{(j)}_{U} + e_{i, t} \lambda_{i, t}\right)^2}
\end{align*}}\hspace{-4.1pt}
for the number of claims and

\vspace{-15pt}{\small\begin{align*}
    H^{(j)}_{X, 1, 1}\left(\bm{\vartheta}^{(j)}_{X}\right) &= \left.\left. \sum_{i = 1}^{M} \sum_{t = 1}^{T_{i}} \gamma^{(j)}_{i, t} \right\{ \frac{\mu^{(j)}_{i, t}}{\varphi^{(j)}} \right[ \frac{N_{i, t}}{\varphi^{(j)}} \left(\mu^{(j)}_{i, t} \left[\psi_{1}\left(\mu^{(j)}_{i, t} + a^{(j)}_{V}\right) - \psi_{1}\left(\mu^{(j)}_{i, t}\right)\right] + 1\right) \\
    &\quad - \left.\left. \sum_{n = 0}^{N_{i, t}} \frac{2 X_{i, t, n}}{a^{(j)}_{V} - 1 + \varphi^{(j)} X_{i, t, n}}\right] + \sum_{n = 0}^{N_{i, t}} \frac{\left[\mu^{(j)}_{i, t} + a^{(j)}_{V}\right] X_{i, t, n}^2}{\left[a^{(j)}_{V} - 1 + \varphi^{(j)} X_{i, t, n}\right]^2} \right\}, \\
    \bm{H}^{(j)}_{X, 1, 2}\left(\bm{\vartheta}^{(j)}_{X}\right) &= \left. \sum_{i = 1}^{M} \sum_{t = 1}^{T_{i}} \gamma^{(j)}_{i, t} \frac{\mu^{(j)}_{i, t}}{\varphi^{(j)}} \right\{ N_{i, t} \left[ \psi\left(\mu^{(j)}_{i, t} + a^{(j)}_{V}\right) - \psi\left(\mu^{(j)}_{i, t}\right) + \mu^{(j)}_{i, t} \left(\psi_{1}\left[\mu^{(j)}_{i, t} + a^{(j)}_{V}\right] - \psi_{1}\left[\mu^{(j)}_{i, t}\right]\right) \right. \\
    &\quad + \left.\left. \ln\left(\varphi^{(j)}\right) + 1 \right] + \sum_{n = 0}^{N_{i, t}} \left[\ln\left(X_{i, t, n}\right) - \ln\left(a^{(j)}_{V} - 1 + \varphi^{(j)} X_{i, t, n}\right) - \frac{\varphi^{(j)} X_{i, t, n}}{a^{(j)}_{V} - 1 + \varphi^{(j)} X_{i, t, n}}\right] \right\} \bm{B}_{i, t}^{\prime}, \\
    H^{(j)}_{X, 1, 3}\left(\bm{\vartheta}^{(j)}_{X}\right) &= \sum_{i = 1}^{M} \sum_{t = 1}^{T_{i}} \frac{\gamma^{(j)}_{i, t}}{\varphi^{(j)}} \left[ N_{i, t} \mu^{(j)}_{i, t} \psi_{1}\left(\mu^{(j)}_{i, t} + a^{(j)}_{V}\right) - \sum_{n = 0}^{N_{i, t}} \frac{\varphi^{(j)} X_{i, t, n} \left(\varphi^{(j)} X_{i, t, n} - 1\right) + \mu^{(j)}_{i, t} \left(a^{(j)}_{V} - 1\right)}{\left(a^{(j)}_{V} - 1 + \varphi^{(j)} X_{i, t, n}\right)^2} \right], \\
\end{align*}}\hspace{-4.1pt}
\vspace{-15pt}{\small\begin{align*}
    \bm{H}^{(j)}_{X, 2, 2}\left(\bm{\vartheta}^{(j)}_{X}\right) &= \left. \sum_{i = 1}^{M} \sum_{t = 1}^{T_{i}} \gamma^{(j)}_{i, t} \mu^{(j)}_{i, t} \right\{N_{i, t} \left[\psi\left(\mu^{(j)}_{i, t} + a^{(j)}_{V}\right) - \psi\left(\mu^{(j)}_{i, t}\right) + \mu^{(j)}_{i, t} \left(\psi_{1}\left[\mu^{(j)}_{i, t} + a^{(j)}_{V}\right] - \psi_{1}\left[\mu^{(j)}_{i, t}\right]\right) \right. \hspace{34pt} \\
    &\quad  + \left.\left. \ln\left(\varphi^{(j)}\right)\right] + \sum_{n = 0}^{N_{i, t}} \left[ \ln\left(X_{i, t, n}\right) - \ln\left(a^{(j)}_{V} - 1 + \varphi^{(j)} X_{i, t, n}\right) \right] \right\} \bm{B}_{i, t} \bm{B}_{i, t}^{\prime}, \\
    \bm{H}^{(j)}_{X, 2, 3}\left(\bm{\vartheta}^{(j)}_{X}\right) &= \sum_{i = 1}^{M} \sum_{t = 1}^{T_{i}} \gamma^{(j)}_{i, t} \mu^{(j)}_{i, t} \left[ N_{i, t} \psi_{1}\left(\mu^{(j)}_{i, t} + a^{(j)}_{V}\right) - \sum_{n = 0}^{N_{i, t}} \frac{1}{a^{(j)}_{V} - 1 + \varphi^{(j)} X_{i, t, n}} \right] \bm{B}_{i, t}^{\prime}, \\
    H^{(j)}_{X, 3, 3}\left(\bm{\vartheta}^{(j)}_{X}\right) &= \sum_{i = 1}^{M} \sum_{t = 1}^{T_{i}} \gamma^{(j)}_{i, t} \left\{ N_{i, t} \left[\psi_{1}\left(\mu^{(j)}_{i, t} + a^{(j)}_{V}\right) - \psi_{1}\left(a^{(j)}_{V}\right) + \frac{a^{(j)}_{V} - 2}{\left(a^{(j)}_{V} - 1\right)^2}\right] \right. \\
    &\quad - \left. \sum_{n = 0}^{N_{i, t}} \frac{a^{(j)}_{V} - 2 + 2\varphi^{(j)} X_{i, t, n} - \mu^{(j)}_{i, t}}{\left[a^{(j)}_{V} - 1 + \varphi^{(j)} X_{i, t, n}\right]^2} \right\}
\end{align*}}\hspace{-4.1pt}
or

\vspace{-15pt}{\small\begin{align*}
    H_{X, 1, 1}\left(\bm{\vartheta}_{X}\right) &= \sum_{i = 1}^{M} \sum_{t = 1}^{T_{i}} \sum_{j = 1}^{K} \gamma^{(j)}_{i, t} \left\{ \frac{\mu_{i, t}}{\varphi} \left[ \frac{N_{i, t}}{\varphi} \left(\mu_{i, t} \left[\psi_{1}\left(\mu_{i, t} + a^{(j)}_{V}\right) - \psi_{1}\left(\mu_{i, t}\right)\right] + 1\right) - \sum_{n = 0}^{N_{i, t}} \frac{X_{i, t, n}}{b^{(j)}_{V} + \varphi X_{i, t, n}}\right] \right. \\
    &\quad - \left. \sum_{n = 0}^{N_{i, t}} \frac{\mu_{i, t} b^{(j)}_{V} - \varphi X_{i, t, n} a^{(j)}_{V}}{\varphi \left[b^{(j)}_{V} + \varphi X_{i, t, n}\right]^2} X_{i, t, n} \right\}, \\
    \bm{H}_{X, 1, 2}\left(\bm{\vartheta}_{X}\right) &= \left. \sum_{i = 1}^{M} \sum_{t = 1}^{T_{i}} \sum_{j = 1}^{K} \gamma^{(j)}_{i, t} \frac{\mu_{i, t}}{\varphi} \right\{ N_{i, t} \left[ \psi\left(\mu_{i, t} + a^{(j)}_{V}\right) - \psi\left(\mu_{i, t}\right) + \mu_{i, t} \left(\psi_{1}\left[\mu_{i, t} + a^{(j)}_{V}\right] - \psi_{1}\left[\mu_{i, t}\right]\right) \right. \\
    &\quad + \left.\left. \ln\left(\varphi\right) + 1 \right] + \sum_{n = 0}^{N_{i, t}} \left[\ln\left(X_{i, t, n}\right) - \ln\left(b^{(j)}_{V} + \varphi X_{i, t, n}\right) - \frac{\varphi X_{i, t, n}}{b^{(j)}_{V} + \varphi X_{i, t, n}}\right] \right\} \bm{B}_{i, t}^{\prime}, \\
    H^{(j)}_{X, 1, 3}\left(\bm{\vartheta}_{X}\right) &= \sum_{i = 1}^{M} \sum_{t = 1}^{T_{i}} \gamma^{(j)}_{i, t} \left[N_{i, t} \frac{\mu_{i, t}}{\varphi} \psi_{1}\left(\mu_{i, t} + a^{(j)}_{V}\right) - \sum_{n = 0}^{N_{i, t}} \frac{X_{i, t, n}}{b^{(j)}_{V} + \varphi X_{i, t, n}} \right], \\
    H^{(j)}_{X, 1, 4}\left(\bm{\vartheta}_{X}\right) &= \sum_{i = 1}^{M} \sum_{t = 1}^{T_{i}} \gamma^{(j)}_{i, t} \sum_{n = 0}^{N_{i, t}} \frac{\varphi X_{i, t, n} a^{(j)}_{V} - \mu_{i, t} b^{(j)}_{V}}{\varphi \left(b^{(j)}_{V} + \varphi X_{i, t, n}\right)^2}, \\
    \bm{H}_{X, 2, 2}\left(\bm{\vartheta}_{X}\right) &= \left. \sum_{i = 1}^{M} \sum_{t = 1}^{T_{i}} \sum_{j = 1}^{K} \gamma^{(j)}_{i, t} \mu_{i, t} \right\{N_{i, t} \left[\psi\left(\mu_{i, t} + a^{(j)}_{V}\right) - \psi\left(\mu_{i, t}\right) + \mu_{i, t} \left(\psi_{1}\left[\mu_{i, t} + a^{(j)}_{V}\right] - \psi_{1}\left[\mu_{i, t}\right]\right) \right. \\
    &\quad + \left.\left. \ln\left(\varphi\right)\right] + \sum_{n = 0}^{N_{i, t}} \left[ \ln\left(X_{i, t, n}\right) - \ln\left(b^{(j)}_{V} + \varphi X_{i, t, n}\right) \right] \right\} \bm{B}_{i, t} \bm{B}_{i, t}^{\prime}, \\
    \bm{H}^{(j)}_{X, 2, 3}\left(\bm{\vartheta}_{X}\right) &= \sum_{i = 1}^{M} \sum_{t = 1}^{T_{i}} \gamma^{(j)}_{i, t} N_{i, t} \mu_{i, t} \psi_{1}\left(\mu_{i, t} + a^{(j)}_{V}\right) \bm{B}_{i, t}^{\prime}, \\
    \bm{H}^{(j)}_{X, 2, 4}\left(\bm{\vartheta}_{X}\right) &= - \sum_{i = 1}^{M} \sum_{t = 1}^{T_{i}} \gamma^{(j)}_{i, t} \sum_{n = 0}^{N_{i, t}} \frac{\mu_{i, t}}{b^{(j)}_{V} + \varphi X_{i, t, n}} \bm{B}_{i, t}^{\prime}, \\
    H^{(h, j)}_{X, 3, 3}\left(\bm{\vartheta}_{X}\right) &= \mathbbm{1}\left[h = j\right] \sum_{i = 1}^{M} \sum_{t = 1}^{T_{i}} \gamma^{(j)}_{i, t} N_{i, t} \left[\psi_{1}\left(\mu_{i, t} + a^{(j)}_{V}\right) - \psi_{1}\left(a^{(j)}_{V}\right)\right], \\
    H^{(h, j)}_{X, 3, 4}\left(\bm{\vartheta}_{X}\right) &= \mathbbm{1}\left[h = j\right] \sum_{i = 1}^{M} \sum_{t = 1}^{T_{i}} \gamma^{(j)}_{i, t} \sum_{n = 0}^{N_{i, t}} \frac{\varphi X_{i, t, n}}{b^{(j)}_{V} \left(b^{(j)}_{V} + \varphi X_{i, t, n}\right)},
\end{align*}}\hspace{-4.1pt}
\vspace{-15pt}{\small\begin{align*}
    H^{(h, j)}_{X, 4, 4}\left(\bm{\vartheta}_{X}\right) &= - \mathbbm{1}\left[h = j\right] \sum_{i = 1}^{M} \sum_{t = 1}^{T_{i}} \gamma^{(j)}_{i, t} \sum_{n = 0}^{N_{i, t}} \frac{\varphi X_{i, t, n} a^{(j)}_{V} \left(\beta X_{i, t, n} + 2 b^{(j)}_{V}\right) - \mu_{i, t} {b^{(j)}_{V}}^2}{{b^{(j)}_{V}}^2 \left(b^{(j)}_{V} + \varphi X_{i, t, n}\right)^2} \hspace{102pt}
\end{align*}}\hspace{-4.1pt}
for the claim sizes, for all $h, j \in \{1, \dots, K\}$ and where $\psi_{1}(x) = \frac{\mathrm{d}}{\mathrm{d}x} \psi(x)$ denotes the trigamma function. Since the Hessians involve complicated expressions of $N_{i, t}$ or $X_{i, t, n}$, it becomes infeasible to calculate the Fisher information, or negative expected Hessian, matrices necessary for the Fisher scoring method. We therefore adopt the Newton-Raphson method, where $\tensor*[^{(s + 1, r)}]{\!\bm{\vartheta}}{} = \tensor*[^{(s, r)}]{\!\bm{\vartheta}}{} - \bm{H}\left(\tensor*[^{(s, r)}]{\!\bm{\vartheta}}{}\right)^{-1} \bm{g}\left(\tensor*[^{(s, r)}]{\!\bm{\vartheta}}{}\right)$ in iteration $s + 1$, to overcome this issue. Finally, the asymptotic variance of the parameter estimates also depends on the Fisher information matrix, but a sample analogue can be determined from the diagonal entries of the inverse of the observed information, or negative Hessian, matrix. 
\section{Supplementary material} \label{AppendixC}

\begin{table}[hb!]
    \caption{Log-likelihood, information criteria and portfolio loss ratio of the MTPL insurance HMMs.}
    \label{TC.1}\vspace{-3pt}\vspace{-3pt}
    \centerline{\scalebox{0.94}{\begin{tabularx}{1.064\textwidth}{l c r c rrr c rr}
		\toprule \addlinespace[1ex] \vspace{1pt}
		&& && && && \multicolumn{2}{c}{\textbf{Loss ratio (\%)}} \\ \cline{9-10} \addlinespace[0.4ex]
		\textbf{Mixture} && \textbf{Parameters} && \textbf{Log-likelihood} & \textbf{AIC} & \textbf{BIC} && \textbf{Prior} & \textbf{Posterior} \\ \hline \addlinespace[0.4ex]
		\textit{Independent}\hspace{-5pt} && 82 && -77{,}384.62 & 154{,}933.23 & 155{,}767.52 && 130.38 & 129.10 \\ 
		\quad ($\mathit{K = 1}$) \\ \hline \addlinespace[0.4ex]
		$\mathit{K = 2}$ && && & & && & \\ 
		\quad - \textit{Full} && 170 && -77{,}014.92 & 154{,}369.85 & 156{,}099.48 && 88.99 & 117.52 \\ 
		\quad - \textit{Sparse} && 92 && -77{,}208.80 & 154{,}601.60 & 155{,}537.60 && 97.20 & 110.18 \\ \hline \addlinespace[0.4ex]
		$\mathit{K = 3}$ && && & & && & \\
		\quad - \textit{Full} && 258 && -76{,}987.57 & 154{,}491.14 & 157{,}116.10 && 117.24 & 118.96 \\ 
		\quad - \textit{Sparse} && 102 && -77{,}119.25 & 154{,}442.49 & \textbf{155{,}480.27} && \textbf{99.12} & 122.32 \\ \hline \addlinespace[0.4ex]
		$\mathit{K = 4}$ && && & & && & \\
		\quad - \textit{Full} && 348 && \textbf{-76{,}755.70} & \textbf{154{,}207.40} & 157{,}748.00 && 96.07 & 117.89 \\ 
		\quad - \textit{Sparse} && 114 && -77{,}060.86 & 154{,}349.71 & 155{,}509.58 && 65.75 & \textbf{92.44} \\
		\bottomrule
	\end{tabularx}}}
\end{table}

\begin{table}[ht!]
    \caption{Estimated prior parameters for each MTPL insurance HMM.}
    \label{TC.2}\vspace{-3pt}\vspace{-3pt}
	\centerline{\scalebox{0.965}{\begin{tabularx}{1.04\textwidth}{l p{1.9em} rr p{2.5em} rrrr}
		\toprule \addlinespace[1ex] \vspace{1pt}
		&& \multicolumn{2}{c}{\textbf{Full representation}} && \multicolumn{4}{c}{\textbf{Sparse representation}} \\ \cline{3-4} \cline{6-9} \addlinespace[0.4ex]
		\textbf{Mixture} && \multicolumn{1}{c}{$\bm{a^{(j)}_{U}}$} & \multicolumn{1}{c}{$\bm{a^{(j)}_{V}}$} && \multicolumn{1}{c}{$\bm{a^{(j)}_{U}}$} & \multicolumn{1}{c}{$\bm{b^{(j)}_{U}}$} & \multicolumn{1}{c}{$\bm{a^{(j)}_{V}}$} & \multicolumn{1}{c}{$\bm{b^{(j)}_{V}}$} \\ \hline \addlinespace[0.4ex]
		\textit{Independent} && 0.2924 & 1.7597 && 0.2924 & 6.1742 & 1.7597 & 1{,}814.0806 \\ 
		\quad ($\mathit{K = 1}$) \\ \hline \addlinespace[0.4ex]
		$\mathit{K = 2}$ \\
		\quad - $\mathit{j = 1}$ && 0.3893 & 1{,}541.9100 && 0.2737 & 5.5105 & 1.5522 & 1{,}213.37 \\
		\quad - $\mathit{j = 2}$ && 0.9713 & 1.4317 && 1.0743 & 40.3253 & 1.0144 & 82.4183 \\ \hline \addlinespace[0.4ex]
		$\mathit{K = 3}$ \\
		\quad - $\mathit{j = 1}$ && 16.0000 & 231.9674 && 0.2717 & 115.2963 & 4.7075 & 423.5172 \\
		\quad - $\mathit{j = 2}$ && 1.2907 & 174.3674 && 0.0763 & 4.9981 & 1.0070 & 41.9193 \\ 
		\quad - $\mathit{j = 3}$ && 0.8251 & 1.4286 && 1.0873 & 10.6704 & 1.5563 & 1{,}166.6399 \\ \hline \addlinespace[0.4ex]
		$\mathit{K = 4}$ \\
		\quad - $\mathit{j = 1}$ && 0.1128 & 1{,}124.7680 && 0.0032 & 46.4526 & 35.0477 & 295.0512 \\
		\quad - $\mathit{j = 2}$ && 77.9492 & 151.8134 && 4.9798 & 136.7600 & 3.5451 & 2{,}038.4914 \\
		\quad - $\mathit{j = 3}$ && 33.9498 & 1{,}263.3680 && 7.9950 & 123.2833 & 1.0703 & 77.2851 \\
		\quad - $\mathit{j = 4}$ && 1.3160 & 1.4524 && 0.6714 & 5.3886 & 1.2336 & 779.7081 \\
		\bottomrule
	\end{tabularx}}}
\end{table}

\begin{table}[ht!]
    \caption{Estimated prior transition probabilities in percentages for each MTPL insurance HMM.}
    \label{TC.3}\vspace{-3pt}\vspace{-3pt}
	\centerline{\scalebox{0.83}{\begin{tabularx}{1.20\textwidth}{l c rrrr c rrrr}
		\toprule \addlinespace[1ex] \vspace{1pt}
		&& \multicolumn{4}{c}{\textbf{Full representation}} && \multicolumn{4}{c}{\textbf{Sparse representation}} \\ \cline{3-6} \cline{8-11} \addlinespace[0.4ex]
		\textbf{Mixture} && \textbf{To }$\bm{j = 1}$ & \textbf{To }$\bm{j = 2}$ & \textbf{To }$\bm{j = 3}$ & \textbf{To }$\bm{j = 4}$ && \textbf{To }$\bm{j = 1}$ & \textbf{To }$\bm{j = 2}$ & \textbf{To }$\bm{j = 3}$ & \textbf{To }$\bm{j = 4}$ \\ \hline \addlinespace[0.4ex]
		$\mathit{K = 2}$ \\
		\quad - \textit{From }$\mathit{h = 0}$ && 64.7900 & 35.2100 &  &  && 89.4018 & 10.5982 &  &  \\
		\quad - \textit{From }$\mathit{h = 1}$ && 91.9563 & 8.0437 &  &  && 99.1273 & 0.8727&  &  \\
		\quad - \textit{From }$\mathit{h = 2}$ && 25.1172 & 74.8828 &  &  && 13.6503 & 86.3497 &  &  \\ \hline \addlinespace[0.4ex]
		$\mathit{K = 3}$ \\
		\quad - \textit{From }$\mathit{h = 0}$ && 28.6298 & 33.2533 & 38.1170 &  && 51.5818 & 4.5401 & 43.8781 &  \\
		\quad - \textit{From }$\mathit{h = 1}$ && 98.7127 & 0.0291 & 1.2582 &  && 89.8854 & 0.2112 & 9.9034 &  \\
		\quad - \textit{From }$\mathit{h = 2}$ && 1.8680 & 67.0011 & 31.1309 &  && 0.1656 & 3.3070 & 96.5274 &  \\
		\quad - \textit{From }$\mathit{h = 3}$ && 3.1558 & 38.4530 & 58.3911 &  && 17.0302 & 10.7982 & 72.1717 &  \\ \hline \addlinespace[0.4ex]
		$\mathit{K = 4}$ \\
		\quad - \textit{From }$\mathit{h = 0}$ && 43.6139 & 16.7018 & 9.8774 & 29.8069 && 34.3907 & 35.7452 & 4.6674 & 25.1968 \\
		\quad - \textit{From }$\mathit{h = 1}$ && 92.1153 & 7.8847 & {\small$<$}0.0000 & {\small$<$}0.0000 && 96.6924 & 3.3076 & {\small$<$}0.0000 & {\small$<$}0.0000 \\
		\quad - \textit{From }$\mathit{h = 2}$ && 15.6389 & 26.3936 & 32.5300 & 25.4375 && 0.0035 & 83.9551 & 1.0175 & 15.0239 \\
		\quad - \textit{From }$\mathit{h = 3}$ && 18.6846 & 13.0733 & {\small$<$}0.0000 & 68.2421 && {\small$<$}0.0000 & 0.0002 & 42.9188 & 57.0810 \\
		\quad - \textit{From }$\mathit{h = 4}$ && {\small$<$}0.0000 & 31.7293 & 24.0639 & 44.2067 && 10.6968 & 22.8832 & 6.9203 & 59.4997 \\
		\bottomrule
	\end{tabularx}}}
\end{table}

\begin{table}[ht!]
    \caption{Mean [median] of MTPL insurance observations and predictions for each sparse mixture component, where the claim frequencies and risk premia are weighted by their corresponding exposures to risk.}
    \label{TC.4}\vspace{-3pt}\vspace{-3pt}
	\centerline{\scalebox{0.76}{\begin{tabularx}{1.32\textwidth}{l c rrr c rrr c rrr}
		\toprule \addlinespace[1ex] \vspace{1pt}
		&& \multicolumn{3}{c}{\textbf{Claim frequency}} && \multicolumn{3}{c}{\textbf{Claim severity}} && \multicolumn{3}{c}{\textbf{Risk premium}} \\ \cline{3-5} \cline{7-9} \cline{11-13} \addlinespace[0.4ex]
		\textbf{Mixture} && \textbf{Observed} & \textbf{Prior} & \textbf{Posterior} && \textbf{Observed} & \textbf{Prior} & \textbf{Posterior} && \textbf{Observed} & \textbf{Prior} & \textbf{Posterior} \\ \hline \addlinespace[0.4ex]
		\textit{Independent}\hspace{-10pt} && 0.0524\phantom{]} & 0.0559\phantom{]} & 0.0562\phantom{]} && 3{,}445.43\phantom{]} & 2{,}475.57\phantom{]} & 2{,}525.16\phantom{]} && 180.48\phantom{]} & 138.43\phantom{]} & 139.80\phantom{]} \\
		\quad ($\mathit{K = 1}$) && [0.0000] & [0.0514] & [0.0443] && [1{,}100.00] & [2{,}462.14] & [2{,}444.35] && [0.00] & [125.21] & [107.47] \\ \hline \addlinespace[0.4ex]
		$\mathit{K = 2}$ && 0.0524\phantom{]} & 0.0479\phantom{]} & 0.0563\phantom{]} && 3{,}445.43\phantom{]} & 4{,}509.14\phantom{]} & 3{,}147.61\phantom{]} && 180.48\phantom{]} & 185.68\phantom{]} & 163.81\phantom{]}  \\
		&& [0.0000] & [0.0438] & [0.0441] && [1{,}100.00] & [3{,}887.16] & [3{,}091.25] && [0.00] & [163.58] & [125.30] \\
		\quad - $\mathit{j = 1}$ &&  & 0.0588\phantom{]} & 0.0586\phantom{]} &&  & 2{,}721.04\phantom{]} & 2{,}756.99\phantom{]} && & 160.22\phantom{]} & 157.20\phantom{]} \\
		&& & [0.0361] & [0.0325] && & [2{,}695.96] & [2{,}669.02] && & [142.40] & [118.29] \\
		\quad - $\mathit{j = 2}$ &&  & 0.0315\phantom{]} & 0.0324\phantom{]} &&  & 7{,}087.61\phantom{]} & 6{,}706.65\phantom{]} && & 223.81\phantom{]} & 211.73\phantom{]} \\
		&& & [0.0289] & [0.0287] && & [7{,}022.28] & [6{,}935.12] && & [198.92] & [186.84] \\ \hline \addlinespace[0.4ex]
		$\mathit{K = 3}$ && 0.0524\phantom{]} & 0.0633\phantom{]} & 0.0551\phantom{]} && 3{,}445.43\phantom{]} & 2{,}462.12\phantom{]} & 1{,}656.95\phantom{]} && 180.48\phantom{]} & 182.09\phantom{]} & 147.55\phantom{]} \\
		&& [0.0000] & [0.0642] & [0.0459] && [1{,}100.00] & [2{,}484.10] & [1{,}489.85] && [0.00] & [175.88] & [116.01] \\
		\quad - $\mathit{j = 1}$ &&  & 0.0030\phantom{]} & 0.0038\phantom{]} && & 147.17\phantom{]} & 427.42\phantom{]} && & 0.45\phantom{]} & 2.92\phantom{]} \\
		&& & [0.0028] & [0.0028] && & [146.07] & [147.87] && & [0.40] & [0.41] \\
		\quad - $\mathit{j = 2}$ &&  & 0.0196\phantom{]} & 0.0275\phantom{]} && & 7{,}686.65\phantom{]} & 7{,}247.72\phantom{]} && & 151.03\phantom{]} & 146.15\phantom{]} \\
		&& & [0.0181] & [0.0150] && & [7{,}629.23] & [7{,}533.46] && & [135.19] & [109.31] \\
		\quad - $\mathit{j = 3}$ &&  & 0.1310\phantom{]} & 0.1220\phantom{]} && & 2{,}701.73\phantom{]} & 2{,}739.94\phantom{]} && & 354.25\phantom{]} & 327.49\phantom{]} \\
		&& & [0.1206] & [0.1093] && & [2{,}681.55] & [2{,}656.25] && & [317.09] & [276.18] \\ \hline \addlinespace[0.4ex]
		$\mathit{K = 4}$ && 0.0524\phantom{]} & 0.0691\phantom{]} & 0.0546\phantom{]} && 3{,}445.43\phantom{]} & 2{,}121.38\phantom{]} & 1{,}836.25\phantom{]} && 180.48\phantom{]} & 274.49\phantom{]} & 195.24\phantom{]} \\
		&& [0.0000] & [0.0618] & [0.0460] && [1{,}100.00] & [2{,}003.32] & [1{,}692.04] && [0.00] & [236.51] & [156.88] \\
		\quad - $\mathit{j = 1}$ &&  & {\small$<$}0.0000\phantom{]} & 0.0020\phantom{]} && & 13.96\phantom{]} & 206.48\phantom{]} && & {\small$<$}0.00\phantom{]} & 3.20\phantom{]} \\
		&& & [{\small$<$}0.0000] & [{\small$<$}0.0000] && & [13.79] & [14.01] && & [{\small$<$}0.00] & [{\small$<$}0.00] \\
		\quad - $\mathit{j = 2}$ &&  & 0.0466\phantom{]} & 0.0467\phantom{]} && & 1{,}290.41\phantom{]} & 1{,}458.20\phantom{]} && & 60.24\phantom{]} & 64.76\phantom{]} \\
		&& & [0.0430] & [0.0430] && & [1{,}275.03] & [1{,}271.20] && & [53.50] & [52.88] \\
		\quad - $\mathit{j = 3}$ &&  & 0.0830\phantom{]} & 0.0826\phantom{]} && & 1{,}772.12\phantom{]} & 1{,}900.42\phantom{]} && & 147.34\phantom{]} & 151.22\phantom{]} \\
		&& & [0.0765] & [0.0761] && & [1{,}751.00] & [1{,}738.05] && & [130.86] & [127.70] \\
		\quad - $\mathit{j = 4}$ &&  & 0.1594\phantom{]} & 0.1400\phantom{]} && & 5{,}377.23\phantom{]} & 5{,}160.02\phantom{]} && & 858.96\phantom{]} & 701.86\phantom{]} \\
		&& & [0.1470] & [0.1231] && & [5{,}313.15] & [5{,}241.05] && & [762.93] & [600.31] \\
		\bottomrule
	\end{tabularx}}}
\end{table}\clearpage

\begin{table}[ht!]
    \caption{Average Bonus-Malus correction of prior risk premium in terms of total previous number of claims and claim amount in independent model ($K = 1$) and each sparse HMM in MTPL insurance. These corrections coincide with those shown in \Cref{F4.6} but are represented in tabular form in this table.}
    \label{TC.5}\vspace{-3pt}\vspace{-3pt}
    \centerline{\scalebox{0.8505}{\begin{tabularx}{1.18\textwidth}{l p{0.5em} rrrrr p{1.5em} rrrrr}
		\toprule \addlinespace[1ex] \vspace{1pt}
		&& \multicolumn{11}{c}{\textbf{Average Bonus-Malus correction of prior risk premium ($\mathbf{\times\textbf{\texteuro}100}$)}} \\ \cline{3-13} \addlinespace[0.4ex]
		&& \multicolumn{5}{c}{\textbf{Total number of claims}} && \multicolumn{5}{c}{\textbf{Total claim amount ($\mathbf{\times\textbf{\texteuro}100}$)}} \\ \cline{3-7} \cline{9-13} \addlinespace[0.4ex]
		\textbf{Independent ($\mathbf{K = 1}$)} && $\mathbf{0}$ & $\mathbf{1}$ & $\mathbf{2}$ & $\mathbf{3}$ & $\mathbf{4}$ && $\mathbf{0}$ & $\mathbf{25}$ & $\mathbf{50}$ & $\mathbf{75}$ & $\mathbf{100}$ \\ \hline \addlinespace[0.4ex]
		\textit{Total exposure to risk} \\
		\quad - \textit{1 year} && -0.16 & 4.96 & 8.25 & 24.15 & 26.07 && -0.16 & 4.87 & 10.65 & 19.28 & 21.68 \\
		\quad - \textit{2 years} && -0.33 & 3.20 & 6.85 & 15.18 & 5.09 && -0.33 & 3.33 & 7.06 & 12.82 & 18.77 \\
		\quad - \textit{3 years} && -0.43 & 3.18 & 3.52 & 9.19 & 12.78 && -0.43 & 2.60 & 5.34 & 7.74 & 11.63 \\
		\quad - \textit{4 years} && -0.51 & 2.19 & 3.26 & 7.64 & 16.33 && -0.51 & 2.19 & 5.08 & 7.34 & 11.60 \\
		\quad - \textit{5 years} && -0.55 & 1.42 & 3.35 & 9.97 & 20.08 && -0.55 & 1.77 & 4.11 & 5.98 & 7.40 \\
		\quad - \textit{6 years} && -0.59 & 1.76 & 12.42 & 4.74 & 2.29 && -0.59 & 1.39 & 3.62 & 6.75 &  \\
		\quad - \textit{7 years} && -0.59 & 1.09 & 8.09 & 3.44 & 2.08 && -0.59 & 1.15 & 2.88 & 6.12 & 6.78 \\
		\quad - \textit{8 years} && -0.60 & 0.80 & 3.46 & 3.86 & 2.118 && -0.60 & 1.16 & 2.72 & 3.93 & 6.53 \\
		\quad - \textit{9 years} && -0.63 & 0.60 & 5.90 & 0.867 &  && -0.63 & 1.04 & 3.89 & 5.90 &  \\
		\quad - \textit{10 years} && -0.44 & -0.21 &  &  &  && -0.44 &  &  &  &  \\ \hline \addlinespace[0.4ex]
		&& \multicolumn{5}{c}{\textbf{Total number of claims}} && \multicolumn{5}{c}{\textbf{Total claim amount ($\mathbf{\times\textbf{\texteuro}100}$)}} \\ \cline{3-7} \cline{9-13} \addlinespace[0.4ex]
		\textbf{Sparse HMM ($\mathbf{K = 2}$)} && $\mathbf{0}$ & $\mathbf{1}$ & $\mathbf{2}$ & $\mathbf{3}$ & $\mathbf{4}$ && $\mathbf{0}$ & $\mathbf{25}$ & $\mathbf{50}$ & $\mathbf{75}$ & $\mathbf{100}$ \\ \hline \addlinespace[0.4ex]
		\textit{Total exposure to risk} \\
		\quad - \textit{1 year} && -0.41 & 4.87 & 8.28 & 26.29 & 27.63 && -0.41 & 4.66 & 12.17 & 23.43 & 25.03 \\
		\quad - \textit{2 years} && -0.55 & 2.88 & 5.9 & 15.73 & 4.85 && -0.56 & 3.02 & 7.87 & 15.32 & 21.55 \\
		\quad - \textit{3 years} && -0.63 & 3.02 & 3.03 & 9.19 & 12.62 && -0.64 & 2.26 & 5.72 & 8.05 & 13.26 \\
		\quad - \textit{4 years} && -0.69 & 1.94 & 2.75 & 7.71 & 16.63 && -0.70 & 1.79 & 5.39 & 7.89 & 12.13 \\
		\quad - \textit{5 years} && -0.71 & 1.07 & 3.01 & 10.03 & 20.40 && -0.71 & 1.44 & 4.24 & 5.53 & 8.85 \\
		\quad - \textit{6 years} && -0.74 & 1.57 & 13.24 & 4.42 & 2.12 && -0.74 & 1.06 & 3.38 & 7.98 \\
		\quad - \textit{7 years} && -0.72 & 0.80 & 8.28 & 3.04 & 1.91 && -0.72 & 0.81 & 2.78 & 6.21 & 7.09 \\
		\quad - \textit{8 years} && -0.72 & 0.57 & 3.35 & 3.37 & 1.95 && -0.72 & 0.92 & 2.60 & 4.31 & 6.77 \\
		\quad - \textit{9 years} && -0.74 & 0.33 & 6.15 & -0.30 &  && -0.75 & 0.81 & 4.00 & 6.04 \\
		\quad - \textit{10 years} && -0.46 & -0.60 &  &  &  && -0.46 \\ \hline \addlinespace[0.4ex]
		&& \multicolumn{5}{c}{\textbf{Total number of claims}} && \multicolumn{5}{c}{\textbf{Total claim amount ($\mathbf{\times\textbf{\texteuro}100}$)}} \\ \cline{3-7} \cline{9-13} \addlinespace[0.4ex]
		\textbf{Sparse HMM ($\mathbf{K = 3}$)} && $\mathbf{0}$ & $\mathbf{1}$ & $\mathbf{2}$ & $\mathbf{3}$ & $\mathbf{4}$ && $\mathbf{0}$ & $\mathbf{25}$ & $\mathbf{50}$ & $\mathbf{75}$ & $\mathbf{100}$ \\ \hline \addlinespace[0.4ex]
		\textit{Total exposure to risk} \\
		\quad - \textit{1 year} && -0.79 & 3.34 & 5.05 & 15.62 & 15.87 && 0.80 & 2.86 & 8.79 & 16.99 & 18.86 \\
		\quad - \textit{2 years} && -0.71 & 1.91 & 4.24 & 9.85 & 2.43 && -0.72 & 1.95 & 5.3 & 12.03 & 16.73 \\
		\quad - \textit{3 years} && -0.67 & 2.19 & 1.89 & 6.11 & 9.54 && -0.68 & 1.33 & 3.83 & 6.2 & 9.09 \\
		\quad - \textit{4 years} && -0.67 & 1.19 & 1.76 & 5.23 & 12.86 && -0.67 & 1.22 & 3.61 & 5.3 & 9.44 \\
		\quad - \textit{5 years} && -0.62 & 0.58 & 1.92 & 6.89 & 14.34 && -0.63 & 0.92 & 2.75 & 3.5 & 8.03 \\
		\quad - \textit{6 years} && -0.64 & 0.85 & 9.67 & 2.17 & 0.60 && -0.65 & 0.87 & 1.63 & 6.42 \\
		\quad - \textit{7 years} && -0.60 & 0.59 & 5.26 & 1.52 & 0.30 && -0.60 & 0.51 & 1.41 & 7.64 & 6.18 \\
		\quad - \textit{8 years} && -0.63 & 0.15 & 2.09 & 1.23 & 0.16 && -0.63 & 0.53 & 0.40 & 3.92 & 5.11 \\
		\quad - \textit{9 years} && -0.69 & 0.08 & 3.77 & -1.32 &  && -0.69 & 0.52 & 1.76 & 3.93 \\
		\quad - \textit{10 years} && -0.25 & -0.81 &  &  &  && -0.25 \\ \bottomrule
	\end{tabularx}}}
\end{table}

\begin{table}[ht!]
    \caption{Mean [median] of MTPL insurance observations and predictions for each full mixture component, where the claim frequencies and risk premia are weighted by their corresponding exposures to risk.}
    \label{TC.6}\vspace{-3pt}\vspace{-3pt}
	\centerline{\scalebox{0.76}{\begin{tabularx}{1.32\textwidth}{l c rrr c rrr c rrr}
		\toprule \addlinespace[1ex] \vspace{1pt}
		&& \multicolumn{3}{c}{\textbf{Claim frequency}} && \multicolumn{3}{c}{\textbf{Claim severity}} && \multicolumn{3}{c}{\textbf{Risk premium}} \\ \cline{3-5} \cline{7-9} \cline{11-13} \addlinespace[0.4ex]
		\textbf{Mixture} && \textbf{Observed} & \textbf{Prior} & \textbf{Posterior} && \textbf{Observed} & \textbf{Prior} & \textbf{Posterior} && \textbf{Observed} & \textbf{Prior} & \textbf{Posterior} \\ \hline \addlinespace[0.4ex]
		\textit{Independent}\hspace{-10pt} && 0.0524\phantom{]} & 0.0559\phantom{]} & 0.0562\phantom{]} && 3{,}445.43\phantom{]} & 2{,}475.57\phantom{]} & 2{,}525.16\phantom{]} && 180.48\phantom{]} & 138.43\phantom{]} & 139.80\phantom{]} \\
		\quad ($\mathit{K = 1}$) && [0.0000] & [0.0514] & [0.0443] && [1{,}100.00] & [2{,}462.14] & [2{,}444.35] && [0.00] & [125.21] & [107.47] \\ \hline \addlinespace[0.4ex]
		$\mathit{K = 2}$ && 0.0524\phantom{]} & 0.0651\phantom{]} & 0.0554\phantom{]} && 3{,}445.43\phantom{]} & 2{,}080.79\phantom{]} & 1{,}776.92\phantom{]} && 180.48\phantom{]} & 202.81\phantom{]} & 153.58\phantom{]} \\
		&& [0.0000] & [0.0555] & [0.0451] && [1{,}100.00] & [1{,}898.45] & [1{,}665.33] && [0.00] & [160.36] & [116.01] \\
		\quad - $\mathit{j = 1}$ &&  & 0.0249\phantom{]} & 0.0263\phantom{]} &&  & 795.53\phantom{]} & 795.75\phantom{]} && & 22.34\phantom{]} & 23.40\phantom{]} \\
		&& & [0.0195] & [0.0192] && & [762.52] & [762.68] && & [15.42] & [15.32] \\
		\quad - $\mathit{j = 2}$ &&  & 0.1250\phantom{]} & 0.1156\phantom{]} &&  & 3{,}977.88\phantom{]} & 3{,}865.56\phantom{]} && & 474.76\phantom{]} & 422.55\phantom{]} \\
		&& & [0.1076] & [0.0963] && & [3{,}702.59] & [3{,}650.74] && & [403.01] & [340.47] \\ \hline \addlinespace[0.4ex]
		$\mathit{K = 3}$ && 0.0524\phantom{]} & 0.0517\phantom{]} & 0.0547\phantom{]} && 3{,}445.43\phantom{]} & 3{,}101.17\phantom{]} & 2{,}984.73\phantom{]} && 180.48\phantom{]} & 148.27\phantom{]} & 150.38\phantom{]} \\
		&& [0.0000] & [0.065] & [0.0450] && [1{,}100.00] & [1{,}841.63] & [1{,}859.48] && [0.00] & [124.94] & [115.47] \\
		\quad - $\mathit{j = 1}$ &&  & {\small$<$}0.0000\phantom{]} & {\small$<$}0.0000\phantom{]} && & 4{,}491.50\phantom{]} & 4{,}328.27\phantom{]} && & {\small$<$}0.00\phantom{]} & {\small$<$}0.00\phantom{]} \\
		&& & [{\small$<$}0.0000] & [{\small$<$}0.0000] && & [1{,}017.98] & [1{,}037.33] && & [{\small$<$}0.00] & [{\small$<$}0.00] \\
		\quad - $\mathit{j = 2}$ &&  & 0.0496\phantom{]} & 0.0487\phantom{]} && & 789.79\phantom{]} & 791.84\phantom{]} && & 44.44\phantom{]} & 43.44\phantom{]} \\
		&& & [0.0455] & [0.0452] && & [757.01] & [757.95] && & [30.77] & [30.61] \\
		\quad - $\mathit{j = 3}$ &&  & 0.1147\phantom{]} & 0.1061\phantom{]} && & 3{,}944.18\phantom{]} & 3{,}862.51\phantom{]} && & 438.99\phantom{]} & 389.72\phantom{]} \\
		&& & [0.0970] & [0.0864] && & [3{,}768.19] & [3{,}716.65] && & [370.57] & [310.80] \\ \hline \addlinespace[0.4ex]
		$\mathit{K = 4}$ && 0.0524\phantom{]} & 0.0638\phantom{]} & 0.0547\phantom{]} && 3{,}445.43\phantom{]} & 4{,}896.41\phantom{]} & 3{,}298.50\phantom{]} && 180.48\phantom{]} & 187.86\phantom{]} & 153.10\phantom{]} \\
		&& [0.0000] & [0.0565] & [0.0467] && [1{,}100.00] & [3{,}273.96] & [2{,}366.81] && [0.00] & [147.18] & [115.13] \\
		\quad - $\mathit{j = 1}$ &&  & 0.0019\phantom{]} & 0.0026\phantom{]} && & 223.48\phantom{]} & 551.60\phantom{]} && & 2.32\phantom{]} & 6.55\phantom{]} \\
		&& & [0.0006] & [0.0007] && & [124.33] & [139.02] && & [0.08] & [0.10] \\
		\quad - $\mathit{j = 2}$ &&  & 0.1070\phantom{]} & 0.1068\phantom{]} && & 862.27\phantom{]} & 864.76\phantom{]} && & 100.62\phantom{]} & 100.56\phantom{]} \\
		&& & [0.0863] & [0.0863] && & [827.04] & [827.88] && & [73.48] & [73.49] \\
		\quad - $\mathit{j = 3}$ &&  & 0.0047\phantom{]} & 0.0047\phantom{]} && & 20{,}125.76\phantom{]} & 18{,}532.41\phantom{]} && & 107.10\phantom{]} & 101.53\phantom{]} \\
		&& & [0.0011] & [0.0011] && & [11{,}989.07] & [10{,}528.14] && & [14.06] & [12.36] \\
		\quad - $\mathit{j = 4}$ &&  & 0.1408\phantom{]} & 0.1304\phantom{]} && & 3{,}912.16\phantom{]} & 3{,}808.69\phantom{]} && & 522.29\phantom{]} & 468.40\phantom{]} \\
		&& & [0.1172] & [0.1071] && & [3{,}642.41] & [3{,}595.60] && & [436.11] & [374.31] \\
		\bottomrule
	\end{tabularx}}}
\end{table}

\begin{figure}[ht!]
    \centering
    \includegraphics[width=0.925\textwidth]{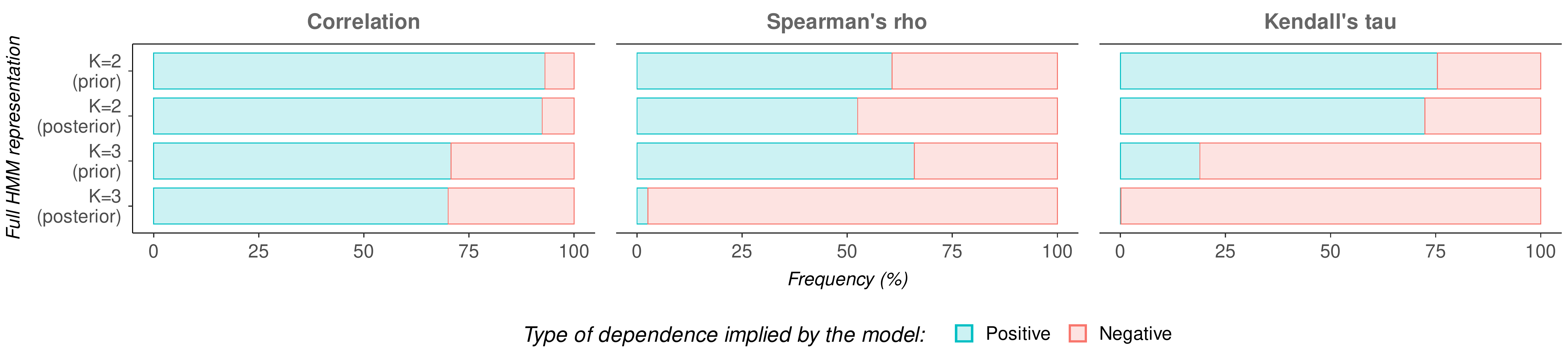}
    \vspace{-3pt}
    \caption{Frequency of the type of dependence, i.e., positive or negative correlation, Spearman's rho, and Kendall's tau, as implied by the estimated full HMM representation in MTPL insurance.}
	\label{FC.1}
\end{figure}

\begin{figure}[ht!]
    \centering
    \begin{subfigure}{\textwidth}
        \centering
        \begin{tabular}{c c c}
            \centering
            \includegraphics[width=0.30\textwidth]{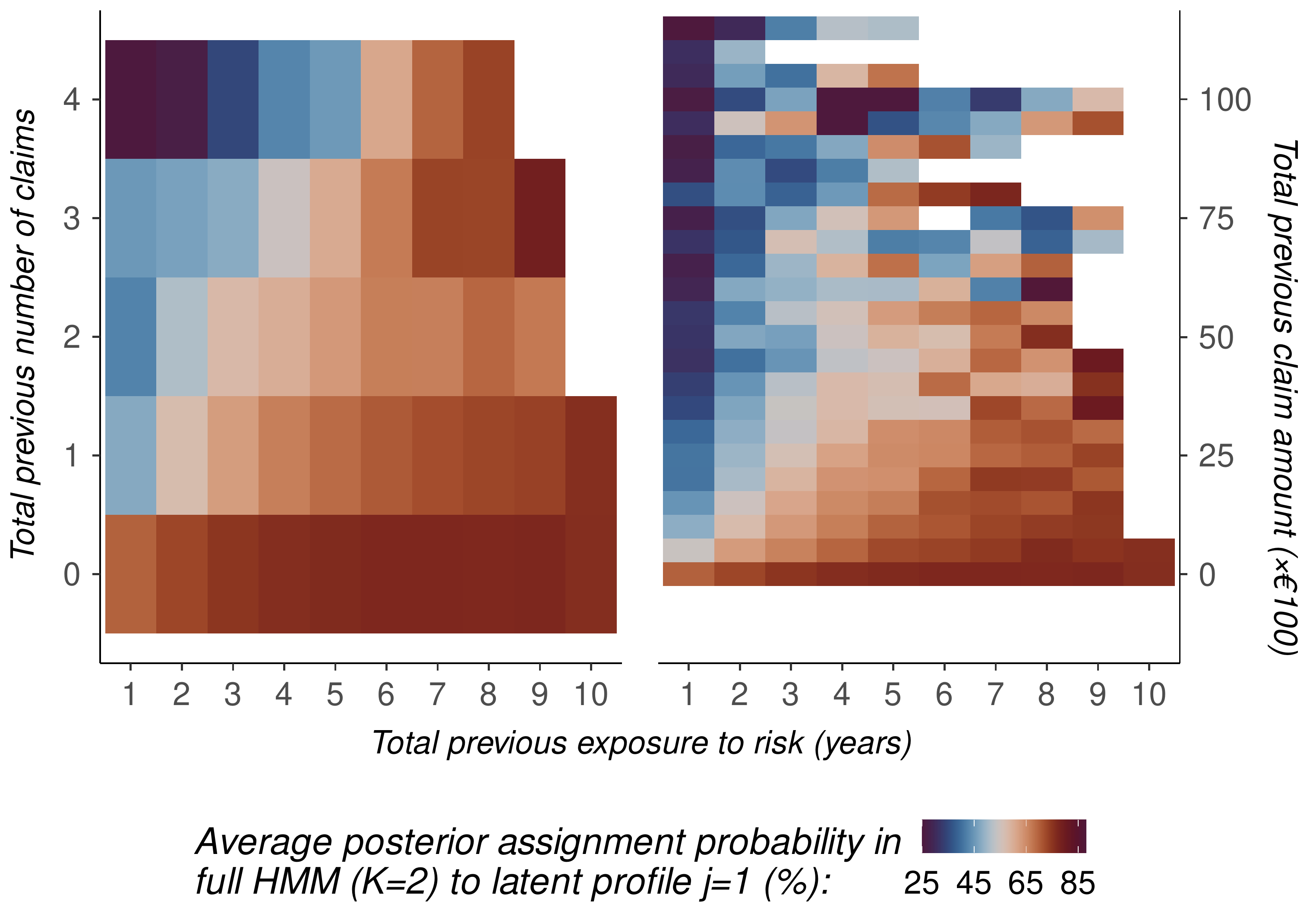}&
            \includegraphics[width=0.30\textwidth]{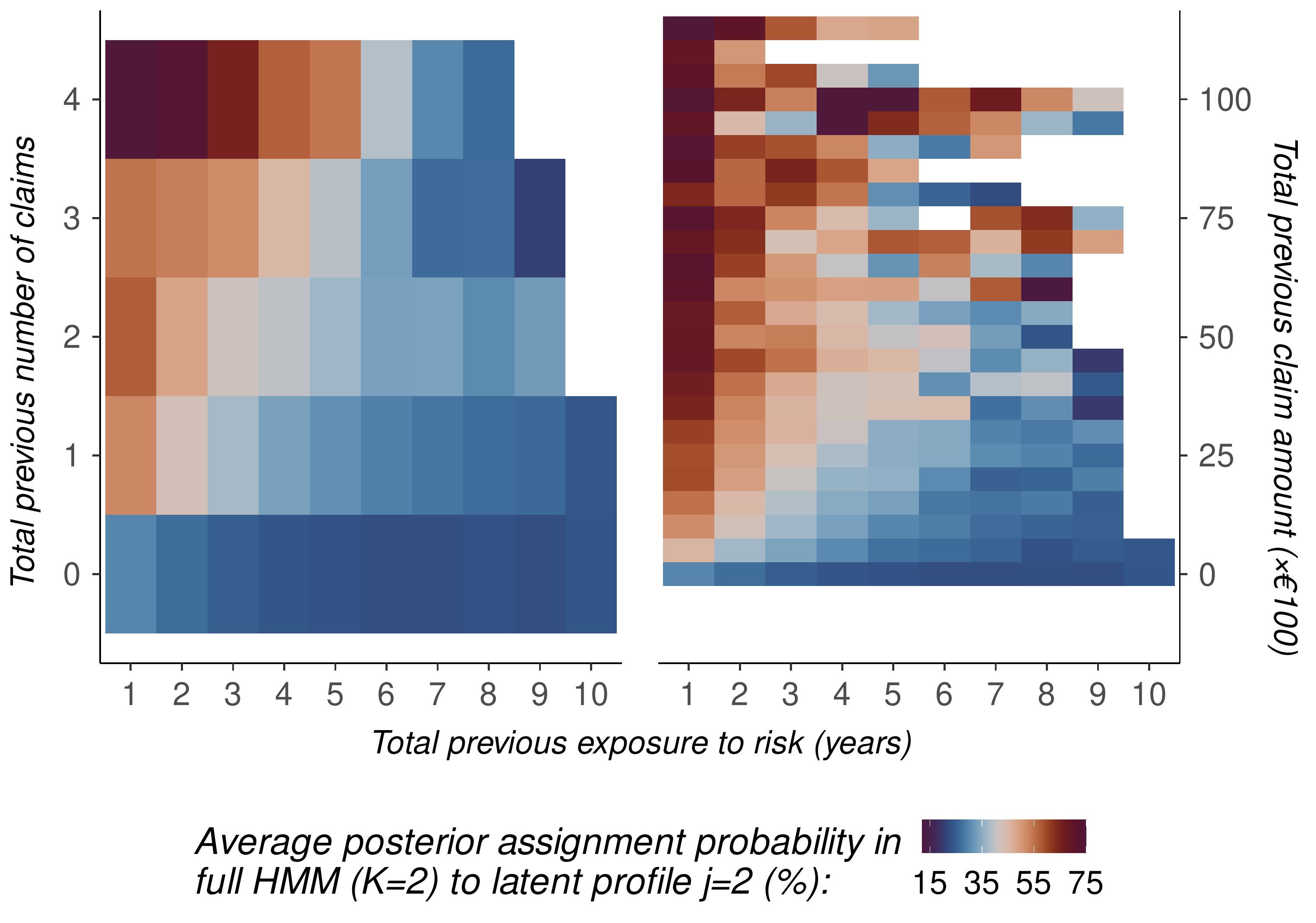}&
            \phantom{\includegraphics[width=0.30\textwidth]{Figures/Figure_C2_a_middle.pdf}}
        \end{tabular}\vspace{-6pt}
        \caption{Based on $K = 2$ latent risk profiles}
        \label{FC.2a}
    \end{subfigure}
    
    \vspace{3pt}
    
    \begin{subfigure}{\textwidth}
        \centering
        \begin{tabular}{c c c}
            \centering
            \includegraphics[width=0.30\textwidth]{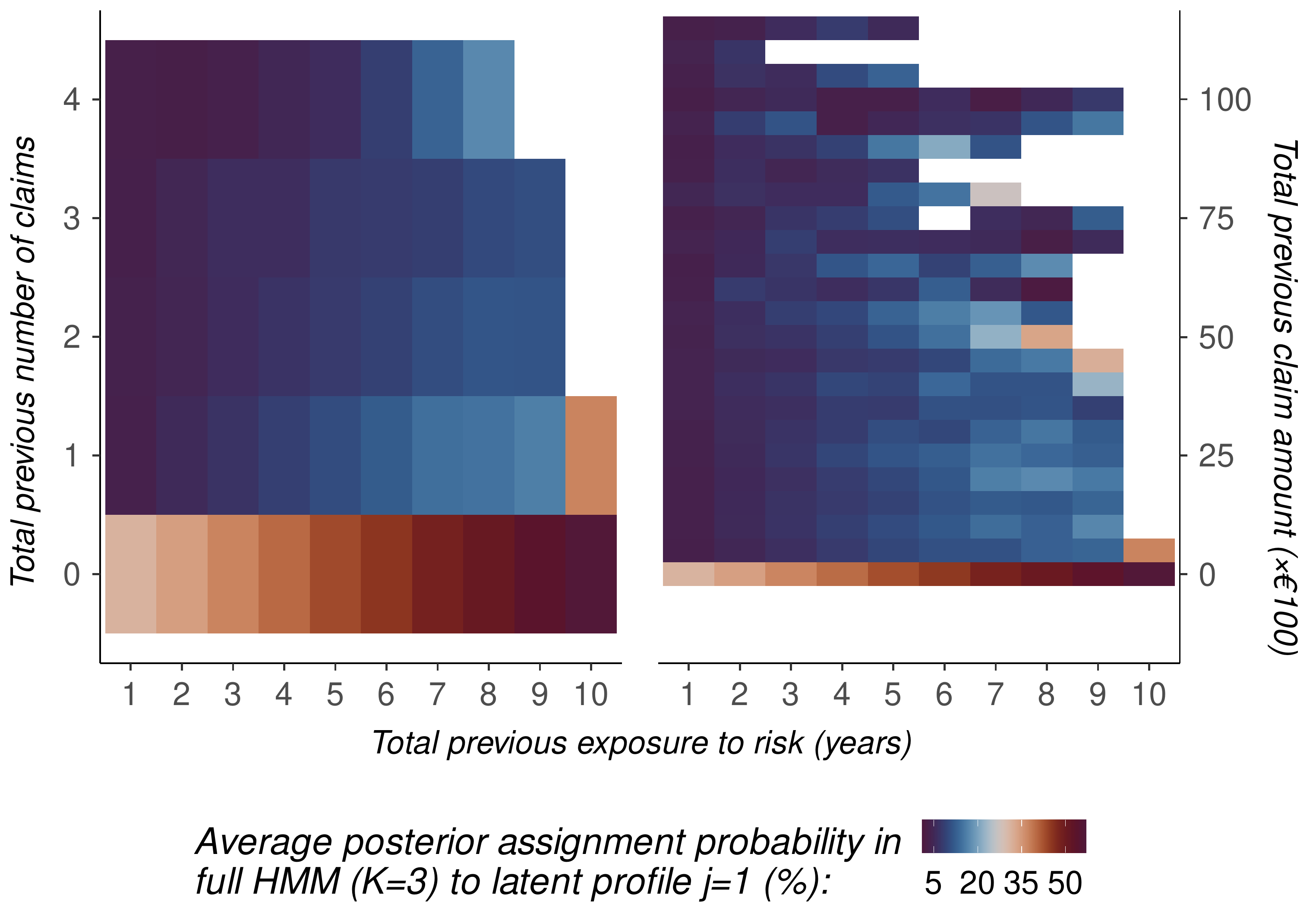}&
            \includegraphics[width=0.30\textwidth]{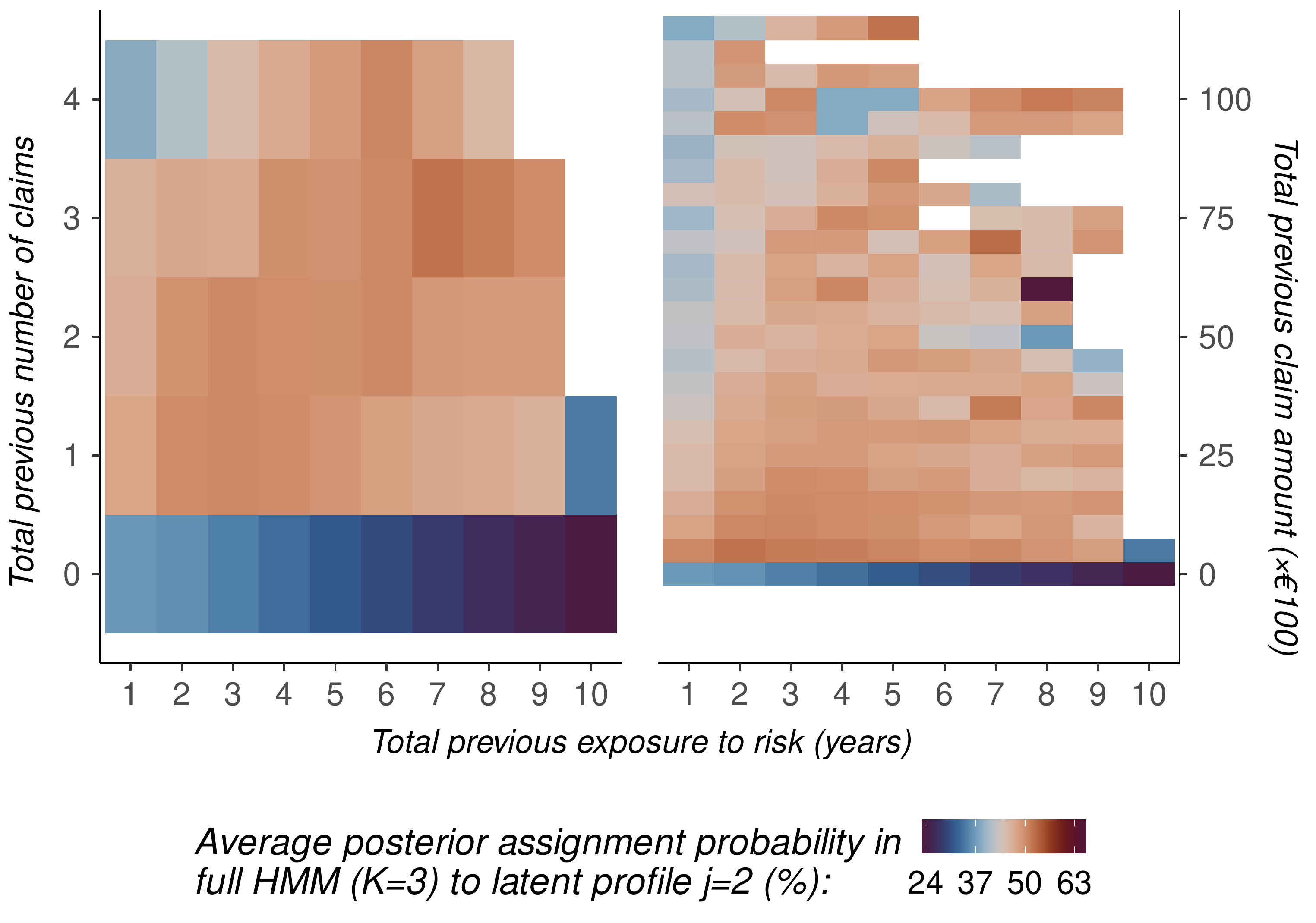}&
            \includegraphics[width=0.30\textwidth]{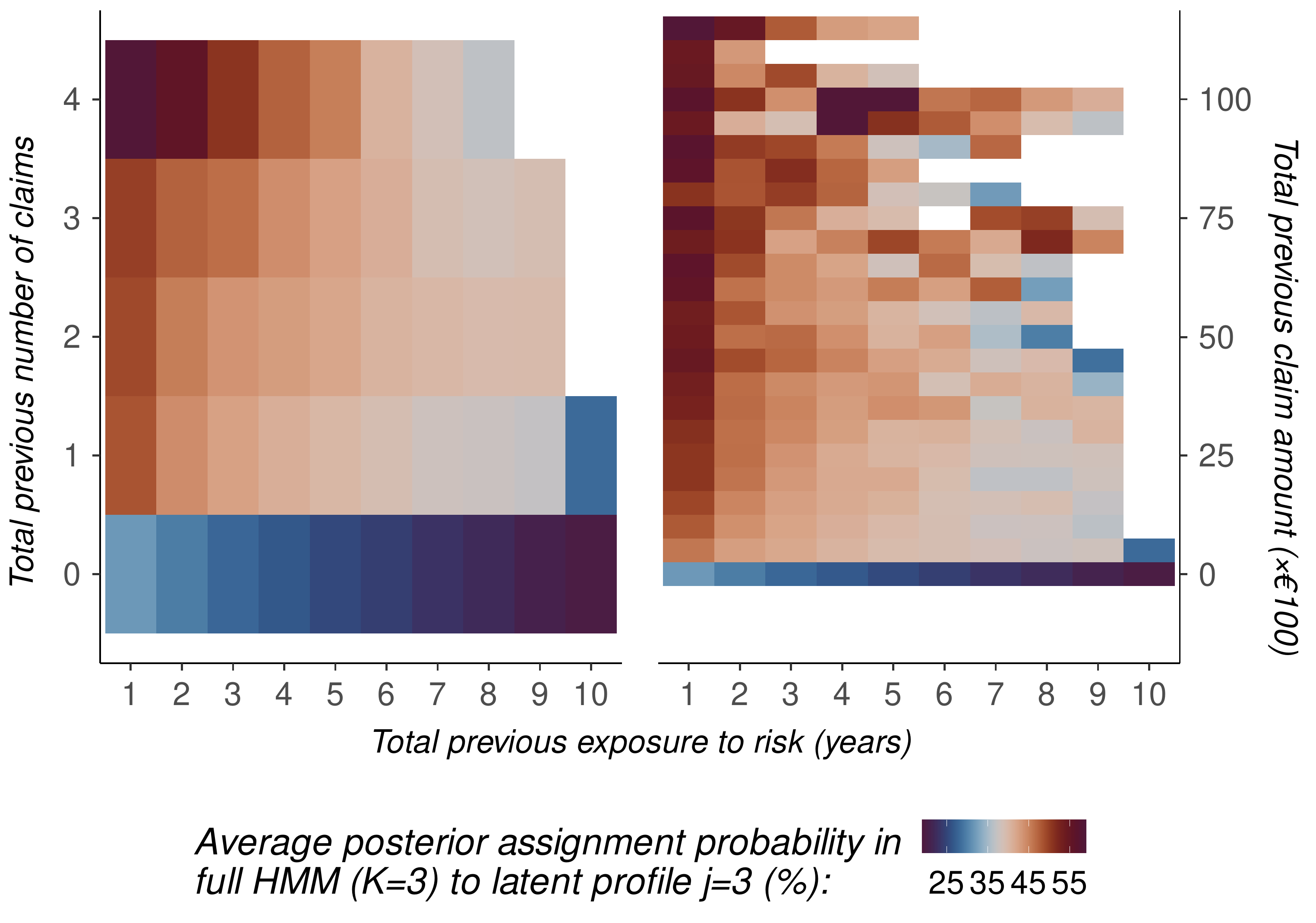}
        \end{tabular}\vspace{-6pt}
        \caption{Based on $K = 3$ latent risk profiles}
        \label{FC.2b}
    \end{subfigure}\vspace{-3pt}
    \caption{Distribution of average posterior assignment probabilities over total previous number of claims and claim amount for profile $j = 1$ (left), $j = 2$ (middle), and $j = 3$ (right) with $K = 2$ (panel (a)) and $K = 3$ (panel (b)) latent risk profiles for the full HMM representation in MTPL insurance.}
	\label{FC.2}
\end{figure}

\begin{figure}[ht!]
    \centering
    \begin{tabular}{c c c}
        \centering
        \includegraphics[width=0.30\textwidth]{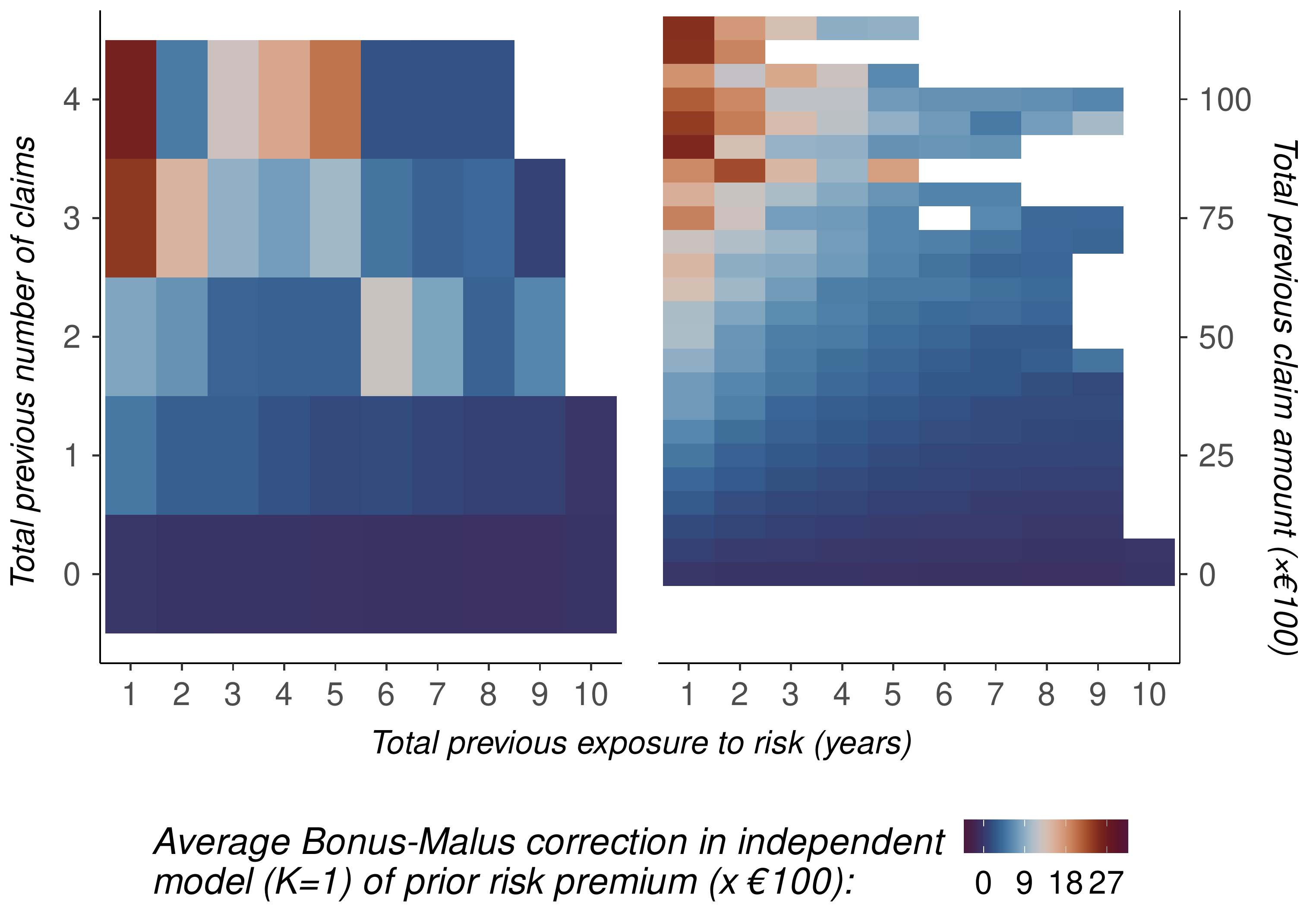}&
        \includegraphics[width=0.30\textwidth]{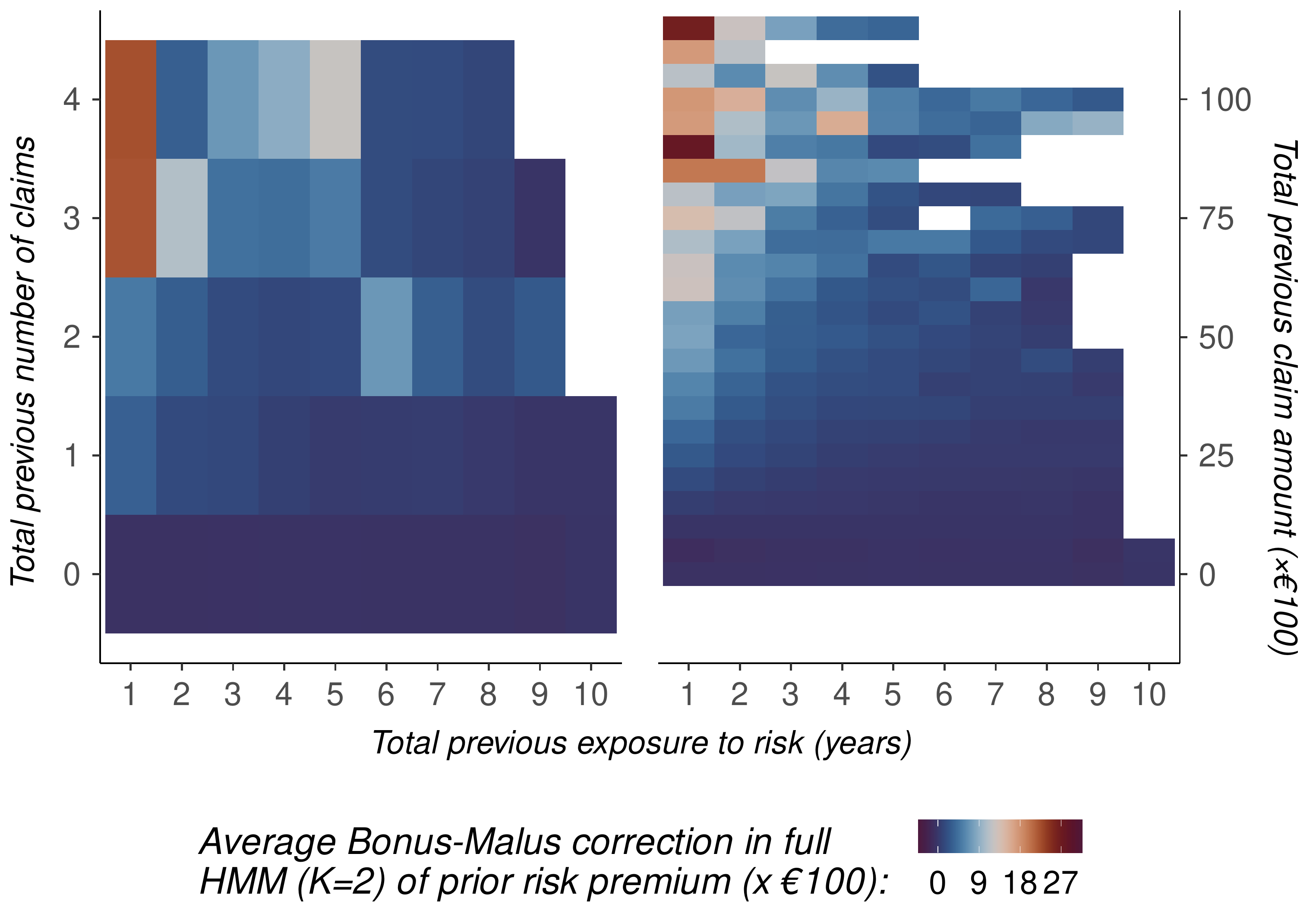}&
        \includegraphics[width=0.30\textwidth]{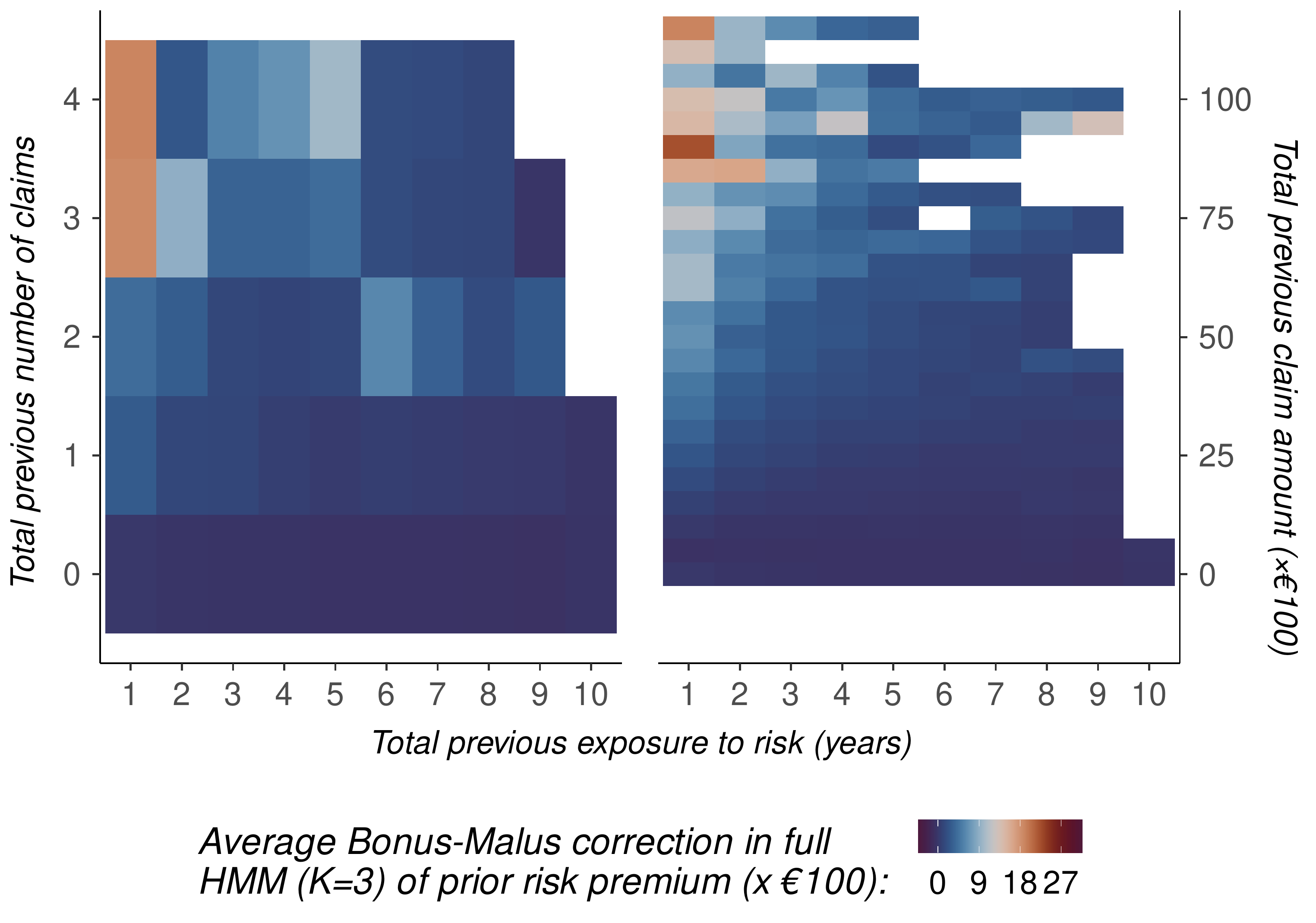}
        \end{tabular}\vspace{-6pt}
    \caption{Average Bonus-Malus correction of prior risk premium in terms of total previous number of claims and claim amount in independent model ($K = 1$, left) and with $K = 2$ (middle) and $K = 3$ (right) latent risk profiles for the full HMM representation in MTPL insurance.}
	\label{FC.3}
\end{figure}

\begin{table}[ht!]
    \caption{Average Bonus-Malus correction of prior risk premium in terms of total previous number of claims and claim amount in independent model ($K = 1$) and each full HMM in MTPL insurance. These corrections coincide with those shown in \Cref{FC.3} but are represented in tabular form in this table.}
    \label{TC.7}\vspace{-3pt}\vspace{-3pt}
    \centerline{\scalebox{0.8505}{\begin{tabularx}{1.18\textwidth}{l p{0.5em} rrrrr p{0.5em} rrrrr}
		\toprule \addlinespace[1ex] \vspace{1pt}
		&& \multicolumn{11}{c}{\textbf{Average Bonus-Malus correction of prior risk premium ($\mathbf{\times\textbf{\texteuro}100}$)}} \\ \cline{3-13} \addlinespace[0.4ex]
		&& \multicolumn{5}{c}{\textbf{Total number of claims}} && \multicolumn{5}{c}{\textbf{Total claim amount ($\mathbf{\times\textbf{\texteuro}100}$)}} \\ \cline{3-7} \cline{9-13} \addlinespace[0.4ex]
		\textbf{Independent ($\mathbf{K = 1}$)} && $\mathbf{0}$ & $\mathbf{1}$ & $\mathbf{2}$ & $\mathbf{3}$ & $\mathbf{4}$ && $\mathbf{0}$ & $\mathbf{25}$ & $\mathbf{50}$ & $\mathbf{75}$ & $\mathbf{100}$ \\ \hline \addlinespace[0.4ex]
		\textit{Total exposure to risk} \\
		\quad - \textit{1 year} && -0.16 & 4.96 & 8.25 & 24.15 & 26.07 && -0.16 & 4.87 & 10.65 & 19.28 & 21.68 \\
		\quad - \textit{2 years} && -0.33 & 3.20 & 6.85 & 15.18 & 5.09 && -0.33 & 3.33 & 7.06 & 12.82 & 18.77 \\
		\quad - \textit{3 years} && -0.43 & 3.18 & 3.52 & 9.19 & 12.78 && -0.43 & 2.60 & 5.34 & 7.74 & 11.63 \\
		\quad - \textit{4 years} && -0.51 & 2.19 & 3.26 & 7.64 & 16.33 && -0.51 & 2.19 & 5.08 & 7.34 & 11.60 \\
		\quad - \textit{5 years} && -0.55 & 1.42 & 3.35 & 9.97 & 20.08 && -0.55 & 1.77 & 4.11 & 5.98 & 7.40 \\
		\quad - \textit{6 years} && -0.59 & 1.76 & 12.42 & 4.74 & 2.29 && -0.59 & 1.39 & 3.62 & 6.75 &  \\
		\quad - \textit{7 years} && -0.59 & 1.09 & 8.09 & 3.44 & 2.08 && -0.59 & 1.15 & 2.88 & 6.12 & 6.78 \\
		\quad - \textit{8 years} && -0.60 & 0.80 & 3.46 & 3.86 & 2.118 && -0.60 & 1.16 & 2.72 & 3.93 & 6.53 \\
		\quad - \textit{9 years} && -0.63 & 0.60 & 5.90 & 0.867 &  && -0.63 & 1.04 & 3.89 & 5.90 &  \\
		\quad - \textit{10 years} && -0.44 & -0.21 &  &  &  && -0.44 &  &  &  &  \\ \hline \addlinespace[0.4ex]
		&& \multicolumn{5}{c}{\textbf{Total number of claims}} && \multicolumn{5}{c}{\textbf{Total claim amount ($\mathbf{\times\textbf{\texteuro}100}$)}} \\ \cline{3-7} \cline{9-13} \addlinespace[0.4ex]
		\textbf{Full HMM ($\mathbf{K = 2}$)} && $\mathbf{0}$ & $\mathbf{1}$ & $\mathbf{2}$ & $\mathbf{3}$ & $\mathbf{4}$ && $\mathbf{0}$ & $\mathbf{25}$ & $\mathbf{50}$ & $\mathbf{75}$ & $\mathbf{100}$ \\ \hline \addlinespace[0.4ex]
		\textit{Total exposure to risk} \\
		\quad - \textit{1 year} && -0.63 & 3.31 & 5.02 & 22.23 & 22.49 && -0.64 & 2.70 & 8.07 & 14.03 & 17.66 \\
		\quad - \textit{2 years} && -0.61 & 1.56 & 3.09 & 10.94 & 3.18 && -0.62 & 1.53 & 3.75 & 11.85 & 15.57 \\
		\quad - \textit{3 years} && -0.56 & 1.42 & 1.65 & 4.42 & 7.14 && -0.57 & 0.99 & 3.09 & 5.28 & 6.46 \\
		\quad - \textit{4 years} && -0.55 & 0.73 & 1.38 & 4.27 & 8.85 && -0.55 & 0.48 & 2.78 & 3.36 & 9.58 \\
		\quad - \textit{5 years} && -0.51 & 0.32 & 1.54 & 5.15 & 12.27 && -0.51 & 0.39 & 2.04 & 1.80 & 5.50 \\
		\quad - \textit{6 years} && -0.53 & 0.4 & 7.15 & 1.75 & 1.76 && -0.54 & 0.23 & 1.55 & 3.89 \\
		\quad - \textit{7 years} && -0.53 & 0.45 & 3.19 & 1.27 & 1.70 && -0.53 & 0.19 & 0.99 & 3.96 & 4.98 \\
		\quad - \textit{8 years} && -0.55 & 0.08 & 1.78 & 0.87 & 1.18 && -0.55 & 0.26 & 0.53 & 3.23 & 3.76 \\
		\quad - \textit{9 years} && -0.68 & -0.22 & 2.69 & -0.33 &  && -0.68 & 0.2 & 1.3 & 2.69 \\
		\quad - \textit{10 years} && -0.35 & -0.20 &  &  &  && -0.35 \\ \hline \addlinespace[0.4ex]
		&& \multicolumn{5}{c}{\textbf{Total number of claims}} && \multicolumn{5}{c}{\textbf{Total claim amount ($\mathbf{\times\textbf{\texteuro}100}$)}} \\ \cline{3-7} \cline{9-13} \addlinespace[0.4ex]
		\textbf{Full HMM ($\mathbf{K = 3}$)} && $\mathbf{0}$ & $\mathbf{1}$ & $\mathbf{2}$ & $\mathbf{3}$ & $\mathbf{4}$ && $\mathbf{0}$ & $\mathbf{25}$ & $\mathbf{50}$ & $\mathbf{75}$ & $\mathbf{100}$ \\ \hline \addlinespace[0.4ex]
		\textit{Total exposure to risk} \\
		\quad - \textit{1 year} && -0.06 & 2.89 & 4.19 & 18.62 & 18.95 && -0.06 & 2.48 & 6.73 & 11.80 & 14.04 \\
		\quad - \textit{2 years} && -0.25 & 1.36 & 3.01 & 9.11 & 2.51 && -0.25 & 1.36 & 3.27 & 9.04 & 12.10 \\
		\quad - \textit{3 years} && -0.35 & 1.20 & 1.34 & 3.45 & 5.66 && -0.35 & 0.94 & 2.61 & 4.48 & 5.03 \\
		\quad - \textit{4 years} && -0.41 & 0.71 & 1.10 & 3.44 & 6.81 && -0.42 & 0.55 & 2.41 & 3.02 & 6.97 \\
		\quad - \textit{5 years} && -0.44 & 0.35 & 1.31 & 4.20 & 9.97 && -0.44 & 0.45 & 1.74 & 1.91 & 4.09 \\
		\quad - \textit{6 years} && -0.48 & 0.50 & 6.07 & 1.75 & 1.72 && -0.48 & 0.25 & 1.32 & 2.94 \\
		\quad - \textit{7 years} && -0.50 & 0.40 & 3.29 & 1.29 & 1.57 && -0.50 & 0.19 & 0.95 & 3.01 & 3.39 \\
		\quad - \textit{8 years} && -0.53 & 0.24 & 1.63 & 1.21 & 1.19 && -0.53 & 0.36 & 0.70 & 2.32 & 3.02 \\
		\quad - \textit{9 years} && -0.63 & 0.05 & 2.64 & -0.20 &  && -0.63 & 0.37 & 1.41 & 2.62 \\
		\quad - \textit{10 years} && -0.29 & -0.25 &  &  &  && -0.29 \\ \bottomrule
	\end{tabularx}}}
\end{table}

\begin{table}[ht!]
    \caption{Distribution of observed claim sizes and GB2 predictions for independent model ($K = 1$) after truncating (top) and removing (bottom) outliers, where $[\mathrm{Q1} - 1.5 \cdot \mathrm{IQR}, \mathrm{Q3} + 1.5 \cdot \mathrm{IQR}] = [-1405, 4110]$.}
    \label{TC.8}\vspace{-3pt}\vspace{-3pt}
    \centerline{\scalebox{0.8505}{\begin{tabularx}{1.18\textwidth}{l p{0.5em} rrrrrr}
		\toprule \addlinespace[1ex] \vspace{1pt}
		&& & & & & & $\mathbf{\in \textbf{\texteuro}[\textbf{Q1} - 1.5 \cdot \textbf{IQR},}$ \\
		\textbf{Outlier truncation} && \textbf{Total} & $\mathbf{\leq \textbf{\texteuro}100{,}000}$ & $\mathbf{\leq \textbf{\texteuro}50{,}000}$ & $\mathbf{\leq \textbf{\texteuro}25{,}000}$ & $\mathbf{\leq \textbf{\texteuro}10{,}000}$ & $\mathbf{\textbf{Q3} + 1.5 \cdot \textbf{IQR}]}$ \\ \hline \addlinespace[0.4ex]
		\textit{Observed} \\
		\quad - \textit{Number of claims} && 5{,}662 & 5{,}662 & 5{,}662 & 5{,}662 & 5{,}662 & 5{,}662 \\
		\quad - \textit{Mean claim size} && 3{,}445.43 & 2{,}936.23 & 2{,}632.56 & 2{,}310.33 & 1{,}903.03 & 1{,}530.26 \\
		\quad - \textit{Median claim size} && 1{,}100.00 & 1{,}100.00 & 1{,}100.00 & 1{,}100.00 & 1{,}100.00 & 1{,}100.00 \\ \hline \addlinespace[0.4ex]
		\textit{Predicted size (GB2)} \\
		\quad - \textit{Min. prediction} && 1{,}415.76 & 1{,}392.29 & 1{,}357.43 & 1{,}290.87 & 1{,}136.62 & 949.71 \\
		\quad - \textit{Mean prediction} && 2{,}475.57 & 2{,}429.55 & 2{,}360.76 & 2{,}228.33 & 1{,}932.21 & 1{,}531.53 \\
		\quad - \textit{Median prediction} && 2{,}462.14 & 2{,}416.38 & 2{,}347.98 & 2{,}215.91 & 1{,}923.20 & 1{,}526.38 \\
		\quad - \textit{Max. prediction} && 3{,}801.38 & 3{,}718.99 & 3{,}595.21 & 3{,}355.67 & 2{,}830.62 & 2{,}083.39 \\ \hline \addlinespace[0.4ex]
		\textbf{Outlier removal} \\ \hline \addlinespace[0.4ex]
		\textit{Observed} \\
		\quad - \textit{Number of claims} && 5{,}662 & 5{,}639 & 5{,}611 & 5{,}556 & 5{,}431 & 5{,}069 \\
		\quad - \textit{Mean claim size} && 3{,}445.43 & 2{,}540.33 & 2{,}202.02 & 1{,}877.45 & 1{,}558.63 & 1{,}228.51 \\
		\quad - \textit{Median claim size} && 1{,}100.00 & 1{,}095.68 & 1{,}090.55 & 1{,}084.11 & 1{,}057.02 & 997.33 \\ \hline \addlinespace[0.4ex]
		\textit{Predicted size (GB2)} \\
		\quad - \textit{Min. prediction} && 1{,}415.76 & 1{,}305.33 & 1{,}220.29 & 895.53 & 776.21 & 634.88 \\
		\quad - \textit{Mean prediction} && 2{,}475.57 & 2{,}244.76 & 2{,}068.49 & 1{,}840.44 & 1{,}559.00 & 1{,}229.35 \\
		\quad - \textit{Median prediction} && 2{,}462.14 & 2{,}232.92 & 2{,}057.80 & 1{,}830.98 & 1{,}553.66 & 1{,}224.61 \\
		\quad - \textit{Max. prediction} && 3{,}801.38 & 3{,}455.81 & 3{,}135.69 & 2{,}789.96 & 2{,}345.96 & 1{,}810.99 \\
		\bottomrule
	\end{tabularx}}}
\end{table}

\begin{table}[ht!]
    \caption{Estimated effects for the risk factors in \Cref{TA.1} in the frequency and severity component of the independent model ($K = 1$) and each sparse HMM in MTPL insurance. Estimates that are statistically significant at a significance level of $5\%$ or less are marked by an asterisk.}
    \label{TC.9}\vspace{-3pt}\vspace{-3pt}
    \centerline{\scalebox{0.82}{\begin{tabularx}{1.22\textwidth}{l p{0.3em} rrrr p{0.3em} rrrr}
		\toprule \addlinespace[1ex] \vspace{1pt}
		&& \multicolumn{4}{c}{\textbf{Frequency component sparse HMM}} && \multicolumn{4}{c}{\textbf{Severity component sparse HMM}} \\ \cline{3-6} \cline{8-11} \addlinespace[0.4ex]
		\textbf{Risk factor} && \textbf{Indep. ($\mathbf{K = 1}$)} & $\mathbf{K = 2}$ & $\mathbf{K = 3}$ & $\mathbf{K = 4}$ && \textbf{Indep. ($\mathbf{K = 1}$)} & $\mathbf{K = 2}$ & $\mathbf{K = 3}$ & $\mathbf{K = 4}$ \\ \hline \addlinespace[0.4ex]
		$\mathit{\varphi}$ &&  &  &  &  && 2.85{ } & 95.37{ } & 297.26{ } & 33.41{ } \\
		\texttt{Cust\_Age} && -0.01$^{*}$ & -0.01$^{*}$ & -0.01$^{*}$ & -0.01$^{*}$ && {\small$<$}0.00$^{*}$ & {\small$<$}0.00$^{*}$ & {\small$<$}0.00$^{*}$ & {\small$<$}0.00$^{*}$ \\
		\texttt{Veh\_Weight} && 0.02$^{*}$ & 0.02$^{*}$ & 0.02$^{*}$ & 0.02$^{*}$ && {\small$<$}0.00$^{*}$ & 0.01$^{*}$ & 0.01$^{*}$ & 0.01$^{*}$ \\
		\texttt{Veh\_Mileage} \\
		\quad - \textit{Category 1} && 0.01$^{*}$ & 0.01$^{*}$ & 0.01$^{*}$ & 0.01$^{*}$ && -0.03$^{*}$ & -0.03$^{*}$ & -0.01$^{*}$ & -0.03$^{*}$ \\
		\quad - \textit{Category 2} && 0.07$^{*}$ & 0.07$^{*}$ & 0.07$^{*}$ & 0.07$^{*}$ && -0.02$^{*}$ & 0.01$^{*}$ & 0.04$^{*}$ & 0.02$^{*}$ \\
		\texttt{Veh\_FuelType} \\
		\quad - \textit{Category 1} && 0.24$^{*}$ & 0.24$^{*}$ & 0.23$^{*}$ & 0.23$^{*}$ && 0.11$^{*}$ & 0.14$^{*}$ & 0.12$^{*}$ & 0.11$^{*}$ \\
		\quad - \textit{Category 2} && 0.30$^{*}$ & 0.30$^{*}$ & 0.28$^{*}$ & 0.29$^{*}$ && -0.04$^{*}$ & -0.05$^{*}$ & -0.07$^{*}$ & -0.05$^{*}$ \\
		\quad - \textit{Category 3} && -0.09{ } & -0.11{ } & -0.11{ } & -0.03{ } && -0.34{ } & -0.77{ } & -0.75{ } & -0.68{ } \\
		\quad - \textit{Category 4} && 0.18{ } & 0.17{ } & 0.23{ } & 0.23{ } && 0.30{ } & 0.22{ } & 0.23{ } & 0.23{ } \\
		\quad - \textit{Category 5} && 32.96$^{*}$ & -17.86$^{*}$ & -8.33$^{*}$ & -8.33$^{*}$ && \\
		\quad - \textit{Category 6} && -33.26$^{*}$ & -18.03$^{*}$ & -6.64$^{*}$ & -6.60$^{*}$ && \\
		\texttt{Veh\_BodyDoors} \\
		\quad - \textit{Category 1} && 0.01$^{*}$ & 0.01$^{*}$ & 0.01$^{*}$ & 0.01$^{*}$ && -0.02$^{*}$ & -0.01$^{*}$ & -0.02$^{*}$ & -0.03$^{*}$ \\
		\quad - \textit{Category 2} && 0.08$^{*}$ & 0.08$^{*}$ & 0.09$^{*}$ & 0.10$^{*}$ && -0.03$^{*}$ & -0.01$^{*}$ & -0.03$^{*}$ & -0.05$^{*}$ \\
		\quad - \textit{Category 3} && 0.18$^{*}$ & 0.18$^{*}$ & 0.16$^{*}$ & 0.15$^{*}$ && 0.05$^{*}$ & 0.01$^{*}$ & 0.04$^{*}$ & {\small$<$}0.00$^{*}$ \\
		\quad - \textit{Category 4} && 0.03$^{*}$ & 0.03$^{*}$ & 0.04$^{*}$ & 0.04$^{*}$ && -0.01$^{*}$ & 0.01$^{*}$ & 0.01$^{*}$ & 0.01$^{*}$ \\
		\quad - \textit{Category 5} && -0.24$^{*}$ & -0.25$^{*}$ & -0.24$^{*}$ & -0.24$^{*}$ && -0.09$^{*}$ & -0.08$^{*}$ & -0.04$^{*}$ & -0.09$^{*}$ \\
		\quad - \textit{Category 6} && 0.30$^{*}$ & -0.31$^{*}$ & -0.31$^{*}$ & -0.28$^{*}$ && -0.01$^{*}$ & {\small$<$}0.00$^{*}$ & {\small$<$}0.00$^{*}$ & -0.05$^{*}$ \\
		\quad - \textit{Category 7} && -0.10$^{*}$ & -0.09$^{*}$ & -0.10$^{*}$ & -0.09$^{*}$ && 0.11$^{*}$ & 0.13$^{*}$ & 0.12$^{*}$ & 0.04$^{*}$ \\
		\quad - \textit{Category 8} && -0.08$^{*}$ & -0.08$^{*}$ & -0.08$^{*}$ & -0.07$^{*}$ && -0.14$^{*}$ & -0.02$^{*}$ & -0.05$^{*}$ & -0.04$^{*}$ \\
		\quad - \textit{Category 9} && -0.24$^{*}$ & -0.25$^{*}$ & -0.24$^{*}$ & -0.23$^{*}$ && 0.15$^{*}$ & 0.05$^{*}$ & 0.04$^{*}$ & 0.01$^{*}$ \\
		\texttt{Veh\_CatValue} && {\small$<$}0.00$^{*}$ & {\small$<$}0.00$^{*}$ & {\small$<$}0.00$^{*}$ & {\small$<$}0.00$^{*}$ && {\small$<$}0.00$^{*}$ & {\small$<$}0.00$^{*}$ & {\small$<$}0.00$^{*}$ & {\small$<$}0.00$^{*}$ \\
		\texttt{Veh\_PowerWeight}\hspace{-20pt} && 0.01$^{*}$ & -0.01$^{*}$ & -0.01$^{*}$ & -0.01$^{*}$ && {\small$<$}0.00$^{*}$ & 0.01$^{*}$ & 0.01$^{*}$ & 0.01$^{*}$ \\
		\texttt{Veh\_Age} && -0.01$^{*}$ & -0.01$^{*}$ & -0.01$^{*}$ & -0.01$^{*}$ && 0.01$^{*}$ & 0.01$^{*}$ & 0.01$^{*}$ & 0.01$^{*}$ \\
		\texttt{Veh\_Region} \\
		\quad - \textit{Category 1} && 0.08$^{*}$ & 0.08$^{*}$ & 0.08$^{*}$ & 0.07$^{*}$ && 0.01$^{*}$ & 0.06$^{*}$ & 0.09$^{*}$ & 0.11$^{*}$ \\
		\quad - \textit{Category 2} && 0.10$^{*}$ & 0.10$^{*}$ & 0.10$^{*}$ & 0.09$^{*}$ && {\small$<$}0.00$^{*}$ & 0.05$^{*}$ & 0.05$^{*}$ & 0.09$^{*}$ \\
		\quad - \textit{Category 3} && 0.04$^{*}$ & 0.04$^{*}$ & 0.03$^{*}$ & 0.03$^{*}$ && 0.01$^{*}$ & 0.07$^{*}$ & 0.07$^{*}$ & 0.10$^{*}$ \\
		\quad - \textit{Category 4} && 0.10$^{*}$ & 0.10$^{*}$ & 0.09$^{*}$ & 0.08$^{*}$ && 0.03$^{*}$ & 0.08$^{*}$ & 0.08$^{*}$ & 0.12$^{*}$ \\
		\quad - \textit{Category 5} && 0.02$^{*}$ & 0.02$^{*}$ & 0.01$^{*}$ & 0.01$^{*}$ && -0.01$^{*}$ & 0.08$^{*}$ & 0.09$^{*}$ & 0.12$^{*}$ \\
		\quad - \textit{Category 6} && -0.01$^{*}$ & -0.01$^{*}$ & -0.01$^{*}$ & -0.01$^{*}$ && {\small$<$}0.00$^{*}$ & 0.11$^{*}$ & 0.12$^{*}$ & 0.15$^{*}$ \\
		\quad - \textit{Category 7} && 0.11$^{*}$ & 0.12$^{*}$ & 0.11$^{*}$ & 0.10$^{*}$ && -0.02$^{*}$ & 0.04$^{*}$ & 0.05$^{*}$ & 0.09$^{*}$ \\
		\quad - \textit{Category 8} && -0.08$^{*}$ & -0.08$^{*}$ & -0.08$^{*}$ & -0.08$^{*}$ && -0.08$^{*}$ & -0.01$^{*}$ & 0.01$^{*}$ & 0.02$^{*}$ \\
		\quad - \textit{Category 9} && 0.03$^{*}$ & 0.03$^{*}$ & 0.02$^{*}$ & 0.02$^{*}$ && -0.04$^{*}$ & 0.04$^{*}$ & 0.06$^{*}$ & 0.10$^{*}$ \\
		\texttt{Cust\_Residence} \\
		\quad - \textit{Category 1} && 0.33$^{*}$ & 0.33$^{*}$ & 0.34$^{*}$ & 0.34$^{*}$ && {\small$<$}0.00$^{*}$ & 0.06$^{*}$ & 0.05$^{*}$ & 0.06$^{*}$ \\
		\quad - \textit{Category 2} && 0.14$^{*}$ & 0.14$^{*}$ & 0.14$^{*}$ & 0.14$^{*}$ && -0.01$^{*}$ & -0.05$^{*}$ & -0.05$^{*}$ & -0.04$^{*}$ \\
		\quad - \textit{Category 3} && 0.37$^{*}$ & 0.37$^{*}$ & 0.37$^{*}$ & 0.37$^{*}$ && 0.02$^{*}$ & -0.02$^{*}$ & -0.02$^{*}$ & {\small$<$}0.00$^{*}$ \\
		\quad - \textit{Category 4} && -0.08$^{*}$ & -0.09$^{*}$ & -0.08$^{*}$ & -0.08$^{*}$ && -0.02$^{*}$ & -0.01$^{*}$ & {\small$<$}0.00$^{*}$ & 0.01$^{*}$ \\
		\quad - \textit{Category 5} && -0.15$^{*}$ & -0.15$^{*}$ & -0.15$^{*}$ & -0.15$^{*}$ && 0.02$^{*}$ & 0.03$^{*}$ & 0.03$^{*}$ & 0.06$^{*}$ \\
		\quad - \textit{Category 6} && 0.63$^{*}$ & 0.63$^{*}$ & 0.63$^{*}$ & 0.63$^{*}$ && -0.07$^{*}$ & -0.06$^{*}$ & -0.06$^{*}$ & -0.05$^{*}$ \\
		\quad - \textit{Category 7} && 0.92$^{*}$ & 0.92$^{*}$ & 0.91$^{*}$ & 0.92$^{*}$ && 0.37{ } & 0.26{ } & 0.27{ } & 0.21{ } \\
		\quad - \textit{Category 8} && -0.73{ } & -0.74{ } & -0.77{ } & -0.77{ } && 0.19$^{*}$ & 0.17$^{*}$ & 0.16$^{*}$ & 0.19$^{*}$ \\
		\quad - \textit{Category 9} && -33.34$^{*}$ & -18.07$^{*}$ & -6.96$^{*}$ & -6.97$^{*}$ &&  \\
		\bottomrule
	\end{tabularx}}}
\end{table}

\begin{table}[ht!]
    \caption{Estimated effects for the risk factors in \Cref{TA.1} in the frequency component of the independent model ($K = 1$) and each full HMM in MTPL insurance. Estimates that are statistically significant at a significance level of $5\%$ or less are marked by an asterisk.}
    \label{TC.10}\vspace{-3pt}\vspace{-3pt}
    \centerline{\scalebox{0.82}{\begin{tabularx}{1.22\textwidth}{l p{0.1em} r p{0.2em} rr p{0.2em} rrr p{0.2em} rrrr}
		\toprule \addlinespace[1ex] \vspace{1pt}
		&& \textbf{Indep.} && \multicolumn{2}{c}{\textbf{Full ($\mathbf{K = 2}$)}} && \multicolumn{3}{c}{\textbf{Full ($\mathbf{K = 3}$)}} && \multicolumn{4}{c}{\textbf{Full ($\mathbf{K = 4}$)}} \\ \cline{5-6} \cline{8-10} \cline{12-15} \addlinespace[0.4ex]
		\textbf{Risk factor} && \textbf{($\mathbf{K = 1}$)} && $\mathbf{j = 1}$ & $\mathbf{j = 2}$ && $\mathbf{j = 1}$ & $\mathbf{j = 2}$ & $\mathbf{j = 3}$ && $\mathbf{j = 1}$ & $\mathbf{j = 2}$ & $\mathbf{j = 3}$ & $\mathbf{j = 4}$ \\ \hline \addlinespace[0.4ex]
		\texttt{(Intercept)} && -3.05$^{*}$ && -5.55$^{*}$ & -1.71$^{*}$ && 7.45{ } & -1.72$^{*}$ & -4.97$^{*}$ && -7.83{ } & -4.01$^{*}$ & -1.44$^{*}$ & -9.43{ } \\
		\texttt{Cust\_Age} && -0.01$^{*}$ && 0.03$^{*}$ & -0.03$^{*}$ && -1.83{ } & -0.03$^{*}$ & 0.03$^{*}$ && 0.01$^{*}$ & 0.03$^{*}$ & -0.03$^{*}$ & 0.05$^{*}$ \\
		\texttt{Veh\_Weight} && 0.02$^{*}$ && 0.02$^{*}$ & 0.03$^{*}$ && 0.56{ } & 0.03$^{*}$ & 0.02$^{*}$ && 0.09$^{*}$ & 0.02$^{*}$ & 0.03$^{*}$ & 0.01$^{*}$ \\
		\texttt{Veh\_Mileage} \\
		\quad - \textit{Category 1} && 0.01$^{*}$ && 0.16$^{*}$ & -0.08$^{*}$ && -7.99{ } & -0.05$^{*}$ & 0.11$^{*}$ && -0.35{ } & 0.16$^{*}$ & -0.06$^{*}$ & -1.32{ } \\
		\quad - \textit{Category 2} && 0.07$^{*}$ && 0.32$^{*}$ & -0.06$^{*}$ && -1.93{ } & -0.06$^{*}$ & 0.30$^{*}$ && 0.73{ } & 0.29$^{*}$ & -0.12$^{*}$ & 0.78{ } \\
		\texttt{Veh\_FuelType} \\
		\quad - \textit{Category 1} && 0.24$^{*}$ && 0.05$^{*}$ & 0.35$^{*}$ && -3.47{ } & 0.36$^{*}$ & 0.02$^{*}$ && -0.07{ } & 0.02$^{*}$ & 0.38$^{*}$ & 1.11{ } \\
		\quad - \textit{Category 2} && 0.30$^{*}$ && 0.34$^{*}$ & 0.29$^{*}$ && -0.10{ } & 0.33$^{*}$ & 0.31$^{*}$ && -0.14{ } & 0.22$^{*}$ & 0.27$^{*}$ & 3.01{ } \\
		\quad - \textit{Category 3} && -0.09{ } && 0.14{ } & -0.43{ } && {\small$<$}0.00{ } & -0.33{ } & 0.14{ } && 4.51{ } & -10.33{ } & -9.93{ } & -6.09{ } \\
		\quad - \textit{Category 4} && 0.18{ } && 0.87{ } & -1.09{ } && {\small$<$}0.00{ } & 0.02{ } & 0.43{ } && 2.93{ } & 0.50{ } & -9.26{ } & 3.98{ } \\
		\quad - \textit{Category 5} && -32.96$^{*}$ && -32.96{ } & -32.96{ } && {\small$<$}0.00{ } & -7.64{ } & -8.40{ } && -3.37{ } & -6.38{ } & -8.38{ } & -2.36{ } \\
		\quad - \textit{Category 6} && -33.26$^{*}$ && -33.26{ } & -33.26{ } && {\small$<$}0.00{ } & -5.69{ } & -5.78{ } && -2.02{ } & -5.12{ } & -6.46{ } & -0.90{ } \\
		\texttt{Veh\_BodyDoors} \\
		\quad - \textit{Category 1} && 0.01$^{*}$ && 0.24$^{*}$ & -0.09$^{*}$ && -2.58{ } & -0.09$^{*}$ & 0.23$^{*}$ && -1.39{ } & 0.25$^{*}$ & -0.06$^{*}$ & -0.62{ } \\
		\quad - \textit{Category 2} && 0.08$^{*}$ && 0.66$^{*}$ & -0.23$^{*}$ && -2.36{ } & -0.20$^{*}$ & 0.64$^{*}$ && -0.74{ } & 0.63$^{*}$ & -0.19$^{*}$ & -0.06{ } \\
		\quad - \textit{Category 3} && 0.18$^{*}$ && 0.49$^{*}$ & -0.03$^{*}$ && -14.49{ } & -0.03$^{*}$ & 0.49$^{*}$ && -0.82{ } & 0.47$^{*}$ & -0.02$^{*}$ & 0.21{ } \\
		\quad - \textit{Category 4} && 0.03$^{*}$ && -0.02$^{*}$ & 0.09$^{*}$ && -7.46{ } & 0.09$^{*}$ & -0.03$^{*}$ && -1.10{ } & -0.03$^{*}$ & 0.12$^{*}$ & 0.30{ } \\
		\quad - \textit{Category 5} && -0.24$^{*}$ && 0.05$^{*}$ & -0.33$^{*}$ && -0.14{ } & -0.34$^{*}$ & 0.07$^{*}$ && 0.79{ } & -0.21$^{*}$ & -0.36$^{*}$ & 1.34{ } \\
		\quad - \textit{Category 6} && -0.30$^{*}$ && 0.12$^{*}$ & -0.52$^{*}$ && -0.01{ } & -0.46$^{*}$ & 0.05$^{*}$ && -3.30{ } & 0.17$^{*}$ & -0.59$^{*}$ & 0.98{ } \\
		\quad - \textit{Category 7} && -0.10$^{*}$ && 0.45$^{*}$ & -0.46$^{*}$ && -0.49{ } & -0.47$^{*}$ & 0.48$^{*}$ && -0.26{ } & 0.44$^{*}$ & -0.54$^{*}$ & 0.84{ } \\
		\quad - \textit{Category 8} && -0.08$^{*}$ && -0.29$^{*}$ & 0.09$^{*}$ && -0.20{ } & 0.11$^{*}$ & -0.31$^{*}$ && -1.36{ } & -0.27$^{*}$ & 0.14$^{*}$ & 0.55{ } \\
		\quad - \textit{Category 9} && -0.24$^{*}$ && -0.29{ } & -0.17$^{*}$ && -2.92{ } & -0.28$^{*}$ & -0.04$^{*}$ && -0.10{ } & -0.08$^{*}$ & -0.33$^{*}$ & 1.50{ } \\
		\texttt{Veh\_CatValue} && {\small$<$}0.00$^{*}$ && {\small$<$}0.00$^{*}$ & {\small$<$}0.00$^{*}$ && -3.72{ } & {\small$<$}0.00$^{*}$ & {\small$<$}0.00$^{*}$ && 0.01$^{*}$ & {\small$<$}0.00$^{*}$ & {\small$<$}0.00$^{*}$ & -0.01$^{*}$ \\
		\texttt{Veh\_PowerWeight}\hspace{-20pt} && -0.01$^{*}$ && -0.01$^{*}$ & -0.01$^{*}$ && 0.17{ } & -0.01$^{*}$ & -0.01$^{*}$ && -0.11$^{*}$ & -0.01$^{*}$ & -0.01$^{*}$ & -0.02$^{*}$ \\
		\texttt{Veh\_Age} && -0.01$^{*}$ && -0.04$^{*}$ & 0.01$^{*}$ && -1.85{ } & {\small$<$}0.00$^{*}$ & -0.03$^{*}$ && -0.03$^{*}$ & -0.03$^{*}$ & 0.01$^{*}$ & -0.04$^{*}$ \\
		\texttt{Veh\_Region} \\
		\quad - \textit{Category 1} && 0.08$^{*}$ && 0.17$^{*}$ & 0.04$^{*}$ && -1.26{ } & 0.09$^{*}$ & 0.07$^{*}$ && 1.94{ } & 0.02$^{*}$ & 0.04$^{*}$ & 0.63{ } \\
		\quad - \textit{Category 2} && 0.10$^{*}$ && 0.02$^{*}$ & 0.14$^{*}$ && -6.26{ } & 0.16$^{*}$ & -0.03$^{*}$ && 1.77{ } & -0.10$^{*}$ & 0.17$^{*}$ & 0.18{ } \\
		\quad - \textit{Category 3} && 0.04$^{*}$ && 0.15$^{*}$ & -0.03$^{*}$ && -0.76{ } & 0.02$^{*}$ & 0.04$^{*}$ && -0.48{ } & {\small$<$}0.00$^{*}$ & 0.02$^{*}$ & 0.62{ } \\
		\quad - \textit{Category 4} && 0.10$^{*}$ && 0.47$^{*}$ & -0.15$^{*}$ && 20.42{ } & -0.18$^{*}$ & 0.44$^{*}$ && 1.34{ } & 0.36$^{*}$ & -0.16$^{*}$ & -1.68{ } \\
		\quad - \textit{Category 5} && 0.02$^{*}$ && 0.09$^{*}$ & -0.03$^{*}$ && -0.80{ } & 0.02$^{*}$ & -0.01$^{*}$ && 1.02{ } & -0.11$^{*}$ & 0.06$^{*}$ & -1.10{ } \\
		\quad - \textit{Category 6} && -0.01$^{*}$ && -0.07$^{*}$ & 0.03$^{*}$ && -0.64{ } & 0.03$^{*}$ & -0.07$^{*}$ && 0.09{ } & -0.15$^{*}$ & 0.07$^{*}$ & -0.69{ } \\
		\quad - \textit{Category 7} && 0.11$^{*}$ && -0.15$^{*}$ & 0.22$^{*}$ && -1.09{ } & 0.26$^{*}$ & -0.22$^{*}$ && -0.30{ } & -0.22$^{*}$ & 0.25$^{*}$ & 1.18{ } \\
		\quad - \textit{Category 8} && -0.08$^{*}$ && -0.32$^{*}$ & 0.02$^{*}$ && -0.68{ } & 0.04$^{*}$ & -0.34$^{*}$ && 1.45{ } & -0.41$^{*}$ & 0.05$^{*}$ & 0.44{ } \\
		\quad - \textit{Category 9} && 0.03$^{*}$ && 0.19$^{*}$ & -0.05$^{*}$ && -1.01{ } & -0.01$^{*}$ & 0.08$^{*}$ && 0.32{ } & 0.09$^{*}$ & -0.03$^{*}$ & 0.34{ } \\
		\texttt{Cust\_Residence} \\
		\quad - \textit{Category 1} && 0.33$^{*}$ && 0.26$^{*}$ & 0.39$^{*}$ && -0.81{ } & 0.41$^{*}$ & 0.25$^{*}$ && -0.32{ } & 0.28$^{*}$ & 0.39$^{*}$ & 2.25{ } \\
		\quad - \textit{Category 2} && 0.14$^{*}$ && -0.02$^{*}$ & 0.24$^{*}$ && 15.27{ } & 0.22$^{*}$ & 0.02$^{*}$ && -1.57{ } & 0.06$^{*}$ & 0.23$^{*}$ & 0.72{ } \\
		\quad - \textit{Category 3} && 0.37$^{*}$ && 0.22$^{*}$ & 0.45$^{*}$ && -1.09{ } & 0.47$^{*}$ & 0.20$^{*}$ && -0.50{ } & 0.21$^{*}$ & 0.47$^{*}$ & 2.03{ } \\
		\quad - \textit{Category 4} && -0.08$^{*}$ && -0.08$^{*}$ & -0.13$^{*}$ && -0.73{ } & -0.12$^{*}$ & -0.08$^{*}$ && -0.57{ } & -0.05$^{*}$ & -0.14$^{*}$ & 0.58{ } \\
		\quad - \textit{Category 5} && -0.15$^{*}$ && -0.34$^{*}$ & -0.06$^{*}$ && -0.26{ } & -0.06$^{*}$ & -0.33$^{*}$ && 0.03{ } & -0.42$^{*}$ & -0.02$^{*}$ & -0.05{ } \\
		\quad - \textit{Category 6} && 0.63$^{*}$ && 0.15$^{*}$ & 0.82$^{*}$ && -0.72{ } & 0.85$^{*}$ & 0.09$^{*}$ && -0.74{ } & 0.23$^{*}$ & 0.85$^{*}$ & -0.06{ } \\
		\quad - \textit{Category 7} && 0.92$^{*}$ && 0.19{ } & 1.14$^{*}$ && -0.12{ } & 1.20$^{*}$ & {\small$<$}0.00{ } && 0.88{ } & -0.17{ } & 1.20$^{*}$ & 3.29{ } \\
		\quad - \textit{Category 8} && -0.73{ } && -0.13{ } & -2.01{ } && {\small$<$}0.00{ } & -2.02{ } & -0.15{ } && 2.66{ } & -1.85{ } & -1.89{ } & -7.21{ } \\
		\quad - \textit{Category 9} && -33.34$^{*}$ && -33.34{ } & -33.34{ } && {\small$<$}0.00{ } & -6.13{ } & -6.49{ } && -2.62{ } & -5.11{ } & -6.68{ } & -0.67{ } \\
		\bottomrule
	\end{tabularx}}}
\end{table}

\begin{table}[ht!]
    \caption{Estimated effects for the risk factors in \Cref{TA.1} in the severity component of the independent model ($K = 1$) and each full HMM in MTPL insurance. Estimates that are statistically significant at a significance level of $5\%$ or less are marked by an asterisk.}
    \label{TC.11}\vspace{-3pt}\vspace{-3pt}
    \centerline{\scalebox{0.82}{\begin{tabularx}{1.22\textwidth}{l p{0.0em} r p{0.05em} rr p{0.05em} rrr p{0.05em} rrrr}
		\toprule \addlinespace[1ex] \vspace{1pt}
		&& \textbf{Indep.} && \multicolumn{2}{c}{\textbf{Full ($\mathbf{K = 2}$)}} && \multicolumn{3}{c}{\textbf{Full ($\mathbf{K = 3}$)}} && \multicolumn{4}{c}{\textbf{Full ($\mathbf{K = 4}$)}} \\ \cline{5-6} \cline{8-10} \cline{12-15} \addlinespace[0.4ex]
		\textbf{Risk factor} && \textbf{($\mathbf{K = 1}$)} && $\mathbf{j = 1}$ & $\mathbf{j = 2}$ && $\mathbf{j = 1}$ & $\mathbf{j = 2}$ & $\mathbf{j = 3}$ && $\mathbf{j = 1}$ & $\mathbf{j = 2}$ & $\mathbf{j = 3}$ & $\mathbf{j = 4}$ \\ \hline \addlinespace[0.4ex]
		$\mathit{\varphi}^{(j)}$ && {\small$<$}0.00$^{*}$ && {\small$<$}0.00$^{*}$ & 38.98{ } && 71.73{ } & 21.22{ } & {\small$<$}0.00$^{*}$ && 65.36{ } & {\small$<$}0.00$^{*}$ & 18.53{ } & 62.9{ } \\
		\texttt{Cust\_Age} && 7.78$^{*}$ && 6.34$^{*}$ & 7.85$^{*}$ && 0.07{ } & 7.81$^{*}$ & 6.33$^{*}$ && 2.57$^{*}$ & 6.53$^{*}$ & 7.76$^{*}$ & 6.51$^{*}$ \\
		\texttt{Veh\_Weight} && {\small$<$}0.00$^{*}$ && 0.01$^{*}$ & {\small$<$}0.00$^{*}$ && 0.08{ } & {\small$<$}0.00$^{*}$ & 0.01$^{*}$ && 0.01$^{*}$ & 0.01$^{*}$ & {\small$<$}0.00$^{*}$ & -0.01$^{*}$ \\
		\texttt{Veh\_Mileage} \\
		\quad - \textit{Category 1} && {\small$<$}0.00$^{*}$ && {\small$<$}0.00$^{*}$ & 0.01$^{*}$ && 0.14{ } & 0.01$^{*}$ & {\small$<$}0.00$^{*}$ && {\small$<$}0.00$^{*}$ & {\small$<$}0.00$^{*}$ & 0.01$^{*}$ & 0.05$^{*}$ \\
		\quad - \textit{Category 2} && -0.03$^{*}$ && 0.01$^{*}$ & 0.02$^{*}$ && {\small$<$}0.00{ } & {\small$<$}0.00$^{*}$ & -0.01$^{*}$ && 0.02$^{*}$ & -0.02$^{*}$ & 0.03$^{*}$ & -0.25$^{*}$ \\
		\texttt{Veh\_FuelType} \\
		\quad - \textit{Category 1} && -0.02$^{*}$ && -0.03$^{*}$ & 0.12$^{*}$ && {\small$<$}0.00{ } & 0.11$^{*}$ & -0.04$^{*}$ && -0.02$^{*}$ & 0.03$^{*}$ & 0.09$^{*}$ & 0.05$^{*}$ \\
		\quad - \textit{Category 2} && 0.11$^{*}$ && 0.01$^{*}$ & 0.10$^{*}$ && -0.02{ } & 0.09$^{*}$ & 0.01$^{*}$ && -0.06$^{*}$ & -0.01$^{*}$ & 0.07$^{*}$ & 0.12$^{*}$ \\
		\quad - \textit{Category 3} && -0.04$^{*}$ && 0.01$^{*}$ & 0.01$^{*}$ && {\small$<$}0.00{ } & {\small$<$}0.00$^{*}$ & -0.02$^{*}$ && -1.99$^{*}$ & -0.02$^{*}$ & -0.09$^{*}$ & -0.73$^{*}$ \\
		\quad - \textit{Category 4} && -0.34{ } && -0.34{ } & 3.11{ } && {\small$<$}0.00{ } & -0.34{ } & -0.41{ } && 0.78$^{*}$ & -0.13{ } & 3.10{ } & 0.10$^{*}$ \\
		\texttt{Veh\_BodyDoors} \\
		\quad - \textit{Category 1} && 0.30{ } && 0.50{ } & 0.06{ } && {\small$<$}0.00{ } & 0.06{ } & 0.52{ } && 2.86$^{*}$ & 0.54{ } & -0.01{ } & {\small$<$}0.00{ } \\
		\quad - \textit{Category 2} && 0.02$^{*}$ && -0.02$^{*}$ & -0.04$^{*}$ && -0.07{ } & -0.05$^{*}$ & -0.03$^{*}$ && 0.31$^{*}$ & -0.08$^{*}$ & -0.06$^{*}$ & 1.19$^{*}$ \\
		\quad - \textit{Category 3} && -0.03$^{*}$ && 0.06$^{*}$ & 0.02$^{*}$ && 0.02{ } & {\small$<$}0.00$^{*}$ & 0.07$^{*}$ && -0.94$^{*}$ & 0.02$^{*}$ & -0.01$^{*}$ & 0.24$^{*}$ \\
		\quad - \textit{Category 4} && 0.05$^{*}$ && 0.09$^{*}$ & 0.19$^{*}$ && {\small$<$}0.00{ } & 0.19$^{*}$ & 0.10$^{*}$ && 0.78$^{*}$ & -0.09$^{*}$ & 0.20$^{*}$ & -1.46$^{*}$ \\
		\quad - \textit{Category 5} && -0.01$^{*}$ && 0.10$^{*}$ & 0.06$^{*}$ && -0.07{ } & 0.07$^{*}$ & 0.10$^{*}$ && 0.21$^{*}$ & 0.04$^{*}$ & 0.07$^{*}$ & 0.08$^{*}$ \\
		\quad - \textit{Category 6} && -0.09$^{*}$ && 0.13$^{*}$ & 0.05$^{*}$ && {\small$<$}0.00{ } & 0.03$^{*}$ & 0.13$^{*}$ && -0.38$^{*}$ & 0.07$^{*}$ & 0.01$^{*}$ & -0.53$^{*}$ \\
		\quad - \textit{Category 7} && -0.01$^{*}$ && -0.06$^{*}$ & -0.06$^{*}$ && 0.01{ } & -0.08$^{*}$ & -0.06$^{*}$ && -0.39$^{*}$ & -0.06$^{*}$ & -0.12$^{*}$ & 0.46$^{*}$ \\
		\quad - \textit{Category 8} && 0.11$^{*}$ && 0.30$^{*}$ & 0.11$^{*}$ && {\small$<$}0.00{ } & 0.09$^{*}$ & 0.29$^{*}$ && -1.52$^{*}$ & 0.28$^{*}$ & 0.05$^{*}$ & -0.55$^{*}$ \\
		\quad - \textit{Category 9} && -0.14$^{*}$ && -0.03$^{*}$ & 0.39$^{*}$ && {\small$<$}0.00{ } & 0.38$^{*}$ & {\small$<$}0.00$^{*}$ && -0.60$^{*}$ & -0.01$^{*}$ & 0.36$^{*}$ & 0.62$^{*}$ \\
		\texttt{Veh\_CatValue} && 0.15$^{*}$ && 0.49$^{*}$ & -0.10$^{*}$ && {\small$<$}0.00{ } & -0.09$^{*}$ & 0.46$^{*}$ && 0.46$^{*}$ & 0.35$^{*}$ & -0.10$^{*}$ & -0.79$^{*}$ \\
		\texttt{Veh\_PowerWeight}\hspace{-20pt} && {\small$<$}0.00$^{*}$ && {\small$<$}0.00$^{*}$ & {\small$<$}0.00$^{*}$ && {\small$<$}0.00{ } & {\small$<$}0.00$^{*}$ & {\small$<$}0.00$^{*}$ && 0.01$^{*}$ & {\small$<$}0.00$^{*}$ & {\small$<$}0.00$^{*}$ & {\small$<$}0.00$^{*}$ \\
		\texttt{Veh\_Age} && {\small$<$}0.00$^{*}$ && -0.01$^{*}$ & 0.01$^{*}$ && -0.03{ } & 0.01$^{*}$ & -0.01$^{*}$ && -0.05$^{*}$ & -0.01$^{*}$ & 0.01$^{*}$ & 0.04$^{*}$ \\
		\texttt{Veh\_Region} \\
		\quad - \textit{Category 1} && 0.01$^{*}$ && {\small$<$}0.00$^{*}$ & 0.01$^{*}$ && 0.02{ } & 0.01$^{*}$ & {\small$<$}0.00$^{*}$ && 0.10$^{*}$ & {\small$<$}0.00$^{*}$ & 0.01$^{*}$ & 0.06$^{*}$ \\
		\quad - \textit{Category 2} && 0.01$^{*}$ && -0.04$^{*}$ & 0.05$^{*}$ && 0.05{ } & 0.06$^{*}$ & -0.04$^{*}$ && 1.19$^{*}$ & -0.04$^{*}$ & 0.07$^{*}$ & 1.16$^{*}$ \\
		\quad - \textit{Category 3} && {\small$<$}0.00$^{*}$ && -0.06$^{*}$ & {\small$<$}0.00$^{*}$ && 0.21{ } & 0.03$^{*}$ & -0.04$^{*}$ && 0.82$^{*}$ & -0.04$^{*}$ & 0.05$^{*}$ & 1.31$^{*}$ \\
		\quad - \textit{Category 4} && 0.01$^{*}$ && -0.06$^{*}$ & -0.02$^{*}$ && {\small$<$}0.00{ } & {\small$<$}0.00$^{*}$ & -0.06$^{*}$ && 1.02$^{*}$ & 0.08$^{*}$ & -0.01$^{*}$ & 1.65$^{*}$ \\
		\quad - \textit{Category 5} && 0.03$^{*}$ && 0.03$^{*}$ & 0.03$^{*}$ && -0.10{ } & 0.03$^{*}$ & 0.02$^{*}$ && 1.90$^{*}$ & 0.03$^{*}$ & 0.06$^{*}$ & 0.85$^{*}$ \\
		\quad - \textit{Category 6} && -0.01$^{*}$ && {\small$<$}0.00$^{*}$ & 0.12$^{*}$ && -0.21{ } & 0.16$^{*}$ & 0.03$^{*}$ && 1.46$^{*}$ & 0.01$^{*}$ & 0.19$^{*}$ & 0.17$^{*}$ \\
		\quad - \textit{Category 7} && {\small$<$}0.00$^{*}$ && -0.21$^{*}$ & 0.06$^{*}$ && 0.04{ } & 0.07$^{*}$ & -0.21$^{*}$ && 0.20$^{*}$ & -0.17$^{*}$ & 0.07$^{*}$ & 0.52$^{*}$ \\
		\quad - \textit{Category 8} && -0.02$^{*}$ && -0.10$^{*}$ & -0.02$^{*}$ && 0.06{ } & {\small$<$}0.00$^{*}$ & -0.07$^{*}$ && 0.26$^{*}$ & -0.07$^{*}$ & 0.02$^{*}$ & 0.50$^{*}$ \\
		\quad - \textit{Category 9} && -0.08$^{*}$ && -0.14$^{*}$ & 0.02$^{*}$ && 0.05{ } & 0.02$^{*}$ & -0.15$^{*}$ && 0.79$^{*}$ & -0.12$^{*}$ & 0.05$^{*}$ & 0.91$^{*}$ \\
		\texttt{Cust\_Residence} \\
		\quad - \textit{Category 1} && -0.04$^{*}$ && -0.20$^{*}$ & -0.02$^{*}$ && 0.04{ } & {\small$<$}0.00$^{*}$ & -0.19$^{*}$ && 0.52$^{*}$ & -0.08$^{*}$ & {\small$<$}0.00$^{*}$ & -0.83$^{*}$ \\
		\quad - \textit{Category 2} && {\small$<$}0.00$^{*}$ && -0.09$^{*}$ & 0.05$^{*}$ && 0.05{ } & 0.04$^{*}$ & -0.09$^{*}$ && 0.20$^{*}$ & -0.12$^{*}$ & 0.02$^{*}$ & 0.10$^{*}$ \\
		\quad - \textit{Category 3} && -0.01$^{*}$ && 0.04$^{*}$ & -0.14$^{*}$ && 0.01{ } & -0.13$^{*}$ & 0.04$^{*}$ && 0.04$^{*}$ & 0.02$^{*}$ & -0.15$^{*}$ & 0.99$^{*}$ \\
		\quad - \textit{Category 4} && -0.02$^{*}$ && {\small$<$}0.00$^{*}$ & -0.12$^{*}$ && 0.07{ } & -0.12$^{*}$ & 0.01$^{*}$ && -0.45$^{*}$ & -0.03$^{*}$ & -0.14$^{*}$ & 0.65$^{*}$ \\
		\quad - \textit{Category 5} && -0.02$^{*}$ && -0.09$^{*}$ & -0.06$^{*}$ && 0.01{ } & -0.07$^{*}$ & -0.10$^{*}$ && -0.37$^{*}$ & -0.11$^{*}$ & -0.10$^{*}$ & 0.40$^{*}$ \\
		\quad - \textit{Category 6} && 0.02$^{*}$ && -0.04$^{*}$ & -0.10$^{*}$ && 0.06{ } & -0.11$^{*}$ & -0.06$^{*}$ && 0.95$^{*}$ & -0.08$^{*}$ & -0.10$^{*}$ & -3.00$^{*}$ \\
		\quad - \textit{Category 7} && -0.07$^{*}$ && -0.12$^{*}$ & -0.13$^{*}$ && -0.05{ } & -0.14$^{*}$ & -0.11$^{*}$ && 0.05$^{*}$ & -0.10$^{*}$ & -0.17$^{*}$ & -0.26$^{*}$ \\
		\quad - \textit{Category 8} && 0.37{ } && 0.58{ } & 2.11{ } && {\small$<$}0.00{ } & 2.11{ } & 0.58{ } && 1.73$^{*}$ & 1.10{ } & 2.09{ } & {\small$<$}0.00{ } \\
		\bottomrule
	\end{tabularx}}}
\end{table} 
}



\end{document}